\pdfoutput=1
\documentclass[11pt,titlepage]{article}

\let\oldmarginpar\marginpar
\renewcommand\marginpar[1]{\-\oldmarginpar[\raggedleft\footnotesize #1]%
{\raggedright\footnotesize #1}}

\makeatletter
\def\@xobeysp{ }
\makeatother

\usepackage[vscale=0.76,hscale=0.76]{geometry}

\usepackage[hyperref,table,fixpdftex,svgnames]{xcolor}

\usepackage{graphicx}
\usepackage{comment}
\usepackage{fancyvrb}
\usepackage{longtable}
\usepackage{rotating}
\usepackage{pdflscape} 
\usepackage{subfig}
\usepackage{attrib}             
\usepackage{currfile}
\usepackage[normalem]{ulem}
\usepackage{bbding}

\usepackage{appendix}
\let\oldapp\appendix
\renewcommand{\appendix}{\oldapp\appendixpage\addappheadtotoc}

\DeclareGraphicsExtensions{.png,.pdf}

\usepackage[margin=10pt,font=small,labelfont=bf,labelsep=period]{caption}

\usepackage[T1]{fontenc}
\usepackage{lmodern}
\usepackage{amsmath}
\usepackage{amsfonts}
\usepackage{fixmath}

\usepackage{fancyhdr}
\fancypagestyle{plain}{%
\fancyhf{}
\lfoot{\today}
\rfoot{\thepage}

}

\usepackage[american]{babel}
\usepackage{csquotes}
\usepackage[style=authoryear-comp,mergedate=maximum,sorting=nyt,maxbibnames=12,maxcitenames=2,natbib=true,firstinits=true,backend=bibtex,isbn=false,mincrossrefs=99
]{biblatex}

\DeclareFieldFormat[article,incollection,unpublished]{title}{#1}
\DeclareFieldFormat[thesis]{title}{\mkbibemph{#1}} 
\DeclareBibliographyDriver{article}{%
  \usebibmacro{bibindex}%
  \usebibmacro{begentry}%
  \usebibmacro{author/translator+others}%
  \setunit{\labelnamepunct}\newblock
  \usebibmacro{title}%
  \newunit
  \printlist{language}%
  \newunit\newblock
  \usebibmacro{byauthor}%
  \newunit\newblock
  \usebibmacro{bytranslator+others}%
  \newunit\newblock
  \printfield{version}%
  \newunit\newblock%
  \usebibmacro{journal+issuetitle}%
  \newunit
  \usebibmacro{byeditor+others}%
  \newunit
  \usebibmacro{note+pages}%
  \newunit\newblock
  \iftoggle{bbx:isbn}
    {\printfield{issn}}
    {}%
  \newunit\newblock
  \usebibmacro{doi+eprint+url}%
  \newunit\newblock
  \usebibmacro{addendum+pubstate}%
  \setunit{\bibpagerefpunct}\newblock
  \usebibmacro{pageref}%
  \usebibmacro{finentry}}

\DeclareNameAlias{sortname}{last-first}

\usepackage[sanitize=none,shortcuts,nonumberlist,numberedsection=autolabel,hyperfirst=false]{glossaries}
\glsdisablehyper

\newcommand{\GinTextWidth}{\setkeys{Gin}{keepaspectratio=true,width=\textwidth}}

\newcommand{\GinKARWidth}[1]{\setkeys{Gin}{keepaspectratio=true,width=#1}}

\usepackage{enumitem}

\setlist{noitemsep}

\setdescription{labelindent=\parindent}

\usepackage{amsthm}

\theoremstyle{plain}
\theoremstyle{plain}
\theoremstyle{plain}
\theoremstyle{plain}
\theoremstyle{definition}
\theoremstyle{remark}
\theoremstyle{definition}

\theoremstyle{plain}\newtheorem*{theorem*}{Theorem}
\theoremstyle{plain}\newtheorem*{proposition*}{Proposition}
\theoremstyle{plain}\newtheorem*{result*}{Result}
\theoremstyle{plain}\newtheorem*{corollary*}{Corollary}
\theoremstyle{definition}\newtheorem*{definition*}{Definition}
\theoremstyle{remark}\newtheorem*{remark*}{Remark}
\theoremstyle{definition}\newtheorem*{example*}{Example}

\newcommand{\cond}{\, \vert\, }

\renewcommand{\vec}[1]{\boldsymbol{#1}}
\newcommand{\mat}[1]{\pmb{\mathrm{#1}}}

\DeclareMathOperator{\logit}{logit}

\DeclareMathOperator{\dNorm}{Normal}

\DeclareMathOperator{\dInvGamma}{InvGamma}

\newcommand{\ltp}[3]{{_{#1}}{#2}_{#3}}

\newcommand{\poscite}[1]{\citeauthor{#1}'s \citeyearpar{#1}}

\newlength{\alertboxtopsep}
\newlength{\alertboxbotsep}
\usepackage{hyperref}

\usepackage{placeins}
\usepackage[nogin,noae]{Sweave}

\fancypagestyle{plain}{%
\fancyhf{}
\cfoot{\thepage}
}
\usepackage[flushmargin,splitrule,multiple]{footmisc}
\AtEveryBibitem{\clearlist{language}} 

\newacronym{CCMPP}{CCMPP}{cohort component model of population projection}
\newacronym{GIP}{GIP}{generalized inverse projection}
\newacronym{SIP}{SIP}{stochastic inverse projection}
\newacronym{TFR}{TFR}{total fertility rate}
\newglossaryentry{LEB}{type=\acronymtype,
name={LEB},
description={life expectancy at birth},
text={LEB},
first={life expectancy at birth (LEB)},
plural={LEBs},
firstplural={life expectancies at birth (LEBs)}}
\newglossaryentry{e0}{type=\acronymtype,
name={e0},
description={life expectancy at birth},
text={$e_0$},
first={life expectancy at birth ($e_0$)},
plural={$e_0$s},
firstplural={life expectancies at birth ($e_0$s)},
long={life expectancy at birth},
short={\protect$e_0\protect$}}
\newglossaryentry{SRTP}{type=\acronymtype,
name={SRTP},
description={sex ratio in the total population},
text={SRTP},
first={sex ratio in the total population (SRTP)},
plural={SRTPs},
firstplural={sex ratios in the total population (SRTPs)},
long={sex ratio in the total population},
short={SRTP}}
\newglossaryentry{SRU5}{type=\acronymtype,
name={SRU5},
description={sex ratio in the population under five},
text={SRU5},
first={sex ratio in the population under five (SRU5)},
plural={SRU5s},
firstplural={sex ratios in the population under five (SRU5s)},
long={sex ratio in the population under five},
short={SRU5}}
\newglossaryentry{SRB}{type=\acronymtype,
name={SRB},
description={sex ratio at birth},
text={SRB},
first={sex ratio at birth (SRB)},
plural={SRBs},
firstplural={sex ratios at birth (SRBs)},
long={sex ratio at birth},
short={SRB}}
\newglossaryentry{SRM}{type=\acronymtype,
name={SRM},
description={sex ratio of mortality},
text={SRM},
first={sex ratio of mortality (SRM)},
plural={SRMs},
firstplural={sex ratios of mortality (SRMs)}}
\newglossaryentry{SRU5MR}{type=\acronymtype,
name={SRU5MR},
description={sex ratio of the under-five mortality rate},
text={SRU5MR},
first={sex ratio of under-five mortality rate (SRU5MR)},
plural={SRU5MRs},
firstplural={sex ratios of under-five mortality rate (SRU5MRs)},
long={sex ratio of under-five mortality rate},
short={SRU5MR}
}
\newglossaryentry{SDe0}{type=\acronymtype,
name={SDe0},
description={sex difference in life expectancy at birth},
text={sex diff.~\protect$e_0\protect$},
first={sex difference in life expectancy at birth (sex diff.~\protect$e_0\protect$)},
plural={sex diffs.~\protect$e_0\protect$},
firstplural={sex differences in life expectancy at birth (sex diff.~\protect$e_0\protect$)},
long={sex difference in life expectancy at birth},
short={sex diff.~\protect$e_0\protect$}}
\newacronym{WFS}{WFS}{World Fertility Survey}
\newacronym{DHS}{DHS}{Demographic and Health Survey}
\newacronym{RHS}{RHS}{Reproductive Health Survey}
\newacronym{MICS}{MICS}{Multiple Indicator Cluster Survey}
\newacronym{vitalrate}{vital rate}{fertility and mortality rate}
\newacronym{Laos}{Laos}{The People's Democratic Republic of Laos}
\newacronym{CEB}{CEB}{children ever born}
\newacronym{CDWest}{CD West}{Coale and Demeny West}
\newacronym{PES}{PES}{post-enumeration survey}
\newacronym{UNSEA}{UNSEA}{United Nations South East Asian}
\newacronym{CME}{CME}{child mortality estimates}
\newacronym{UNICEF}{UNICEF}{the United Nations Children's Fund}
\newacronym{PLT}{PLT}{permanent and long-term migrants}
\newacronym{IMR}{IMR}{infant mortality rate}
\newacronym{U5MR}{U5MR}{under-five mortality rate}

\newacronym{SRS}{SRS}{Sample Registration System} 
\newacronym{NFHS}{NFHS}{National Family Health Survey} 
\newacronym{RCHS}{RCHS}{Reproductive and Child Health Survey} 
\newacronym{DLHS}{DLHS}{District Level Household Survey} 
\newacronym{CPS}{CPS}{Contraceptive Prevalence Survey} 

\newacronym{pdf}{pdf}{probability density function}
\newacronym{MAE}{MAE}{mean absolute error}
\newacronym{MARE}{MARE}{mean absolute relative error}
\newacronym{MSE}{MSE}{mean square error}
\newacronym{RMSE}{RMSE}{root mean square error}
\newacronym{RMSRE}{RMSRE}{root mean square relative error}
\newacronym{MAD}{MAD}{median absolute deviation}
\newacronym{MxAD}{MxAD}{maximum absolute difference}
\newacronym{MARD}{MARD}{mean absolute relative difference}
\newacronym{MxARD}{MxARD}{maximum absolute relative difference}
\newacronym{ARD}{ARD}{absolute relative difference}
\newacronym{ARE}{ARE}{absolute relative error}

\newacronym{MCMC}{MCMC}{Markov chain Monte Carlo}
\newacronym{aMCMC}{an MCMC}{a Markov chain Monte Carlo\protect\glsunset{MCMC}}

\newacronym{M-H}{M-H}{Metropolis-Hastings}
\newacronym{OLS}{OLS}{ordinary least squares}

\newacronym{IHPP}{IHPP}{inhomogeneous Poisson process}

\newacronym{US}{US}{United States}
\newacronym{USCB}{USCB}{United States Census Bureau}

\newacronym{UN}{UN}{United Nations}
\newacronym{UNPD}{UNPD}{United Nations Population Division}
\newacronym{UNFPA}{UNFPA}{United Nations Population Fund}
\newacronym{UNWPP}{UNWPP}{\protect{\emph{United Nations World Population Prospects}}}
\newacronym{WPP}{WPP}{\protect{\emph{World Population Prospects}}}
\newacronym{IGME}{IGME}{Interagency Group for Child Mortality Estimation}
\newacronym{UNIGME}{UN IGME}{United Nations Interagency Group for Child Mortality Estimation}

\newacronym{NRC}{NRC}{National Research Council}

\newacronym{NICHD}{NICHD}{National Institute of Child Health and Human Development}

\newacronym{NSO}{NSO}{National Statistics Office}
\newacronym{SNZ}{SNZ}{Statistics New Zealand}
\newacronym{DCSSL}{DCS Sri Lanka}{Department of Census and Statistics, Sri Lanka}

\newcommand{\mathsrb}{\text{\textit{\footnotesize{SRB}}}}

\usepackage{framed}

\def\keywords#1{{\vskip4pt
\noindent
\hbox to59.5pt{KEY\enspace WORDS:\quad\hss}\vtop{\advance \hsize by -59.5pt
\leftskip=28pt \rightskip=0pt
\noindent\ignorespaces#1\vskip8pt}}}

\title{Bayesian Reconstruction of Two-Sex Populations by Age: Estimating Sex Ratios at Birth and Sex Ratios of Mortality\thanks{\scriptsize{Mark C. Wheldon is Lecturer in the Biostatistics Unit, Faculty of Health and Environmental Sciences, Auckland University of Technology, and Biostatistician, Centre for Clinical Research and Effective Practice, Middlemore Hospital, both in Auckland, New Zealand (e-mail: \url{mwheldon@aut.ac.nz}, web: \url{http://www.aut.ac.nz/~profiles/mwheldon}). Part of this work was done while he was a Ph.D. student in the Department of Statistics at the University of Washington, Seattle, WA, USA. Adrian E. Raftery is Professor of Statistics and Sociology, University of Washington (e-mail: \url{raftery@uw.edu}, web: \url{http://www.stat.washington.edu/~raftery}). Samuel J. Clark is Associate Professor of Sociology, University of Washington, and Researcher at the Institute of Behavioral Science (IBS), University of Colorado at Boulder, CO, USA and at MRC/Wits Rural Public Health and Health Transitions Research Unit (Agincourt), School of Public Health, Faculty of Health Sciences, University of the Witwatersrand, Johannesburg, South Africa (e-mail: \url{samclark@uw.edu}, web: \url{http://www.soc.washington.edu/faculty-details/samclark}). Patrick Gerland is Population Affairs Officer, United Nations Department of Economic and Social Affairs, Population Division, New York, NY, USA (e-mail: \url{gerland@un.org}). This work was supported by grants R01 HD054511, R01 HD070936, K01 HD057246 and 5R24HD042828 from the Eunice Kennedy Shriver National Institute of Child Health and Human Development (NICHD). Wheldon also received partial support from a Shanahan Endowment Fellowship to the Center for Studies in Demography \& Ecology at the University of Washington, and from the Faculty of Health and Environmental Sciences, Auckland University of Technology. Raftery's research was also supported by a Science Foundation Ireland ETS Walton visitor award, grant reference 11/W.1/I2079. The views and opinions expressed in this paper are those of the authors and do not necessarily represent those of the United Nations. This paper has not been formally edited and cleared by the United Nations.}}}
\author{Mark C. Wheldon \\
  Auckland University of Technology \and
  Adrian E. Raftery, Samuel J. Clark \\
  University of Washington \and
  Patrick Gerland \\
  United Nations Population Division}

\hypersetup{pdftitle={Bayesian Reconstruction of Two-Sex Populations by Age:  Estimating Sex Ratios at Birth and Sex Ratios of Mortality},%
pdfauthor={Mark C. Wheldon, Adrian E. Raftery, Samuel J. Clark, and Patrick Gerland},%
pdfcreator={Mark C. Wheldon},%
pdfsubject={Statistics, Applications},%
pdfkeywords={Bayesian hierarchical model, Two-sex model, Population projection, Vital Rates, Sex ratio at birth, Sex ratio of mortality} ,bookmarks,colorlinks,citecolor=DarkBlue,linkcolor=DarkBlue,menucolor=DarkBlue,urlcolor=DarkBlue
}

\begin{document}
\pagenumbering{roman}



\setkeys{Gin}{width=0.6\textwidth}


\DefineVerbatimEnvironment{Sinput}{Verbatim}{xleftmargin=2em}
\DefineVerbatimEnvironment{Soutput}{Verbatim}{xleftmargin=2em}
\DefineVerbatimEnvironment{Scode}{Verbatim}{xleftmargin=2em}
\fvset{listparameters={\setlength{\topsep}{0pt}}}
\renewenvironment{Schunk}{\vspace{\topsep}}{\vspace{\topsep}}




\thispagestyle{plain}
\maketitle

\begin{abstract}
The original version of Bayesian reconstruction \citep{WheldonRafteryClarkEtAl_Reconstructing_JotASA}, a method for estimating age-specific fertility, mortality, migration and population counts of the recent past with uncertainty, produced estimates for female-only populations. Here we show how two-sex populations can be similarly reconstructed and probabilistic estimates of various sex ratio quantities obtained. We demonstrate the method by reconstructing the populations of India from 1971 to 2001, Thailand from 1960 to 2000, and Laos from 1985 to 2005.
We found evidence that in India, \ac{SRB} exceeded its conventional
upper limit of 1.06, and, further, increased over the period of study, with
posterior probability above 0.9. In addition, almost uniquely, we found
evidence that \ac{e0} was lower for females than for males in India
(posterior probability for 1971--1976 %
0.79%
), although there was strong evidence for a narrowing of the gap through to 2001. In both Thailand and Laos, we found strong evidence for the more usual result that \ac{e0} was greater for females and, in Thailand, that the difference increased over the period of study.

  \keywords{Bayesian hierarchical model; Two-sex model; Population projection;\\ Vital Rates; Sex ratio at birth; Sex ratio of mortality}
\end{abstract}

\begin{footnotesize}
\tableofcontents
\end{footnotesize}

\glsresetall

\clearpage

\pagenumbering{arabic}


\section{Introduction}
\label{sec:introduction}


The past, present and future dynamics of human populations at the country level are highly relevant to the work of social scientists in many disciplines as well as planners and evaluators of public policy. These dynamics are driven by population counts, \acp{vitalrate}, and net international migration. Demographers at the \ac{UNPD} are tasked with producing detailed information on these quantities, published biennially in the \ac{WPP} \citep[e.g.,][]{Nations2011World}. Estimates for each country are provided, for periods stretching back from the present to about 1950. Currently, however, \ac{WPP} estimates are not accompanied by any quantitative estimate of uncertainty. Uncertainty should also be measured because the availability, coverage and reliability of data used to derive the estimates differ greatly among countries. Developing countries, in particular, often lack the extensive registration and census-taking systems developed countries maintain, so that estimates in these cases are subject to greater uncertainty. Estimates are not error-free even for developed countries. While estimates of population counts and \acp{vitalrate} are likely to be very accurate, uncertainty about net international migration can be quite substantial. \citet{BeerRaymerErfEtAl_Overcoming_EJoPedD_2010} found this to be the case for Europe.

In response to the need for uncertainty quantification in estimates of the key parameters driving human population dynamics, \citet{WheldonRafteryClarkEtAl_Reconstructing_JotASA} proposed Bayesian population reconstruction (Bayesian reconstruction for short), a method of simultaneously estimating population counts, \acp{vitalrate} and net international migration at the country level, by age, together with uncertainty. The original formulation was able to reconstruct female-only populations. In this article, we describe a major extension to two-sex populations. This allows us to estimate age- and time-specific indicators of fertility, mortality and migration separately for females and males and, importantly, sex ratios of these quantities, all with probabilistic measures of uncertainty. In addition, we also show how Bayesian reconstruction can be used to derive probabilities of change over time in these quantities. To demonstrate the method, we reconstruct the full populations of India from 1971--2001, Thailand from 1960--2000 and Laos from 1985--2005. These countries were selected because, in all cases, the available data are fragmentary, which makes population reconstruction challenging.

Bayesian reconstruction embeds a standard demographic projection model in a hierarchical statistical model. As inputs, it takes bias-reduced initial estimates of age-specific fertility rates, survival proportions (a measure of mortality), net international migration and census-based population counts. Also required is expert opinion about the measurement error of these quantities, informed by data if available. The output is a joint posterior probability distribution on the inputs, allowing all parameters to be estimated simultaneously, together with fully probabilistic posterior estimates of measurement error. \citet{WheldonRafteryClarkEtAl_Reconstructing_JotASA} showed that marginal credible intervals were well calibrated. They demonstrated the method by reconstructing the female population of Burkina Faso from 1960 to 2000. \citet{WheldonRafteryClarkEtAl_Bayesian__PAA2013} extended Bayesian reconstruction to countries with censuses at irregular intervals and showed that it works well across a wide range of data quality contexts by reconstructing the female populations of Laos, Sri Lanka, and New Zealand. Laos is a country with very little vital registration data where population estimation depends largely on surveys, Sri Lanka has some vital registration data, and New Zealand is a country with high-quality registration data. In this paper we focus on countries which lack good vital registration data or for which there are gaps in, or inconsistencies among, the available data sources.

Global \ac{SRTP}, defined as the ratio of the number of males per female, has risen slightly from about 1.00 in 1950 to 1.02 in 2010. There is a great deal of variation among regions, however. For instance, \ac{SRTP} in the more developed regions ranged from 0.91 to 0.95 over this period, while in less developed regions it remained constant at about 1.04. In Eastern Asia, which includes China, and in Southern Asia, which includes India, \acp{SRTP} ranged from 1.05 to 1.06 and from 1.09 to 1.06, respectively \citep{Nations2011World}. \citet{GuilmotoSex_2007} claimed that the population of Asia underwent ``masculinization'' during the latter half of the twentieth century, with one likely consequence being a ``marriage squeeze'' \citep{Guilmoto_Sex_PaDR_2009,Guilmoto_Skewed_D_2012} wherein many males will be unable to find a wife. Imbalances in population sex ratios are caused by imbalances in \acp{SRB} and \acp{SRM} \citep{Guillot_Dynamics_PS_2002}. These quantities have received considerable attention in the literature on the demography of Asia \citep[e.g.][]{Sen_More_NYRoB_1990,Coale_Excess_PaDR_1991,Mayer_Indias_PaDR_1999,Bongaarts_Fertility_PaDR_2001,Bhat_Trail_EaPW_2002,Bhat_Trail_EaPW_2002a,Gupta_Explaining_PaDR_2005,Guilmoto_Sex_PaDR_2009}. \citet{Sawyer_Child_PM_2012} called for further work to quantify uncertainty in estimates of \acp{SRM}. Estimates of the \ac{SRB} are subject to a large amount of uncertainty, especially in India \citep{Bhat_Trail_EaPW_2002,Bhat_Trail_EaPW_2002a,Guillot_Dynamics_PS_2002,Guilmoto_Sex_PaDR_2009}. Here, we respond by quantifying uncertainty in these parameters.

The article is organized as follows. In the remainder of this section we provide some background on existing methods of population reconstruction and the demography of sex ratios in Asia. In Section~\ref{sec:method} we describe the two-sex version of Bayesian reconstruction. In Section~\ref{sec:application} we present results from our case studies of India, Thailand and Laos. We focus mainly on posterior distributions of \ac{TFR}, \acp{SRB} and the sex difference in life expectancy. Certain sex ratios in India are widely believed to be atypical so we devote more attention to this case. Sensitivity to aspects of the prior based primarily on expert opinion is investigated. We end with a discussion in Section~\ref{sec:discussion} which provides further demographic context and an overall conclusion. Selected mathematical derivations, 
further details about data sources and 
results for additional parameters such as sex ratio of under-five mortality rate,
population sex ratios and net international migration, are in the appedices. Bayesian reconstruction is implemented in the ``popReconstruct'' package for the R environment for statistical computing \citep{R2013_R_}.

\subsection[Methods of Population Reconstruction]{Methods of Population Reconstruction}
\label{sec:meth-popul-reconstr}

In human demography, population reconstruction is often referred to simply as ``estimation'' to distinguish it from forecasts of future population counts and \acp{vitalrate}. Instead we use ``reconstruction'' to avoid ambiguity. Nevertheless, use of ``estimation'' agrees with its usage in statistics, namely the estimation of the values of some unknown quantities from data. Reviews of existing methods of population reconstruction are given by  \citet{Oeppen_Back_PS1993}, \citet{BarbiBertinoSonnino_Inverse_2004}, and \citet{WheldonRafteryClarkEtAl_Reconstructing_JotASA}. Many were developed for the purpose of reconstructing populations of the distant past from data on births, deaths and marriages recorded in parish registers \citep[e.g,.][]{WrigleySchofield_Population_1981,BertinoSonnino_Stochastic_IPTOaNA2004,Walters_Fertility__2008}, or in counterfactual exercises to estimate the excess of mortality due to extreme events such as famine or genocide \citep[e.g.,][]{BoyleGrada_Fertility_D1986,DaponteKadaneWolfson_Bayesian_JotASA1997,Heuveline_Between_PS1998,Merli_Mortality_D1998,GoodkindWest_North_PaDR2001} or the number of ``missing women'' due to male-dominated sex ratios in Asia \citep{Sen_More_NYRoB_1990,Coale_Excess_PaDR_1991,Gupta_Explaining_PaDR_2005}. Purely deterministic reconstruction methods used in some of these studies include ``inverse projection'' \citep{Lee_Econometric_1971,Lee_Estimating_PS1974}, ``back projection'' \citep{WrigleySchofield_Population_1981} and ``generalized back projection'' \citep{Oeppen_Generalized_OaNMiHD1993}. \citet{BertinoSonnino_Stochastic_IPTOaNA2004} proposed ``stochastic inverse projection''. This is a non-deterministic method but the only form of uncertainty is that which comes from treating birth and death as stochastic processes at the individual level. Counts of births and deaths are assumed to be known without error and age patterns are fixed. In the cases we treat, accurate data on births and deaths of the parish register kind are often unavailable and uncertainty due to stochastic vital rates is likely to be small relative to uncertainty due to measurement error \citep{Pollard_Note_B1968,Lee_Reflections_BarbiSonino_2004,Cohen_Stochastic_EoSS2006}. Moreover, it is designed to work with the kind of data commonly available for developing countries and does not rely on the existence of detailed births and deaths registers, although this information can be used when available \citep{WheldonRafteryClarkEtAl_Bayesian__PAA2013}.

\citet{DaponteKadaneWolfson_Bayesian_JotASA1997} took a fully Bayesian approach to constructing a counterfactual history of the Iraqi Kurdish population from 1977 to 1990. They constructed prior distributions for fertility and mortality rates using survey data and expert opinion about uncertainty based on historical information and knowledge of demographic processes. Measurement error in the available, fragmentary data was accounted for. However, there were some restrictions such as holding the age pattern of fertility fixed and allowing for mortality variation only through infant mortality. Rural to urban migration was accounted for by treating these populations separately; international migration was assumed to be negligible. Bayesian reconstruction is similar, but no model age patterns are assumed to hold and international migration is explicitly estimated along with fertility, mortality and population counts.

Most methods of population reconstruction in demography employ the \ac{CCMPP} in some form. Population projection uses \acp{vitalrate} and migration to project a set of age-specific population counts in the baseline year, denoted $t_0$, forward in time to the end year, denoted $T$. In its simplest form, the population in year $t+\delta$, $t_0 \leq t \leq T-\delta$, equals the population in year $t$ plus the intervening births and net migration, minus the intervening deaths \citep[e.g.,][]{Whelpton_Empirical_JotASA1936,PrestonHeuvelineGuillot2001}. This is known as the demographic balancing equation. Population projection is distinct from population forecasting since it merely entails evolving a population forward in time from some given baseline under assumptions about prevailing \acp{vitalrate} and migration \citep{KeyfitzCaswell_Applied__2005}. The period of projection may be in the future or in the past. \citet{WheldonRafteryClarkEtAl_Reconstructing_JotASA,WheldonRafteryClarkEtAl_Bayesian__PAA2013} employed single-sex population projection in this way to reconstruct female-only populations taking account of measurement error. In this article we show how two-sex population projection can be used to reconstruct full populations, thereby providing estimates of population sex ratios, \acp{SRB} and \acp{SRM} simultaneously while accounting for measurement error.

\subsection{Estimating Sex Ratios}
\label{sec:sex-ratios-asian}

Methods of reporting sex ratios are not standardized. Here we adopt the convention that all ratios are ``male-per-female''; in the Indian literature the inverse is more common. Hence, \ac{SRTP} is the total number of males per female in the population and \ac{SRB} is the number of male births per female birth. The \ac{SRM} can be expressed using various mortality indicators. We will use the \ac{U5MR} exclusively (see the Appendix for a formal definition). A low \ac{SRM} means that mortality is lower among males than among females. All-age mortality is summarized by \acrlong{e0}\glsunset{e0}, for which the standard demographic abbreviation is \acrshort{e0}. Comparison of \ac{e0} by sex is more commonly done using the difference than the ratio and we adopt that convention here. Male \ac{e0} is subtracted from female \ac{e0} to obtain the difference.

Under typical conditions, \acp{SRB} for most countries are in the range 1.04--1.06 \citep{Fund2010UNFPA}. Estimates of \ac{SRB} in some regions in Asia are higher; the National Family Health Survey in India estimated \ac{SRB} over 2000--2006 to be 1.09, for example \citep{Guilmoto_Sex_PaDR_2009}. For almost all countries, \ac{e0} is higher for females than males. This is thought to be due to a range of biological and environmental factors, with the relative contribution of each class of factor varying among countries \citep{Waldron_What_PBotUN_1985,Waldron_What_PBotUN_1985}. Age-specific \acp{SRM} are more variable as they are affected by sex-specific causes of death such as those associated with child birth. The preferred way of estimating \acp{SRB} and \acp{SRM} at the national level is from counts of births and deaths recorded in official registers (vital registration) together with total population counts from censuses. In many countries where such registers are not kept, surveys such as the Demographic and Health Surveys and World Fertility Surveys must be used. These typically ask a sample of women about their birth histories. Full birth histories collect information about the times of each birth and, if the child subsequently died, the time of the death. Summary birth histories ask only about the total number of births and child deaths the respondent has ever experienced \citep{United_Manual_1983,PrestonHeuvelineGuillot2001}.

Estimates based on both vital registration and surveys are susceptible to systematic biases and non-systematic measurement error. Counts of births or deaths from vital registration may be biased downward by the omission of events from the register or under-coverage of the target population. Full birth histories are susceptible to biases caused by omission of births or misreporting of the timing of events. Some omissions may be deliberate in order to avoid lengthy subsections of the survey \citep{HillYouInoueEtAl_Child_PM_2012}. Fertility and mortality estimates from summary birth histories are derived using so-called indirect techniques such as the Brass $P/F$ ratio method \citep{Brass_Uses_ASoVS1964,United_Manual_1983,Feeney_New_1996}. In addition to the biases affecting full birth histories, estimates based on summary birth histories can also be affected when the assumptions behind the indirect methods are not satisfied. These assumptions concern the pattern of mortality by age and the association between mother and child mortality. They often do not hold, for example, in populations experiencing rapid mortality decline \citep{Silva_Child_PM_2012}.

In the absence of vital registration, estimates of adult mortality may be based on reports of sibling survival histories collected in surveys. Often however, the only data available are on child mortality collected from surveys of women. In such cases, estimates of adult mortality are extrapolations based on model life tables  \citep{United_Manual_1983,PrestonHeuvelineGuillot2001}. Model life tables are families of life tables generated from mortality data collected from a wide range of countries over a long period of time. They are indexed by a summary parameter such as \ac{e0} or \ac{U5MR} and are grouped into regions. The Coale and Demeny system \citep{CoaleDemeny_Regional_1983,Preston1976Mortality} and the \ac{UN} system for developing countries \citep{United_Model_1982} both have five families. Errors in estimates of adult mortality derived in this way come from errors in the survey-based estimates of under five mortality and the inability of the model life table family to capture the true mortality patterns in the population of interest.

Concerns about the accuracy of \ac{SRB} estimates, particularly for periods between 1950 and 1970, have led some authors to suggest using age-specific population sex ratios as a proxy for \acp{SRB}. \citet{Guilmoto_Sex_PaDR_2009} suggests the male-to-female ratio among those aged 0--4 (the ``child sex ratio'') and \citet{Bhat_Trail_EaPW_2002} suggests the ratio among those aged 0--14 for India (the ``juvenile sex ratio''). Such ratios must be estimated from census data which is probably more reliable than survey data, but still subject to age misreporting and underreporting of certain groups. For example, there appears to have been under counting of females in censuses of India \citep{Bhat_Trail_EaPW_2002,Bhat_Trail_EaPW_2002a,Guillot_Dynamics_PS_2002}. In our case study, we estimate the child sex ratio for India between 1971 and 2001.

Estimates of fertility, mortality, migration and population counts, and the implied sex ratios for successive quinquennia are all related to one another via the demographic balancing equation that underlies the \ac{CCMPP}. The estimates of these quantities published in \ac{WPP} must be ``projection consistent'' in that the age-specific population counts for year $t$ must be the counts one gets by projecting the published counts for year $t-5$ forward using the published fertility, mortality and migration rates.

Bias reduction techniques are source- and parameter-specific. For birth history data, these might involve omitting responses of very old women, or responses pertaining to events in the distant past. For census counts, adjustments may be made to compensate for well-known undercounts in certain age-sex groups. In other cases, parametric models of life tables, or specially constructed life tables, may be used if available. For this reason we do not propose a generally applicable method of bias reduction, one which would work well for all parameters and data sources, since many specialized ones already exist \citep[e.g.,][]{United_Manual_1983,MurrayRajaratnamMarcusEtAl_What_PM_2010,Murray2003,AlkemaRafteryGerlandEtAl_Estimating_DR_2012}. Bayesian reconstruction takes as input bias-reduced initial estimates of age-specific fertility rates and age- and sex-specific initial estimates of mortality, international migration and population counts. Measurement error is accounted for by modelling these quantities as probability distributions. Projection consistency is achieved by embedding the \ac{CCMPP} in a Bayesian hierarchical model. Inference is based on the joint posterior distribution of the input parameters, which is sampled from using \ac{MCMC}.

Under the current \ac{UN} procedure, all available representative data sources for a given country are considered and techniques to reduce bias are applied where \ac{UN} analysts deem them appropriate. Projection consistency is achieved through an iterative ``project-and-adjust'' process. \Acp{SRB} and \acp{SRM} are inputs to the procedure, while population sex ratios are calculated using estimate population counts, which are an output.

\section{Method}
\label{sec:method}

\subsection{Notation and Parameters}
\label{sec:notation}

The parameters of interest are age- and time-specific \acp{vitalrate}, net international migration flows, population counts and \ac{SRB}.
The symbols $n$, $s$, $g$ and $f$ denote population counts, survival (a measure of mortality), net migration (immigrants minus emigrants) and fertility, respectively. All of these parameters will be indexed by five-year increments of age, denoted by $a$, and time, denoted by $t$. The parameters $n$, $s$ and $g$ will also be indexed by sex, denoted by $l =F,M$, where $F$ and $M$ indicate female and male, respectively. \Ac{SRB} is defined as the number of male births for every female birth. It will be indexed by time.

Reconstruction will be done over the time interval $[t_0,T]$. The age scale runs from 0 to $A>0$; in our applications $A$ is 80. The total number of age-groups is denoted $K$. To model fertility, we define $a^{[\text{fert}]}_L \le a^{[\text{fert}]}_U$, where fertility is assumed to be zero at ages outside the range $[a^{[\text{fert}]}_L, a^{[\text{fert}]}_U+5)$. Throughout, a prime indicates vector transpose. We will use boldface for vectors and a ``$\cdot$'' to indicate the indices whose entire range is contained therein. Multiple indices are stacked in the order $a$, $t$, $l$. For example, the vector of age-specific female population counts in exact year $t_0$ is $\vec{n}_{\cdot,t_0,F} [n_{0,t_0,F},\ \cdots,\ n_{A,t_0,F}]^\prime$, and
\begin{multline*}
  \vec{n}_{\cdot,\cdot,\cdot} = [n_{0,t_0,F},\  \cdots,\  n_{A,t_0,F},\  n_{0,t_0+5,F},\  \cdots,\  n_{A,t_0+5,F},\  \cdots, \\ n_{0,T,F},\  \cdots,\  n_{A,T,F},\  \cdots,\ n_{0,T,M},\  \cdots,\  n_{A,T,M}]^\prime .
\end{multline*}

The parameters are the standard demographic parameters used for projection. The fertility parameters, $f_{a,t}$, are age-, time-specific occurrence/exposure rates. They give the ratio of the number of babies born over the period $[t,t+5)$ to the number of person-years lived over this period by women in the age range $[a,a+5)$. If a woman survives for the whole quinquennium she contributes five person-years to the denominator; if she survives only for the first year and a half she contributes 1.5 person-years, and so forth. The survival parameters, $s_{a,t,l}$, are age-, time-, and sex-specific proportions. They give the proportion of those alive at time $t$ that survive for five years. The age subscript on the survival parameters indicates the age-range the women will survive \emph{into}. For example, the number of females aged $[15,20)$ alive in 1965 would be the product $(n_{10,1960,F})(s_{15,1960,F})$ (ignoring migration for simplicity). It also means that $s_{0,1960.F}$ is the proportion of female births born during 1960--1965 that are alive in 1965, hence aged 0--5. The oldest age-group is open-ended and we must allow for survival in this age group. Thus, the proportion aged $[A,\infty)$ at time $t$ that survives through the interval $[t,t+5)$ is denoted by $s_{A+5,t,l}$. Migration is also expressed as a proportion. The net number of male migrants aged $[a,a+5)$ over the interval $[t,t+5)$ is $(n_{a,t,M})(g_{a,t,M})$.

\subsection{Projection of Two-Sex Populations}
\label{sec:popul-proj}

The \ac{CCMPP} allows one to calculate the number alive by age and sex at any time, $t=t_0+5, \ldots,T$, using $\vec{n}_{\cdot,t_0,\cdot}$, the vector of age- and sex-specific female and male population counts at baseline $t_0$, and the age-, time-, sex-specific \acp{vitalrate} and migration up to time $t$. The vector of counts $\vec{n}_{\cdot,t,\cdot}$ is simply $\vec{n}_{\cdot,t_0,\cdot}$ plus the intervening births, minus the deaths, plus net migration. Projection is a discrete time approximation to a continuous time process, and several adjustments are made to improve accuracy. The form we use has two-steps; projection is done first for those aged 5 and above, $\vec{n}_{5+,t,l}$ and then for those under five, $n_{0,t,l}$. To this end, let us write
\begin{equation}
  \label{eq:1}
  \vec{n}_{\cdot,t,\cdot} =
  \left[
    \begin{array}{cc|cc}
n_{0,t,F},& \vec{n}_{5^+,t,F}& n_{0,t,M},& \vec{n}_{5^+,t,M}
    \end{array}
  \right]^\prime,
\end{equation}
where vectors and matrices are partitioned according to sex for clarity.

The number alive at exact time $t+5$ aged 5 and above is then given by the following matrix multiplication:
\begin{equation}
  \label{eq:ccmpp-over-5-F-M}
  \left[
    \begin{array}{c}
      \vec{n}_{5^+,t+5,F} \\\hline
      \vec{n}_{5^+,t+5,M}\\
    \end{array}
  \right] =
  \left[
    \begin{array}{cc}
      \mat{L}_{5^+,t,F} & \mat{0} \\\hline
      \mat{0} & \mat{L}_{5^+,t,M} \\
    \end{array}
  \right]
  \left[
    \begin{array}{c}
      \vec{n}_{\cdot,t,F} + \vec{n}_{\cdot,t,F} \circ (\vec{g}_{\cdot,t,F})/2\\\hline
      \vec{n}_{\cdot,t,M} + \vec{n}_{\cdot,t,M} \circ (\vec{g}_{\cdot,t,M})/2\\
    \end{array}
  \right] +
  \left[
    \begin{array}{c}
      (\vec{g}_{5^+,t,F})/2\\\hline
      (\vec{g}_{5^+,t,M})/2\\
    \end{array}
  \right].
  \end{equation}
The symbol ``$\circ$'' indicates the Hadamard (or element-wise) product; $\vec{n}_{5^+,t,l}$, $l=F,M$, are $(K-1)\times 1$ vectors containing the age-specific female and male population counts at exact time $t$; and $\vec{g}_{5^+,t,l}$, $l=F,M$, are $(K-1)\times 1$ vectors of age-specific female and male net migration expressed as a proportion of the population. The matrices $\mat{L}_{5^+,t,F}$ and $\mat{L}_{5^+,t,M}$ are $(K-1)\times K$ matrices of survival proportions for females and males at ages 5 and above, and $\mat{0}$ is a $(K-1)\times K$ matrix of zeros (the ``L'' is for \citealt{Leslie_Use_B1945,Leslie_Some_B1948}). The female and male survival matrices have the same form:
\begin{gather}
  \label{eq:leslie-matrix-F-M}
  \mat{L}_{5^+,t,l} =
  \begin{bmatrix}
         s_{5,t,l}              & 0                   &        & 0                      & 0 \\
         0                    & s_{10,t,l}            & \ddots & 0                      & 0 \\
         0                    & 0                   &        & 0                      & 0 \\
         0                    & 0                   & \cdots & s_{A,t,l}                & s_{A+5,t,l}
  \end{bmatrix},\ l=F,M.
\end{gather}
Splitting migration in half and adding the first half at the beginning of the projection interval and the second half at the end is a standard approximation to improve the discrete time approximation \citep{PrestonHeuvelineGuillot2001}.

The number of females and males alive aged $[0,5)$ in exact year $t+5$ is derived from the total number of births over the interval, $b_t$, where
\begin{equation}
  \label{eq:tot-numb-births}
b_t =  \sum_{a=a^{[\text{fert}]}_L}^{a^{[\text{fert}]}_U}5f_{a,t}\left\{\frac{n_{a,t,F} + (n_{a-5,t,F})(s_{a,t,F})}{2}\right\} .
\end{equation}
The term in braces is an approximation to the number of person-years lived by women of child-bearing age over the projection interval. The total number of each sex aged $[0,5)$ alive at the end of the interval is computed from $b_t$ using under five mortality, migration and \ac{SRB}:
\begin{align}
  \label{eq:f-tilde}
n_{0,t+5,F}&=b_t\frac{1}{1+\mathsrb_t} \left\{s_{0,t,F}\left(1+(g_{0,t,F})/2\right)+(g_{0,t,F})/2\right\}, \\
n_{0,t+5,M}&=b_t\frac{\mathsrb_t}{1+\mathsrb_t} \left\{s_{0,t,M}\left(1+(g_{0,t,M})/2\right)+(g_{0,t,M})/2\right\}.
\end{align}

Note that the $f_{a,t}$ in (\ref{eq:f-tilde}) have only two subscripts; they are the age-specific (female) fertility rates introduced above. Thus the total number of births in the projection interval is a function of the number of females of reproductive age, but not of the number of males of any age, or of females of other ages. This is called ``female dominant projection''.  This approach is preferred to alternatives, such as basing fertility on the number of male person-years lived, because survey-based fertility data are often collected by interviewing mothers, not fathers. All-sex births are computed first and then decomposed because \ac{SRB} is often a parameter of interest to demographers, as it is to us here \citep{PrestonHeuvelineGuillot2001}.

\subsection{Modelling Uncertainty}
\label{sec:modell-uncert}

In many countries, the available data on \acp{vitalrate} and migration are fragmentary and subject to systematic biases and non-systematic measurement error. \citet{WheldonRafteryClarkEtAl_Reconstructing_JotASA} proposed Bayesian reconstruction as a way of estimating past \acp{vitalrate}, migration and population counts for a single-sex population, which accounts for measurement error. Systematic biases are treated in a pre-processing step which yields a set of bias-reduced ``initial estimates'' for each age-, time-specific fertility rate, survival and migration proportion and population counts. We use an asterisk (``${}^*$'') to denote initial estimates. Hence $f_{a,t}^*$ is the initial estimate of $f_{a,t}$. At the heart of Bayesian reconstruction is a hierarchical model which takes the initial estimates as inputs. Here, we present a substantial development of the model given in \citet{WheldonRafteryClarkEtAl_Reconstructing_JotASA} which allows estimation of two-sex populations.

Take $t_0$ and $T$ to be the years for which the earliest and most recent bias adjusted census-based population counts are available (henceforth, we refer to these simply as census counts). Years following $t_0$ for which census counts are also available are denoted by $t_0 <t_L^{\text{[cen]}}, \ldots, t_U^{\text{[cen]}}<T$. Let $\vec{\theta}$ be the vector of all age-, time- and sex-specific fertility rates, survival and migration proportions over the period $[t_0,T)$, the \acp{SRB}, and the age- and sex-specific census counts in year $t_0$. These are the inputs required by the \ac{CCMPP}. We abbreviate \ac{CCMPP} by $M(\cdot)$. Let $\vec{\psi}_t$ be the components of $\vec{\theta}$ corresponding to time $t$, excluding $\vec{n}_{t_0}$. Therefore, $\vec{\theta} = [\vec{n}_{\cdot,t_0,\cdot}^\prime,\ \vec{f}_{\cdot,\cdot,\cdot}^\prime,\  \vec{s}_{\cdot,\cdot,\cdot}^\prime,\ \vec{g}_{\cdot,\cdot,\cdot}^\prime,\ \text{\textbf{\textit{SRB}}}_{\cdot}^\prime]^\prime$ and $\vec{\psi}_t = [\vec{f}_{\cdot,t,\cdot}^\prime,\  \vec{s}_{\cdot,t,\cdot}^\prime,\ \vec{g}_{\cdot,t,\cdot}^\prime,\ \text{\textbf{\textit{SRB}}}_{t}^\prime]^\prime$. Reconstruction requires estimation of $\vec{\theta}$ which we do using the following hierarchical model:
\begin{align}
  \label{eq:method-likelihood}
  \text{Level }1: && \log n^{*}_{a,t,l} \cond n_{a,t,l}, \sigma^2_n &\sim \dNorm\left(\log n_{a,t,l},
    \sigma^2_n\right)\\\notag
      && &t = t^{[\text{cen}]}_L, \ldots,t^{[\text{cen}]}_{U} \displaybreak[0]\\
  \label{eq:method-estimate}
  \text{Level }2: &&  n_{a,t,l} \cond \vec{n}_{\cdot,t-5,\cdot},\ \vec{\psi}_{t-5} &=
    M\left(\vec{n}_{\cdot,t-5,\cdot},\ \vec{\psi}_{t-5}\right)\\\notag
    && & t = t_0+5, \ldots, T \\[1.5em]
    \text{Level }3:\label{eq:prior-SRB}
&& \log \mathsrb_t \cond \mathsrb^*_t, \sigma^2_{SRB} &\sim \dNorm\left(\log \mathsrb^*_t, \sigma^2_{SRB}\right)\\
\label{eq:prior-fertility-rate}
    &&  \log f_{a,t} \cond f_{a,t}^{*}, \sigma^2_f &\sim
      \begin{cases}
        \dNorm\left(\log f^{*}_{a,t}, \sigma^2_f\right),& a =
        a^{[\text{fert}]}_L,\ldots, a^{[\text{fert}]}_U \\
        \text{undefined} ,& \text{otherwise}
      \end{cases} \displaybreak[0]\\
\label{eq:prior-baseline-count}
      &&   \log n_{a,t_0,l} \cond n_{a,t_0,l}^{*}, \sigma^2_n &\sim \dNorm\left(\log
        n^{*}_{a,t_0,l}, \sigma^2_n\right) \\
      \label{eq:prior-survival-proportion}
     && \logit s_{a,t,l} \cond s_{a,t,l}^{*}, \sigma^2_s &\sim \dNorm\left(\logit
        s^{*}_{a,t,l}, \sigma^2_s\right),\ a=0,5,\ldots,A+5\\\notag
     \displaybreak[0]\\
      \label{eq:prior-migration-proportion}
     && g_{a,t,l} \cond g_{a,t,l}^{*}, \sigma^2_g &\sim \dNorm\left(g^{*}_{a,t,l}, \sigma^2_g\right)\\\notag
     \displaybreak[0]\\\intertext{(where $a=0,5,\ldots,A$; $t=t_0, t_0+5, \ldots,T$; $l =F,M$ in (\ref{eq:method-likelihood})--(\ref{eq:prior-migration-proportion}) unless otherwise specified)}
  \label{eq:prior-sigmasq-params}
  \text{Level }4: &&\sigma^2_v &\sim \dInvGamma(\alpha_v,\beta_v),\ v =n,f,s,g,\mathsrb.
\end{align}
For $x <0<1$, $\logit(x) \equiv \log(x/(1-x))$. The joint prior at time $t$ is multiplied by
\begin{equation}
  \label{eq:method-pos-restriction}
    I\left\{M\left(\vec{n}_t,\vec{\psi}_t\right)>0\right\}
    \equiv
    \begin{cases}
      1 & \text{if, for all $a=0,\ldots,A$ and $l=F,M$, }n_{a,t+5,l} \geq 0\\
      0 & \text{otherwise}.
    \end{cases}
  \end{equation}
to ensure a non-negative population. \poscite{WheldonRafteryClarkEtAl_Reconstructing_JotASA} female-only model had $\mathsrb$ fixed at 1.05 and $l=F$. \Ac{SRB} can be interpreted as the odds that a birth is male, so (\ref{eq:prior-SRB}) is a model for the log-odds that a birth is male.

The hyper parameters $\alpha_v,\beta_v$, $v=n,f,s,g,\text{\textit{\small{SRB}}}$ define the distribution of the variance parameters that represent measurement error in the initial estimates. We set these parameters based on the expert opinion of \ac{UNPD} analysts by eliciting liberal, but realistic, estimates of initial estimate accuracy. We elicit on the observable marginal quantities, $f_{a,t}$, $s_{a,t,l}$, $g_{a,t,l}$, and $n_{a,t,l}$. On their respective transformed scales, these have Student's $t$ distributions centred at the initial point estimates and variance and degrees of freedom dependent on $\alpha$ and $\beta$. We set $\alpha_v = 0.5$, $v=f,s,n,g,\text{\textit{\small{SRB}}}$, which gives the initial estimates a weight equivalent to a single data point. The $\beta_v$ are then determined by specifying the limits of the central 90percent probability interval of the untransformed marginal distributions. Population counts, fertility rates and \ac{SRB} are modelled on the log scale so this amounts to making a statement of the form ``the probability that the true parameter values are within $\eta_v \times 100$ percent of the initial point estimates is 90 percent'', $v=f,n,\text{\textit{\small{SRB}}}$. Migration is explicitly modelled as a proportion so this interpretation is direct for migration. The survival parameters are also proportions but they are modelled on the logit scale. We set $\beta_s$ such that the untransformed $s_{a,t,l}$ lie within the elicited intervals. In all cases, we call $\eta_v$, $v=f,s,n,g,\text{\textit{\small{SRB}}}$, the \emph{elicited relative error}.

Bayesian reconstruction defines a joint prior distribution over the input parameters (\ref{eq:prior-SRB})--(\ref{eq:method-pos-restriction}) which induces a prior on the population counts after the baseline via \ac{CCMPP}. This is ``updated'' using the census counts for which a likelihood is given in (\ref{eq:method-likelihood}). Some methods of estimating migration rely on ``residual'' counts; projected counts based only on \acp{vitalrate} are compared with census counts and the difference attributed to international migration. Methods of adjusting vital rates and census counts to ensure mutual consistency have also been proposed that use a similar approach  \citep[e.g.,][]{LutherRetherford_Consistent_MPS1988,LutherDhanasakdiArnold_Consistent__1986}. Initial estimates of $f_{a,t}^*$, $s_{a,t,l}^*$, $g_{a,t,l}^*$ should not be based on such methods since this would amount to using the data twice and uncertainty would be underestimated in the posterior.

\section{Application}
\label{sec:application}

We apply two-sex Bayesian reconstruction to the populations of India from 1971--2001, Thailand from 1960--2000 and Laos from 1985--2005. The periods of reconstruction are determined by the available data. Laos has no vital registration data. Initial estimates of fertility are based on surveys of women and the only mortality estimates are for ages under five derived from these same surveys. Thailand and India have acceptable vital registration data for these periods which provide information about fertility and mortality at all ages. Nevertheless, adjustments are necessary to reduce bias due to undercount of certain groups. For example, vital registration is thought to have underestimated \ac{U5MR} in Thailand \citep{VapattanawongPrasartkul_Under__2011,HillVapattanawongPrasartkulEtAl_Epidemiologic_IJoE_2007} and in India 50--60 percent of children are born at home which increases the likelihood of omission from the register \citep{Fund2010UNFPA}.

Estimates of population sex ratios in India have been relatively high throughout the twentieth century. Prior to the late 1970s, these were thought to have been caused by an excess of female mortality (high \acp{SRM}), and from the late 1970s onward by high \acp{SRB}. Both of these phenomena have been linked to cultural preferences for sons over daughters which were intensified by a rapid fall in fertility rates  \citep{Visaria_Sex__1971,Bhat_Trail_EaPW_2002,Bhat_Trail_EaPW_2002a,Gupta_Explaining_PaDR_2005,GuilmotoCharacteristicsIndia2007}. Concern over the accuracy of certain estimates of \ac{SRB} has led some authors to suggest using the \ac{SRTP} and sex ratios for young age groups as proxies for \ac{SRB} and \acp{SRM} \citep{Bhat_Trail_EaPW_2002,Bhat_Trail_EaPW_2002a,Guilmoto_Sex_PaDR_2009}. We use Bayesian reconstruction to derive credible intervals for the \ac{SRTP} and the sex ratio among those aged 0--5 for India.

Thailand experienced an even more rapid decline in \ac{TFR} between 1960 and about 1980 \citep{KamnuansilpaChamratrithirongKnodel_Thailands_IFPP_1982}. Estimates of Thailand's \acp{SRB} between 1960 and about 1970 are relatively high, but are within the typical range from about 1970 to 2000. Surveys of Thai families in the 1970s found that girls and boys were desired about equally \citep{KnodelRuffoloRatanalangkarnEtAl_Reproductive_SiFP_1996,Guilmoto_Sex_PaDR_2009}.

Fertility rates in Laos have fallen since 1985 but remain high relative to other Asian countries. Very little has been written about sex ratios for this country \citetext{but see \citealp{Frisen1991PopulationA}}.

In the remainder of this section, we briefly describe the data sources for each country and the method used to derive initial estimates. More details are in the appendices. These are followed by results for selected parameters. We focus on key details and the most interesting outputs; further results, including those for migration, can be found in the appendices. All computations were done using the R environment for statistical computing \citep{R2013_R_}; Bayesian reconstruction is implemented in the package ``popReconstruct''. The method of \citet{raftery96:_implem_mcmc} was used to select the length of \ac{MCMC} chains.

\subsection{Data Sources and Initial Estimates}
\label{sec:data-sources-1}

\subsubsection{India, 1971--2001}
\label{sec:india-}

Censuses have been taken roughly every 10 years in India since 1871. We begin our period of reconstruction in 1971. This is the first census year for which \ac{vitalrate} data independent of the censuses are available, collected by the Indian Sample Registration System. Subsequent censuses were taken in 1981, 1991 and 2001 (sufficiently detailed results from the 2011 census were not available at the time of writing). Counts in \ac{WPP} 2010 were used as these were adjusted to reduce bias. Estimates of \ac{SRB}, fertility and survival were based on data from the Sample Registration System \citep{The_Vital__2011}, the National Family Health Surveys conducted between 1992 and 2006 \citep{Population_National__2009} and the 2002--04 Reproductive Child Health Survey. Weighted cubic splines were used to smooth estimates of \ac{SRB} and fertility.
The same initial estimates for migration were used for India as for Laos and Thailand and the elicited relative errors were also the same; see below. The elicited relative error of 10 percent for the \acp{vitalrate} and \ac{SRB} is consistent with independent assessments of the coverage of the Sample Registration System \citep{Bhat2002CompletenessPS,Mahapatr2010Overview}.

\subsubsection{Thailand, 1960--2000}
\label{sec:thailand-2}

Censuses were conducted in 1960, 1970, 1980, 1990 and 2000 (detailed results from the census conducted in 2010 were not available at the time of writing). We used the counts in \ac{WPP} 2010 which were adjusted for known biases such as undercount. Initial estimates of sex ratio at birth were taken from current fertility based on vital registration. The elicited relative error was set to 10 percent. Initial estimates of age-specific fertility were based on direct and indirect estimates of current fertility and \ac{CEB} based on the available data including surveys and vital registration. Each data series was normalized to give the age pattern and summed to give \ac{TFR}. These were smoothed separately using weighted cubic splines and the resulting estimates combined to yield a single series of initial estimates of age-specific fertility rates, in the same manner as for India. The weights were determined by \ac{UN} analysts based on their expert judgement about the relative reliability of each source. The elicited relative error was set to 10 percent. Initial estimates of survival for both sexes were based on life tables calculated from vital registration, adjusted for undercount using data from surveys. We used the same initial estimates of international migration as for Laos; see below.

\subsubsection{Laos, 1985--2004}
\label{sec:laos-2}

National censuses were conducted in 1985, 1995 and 2005, so we reconstruct the whole population between 1985 and 2005. We used \poscite{WheldonRafteryClarkEtAl_Bayesian__PAA2013} initial estimates for fertility, female mortality, migration and population counts. In these, migration was centred at zero for all sexes, ages and quinquennia, with a large relative error of 20 percent. Initial estimates for males were derived in an analogous manner. There was very little information about the sex ratio at birth, so initial estimates were set at 1.05, a demographic convention \citep{PrestonHeuvelineGuillot2001}, with a large elicited relative error of 20 percent.

\subsection{Results}
\label{sec:results}

Key results are given by country; more results are presented in the appendices 
. We show the limits of central 95 percent credible intervals for the marginal prior and posterior distributions of selected parameters. The magnitude of uncertainty will be summarized using half-widths of these intervals, averaged over age, time, and sex. We compare our results to those published in \ac{WPP} 2010 for years with comparable estimates. \ac{WPP} 2010 did not use Bayesian reconstruction but it is based on the same data, so the comparison is useful.

\subsubsection{India, 1971--2001}
\label{sec:india-1}

\begin{figure}[tbph]
  \centering
\subfloat[]{
\GinKARWidth{0.5\textwidth}
\includegraphics{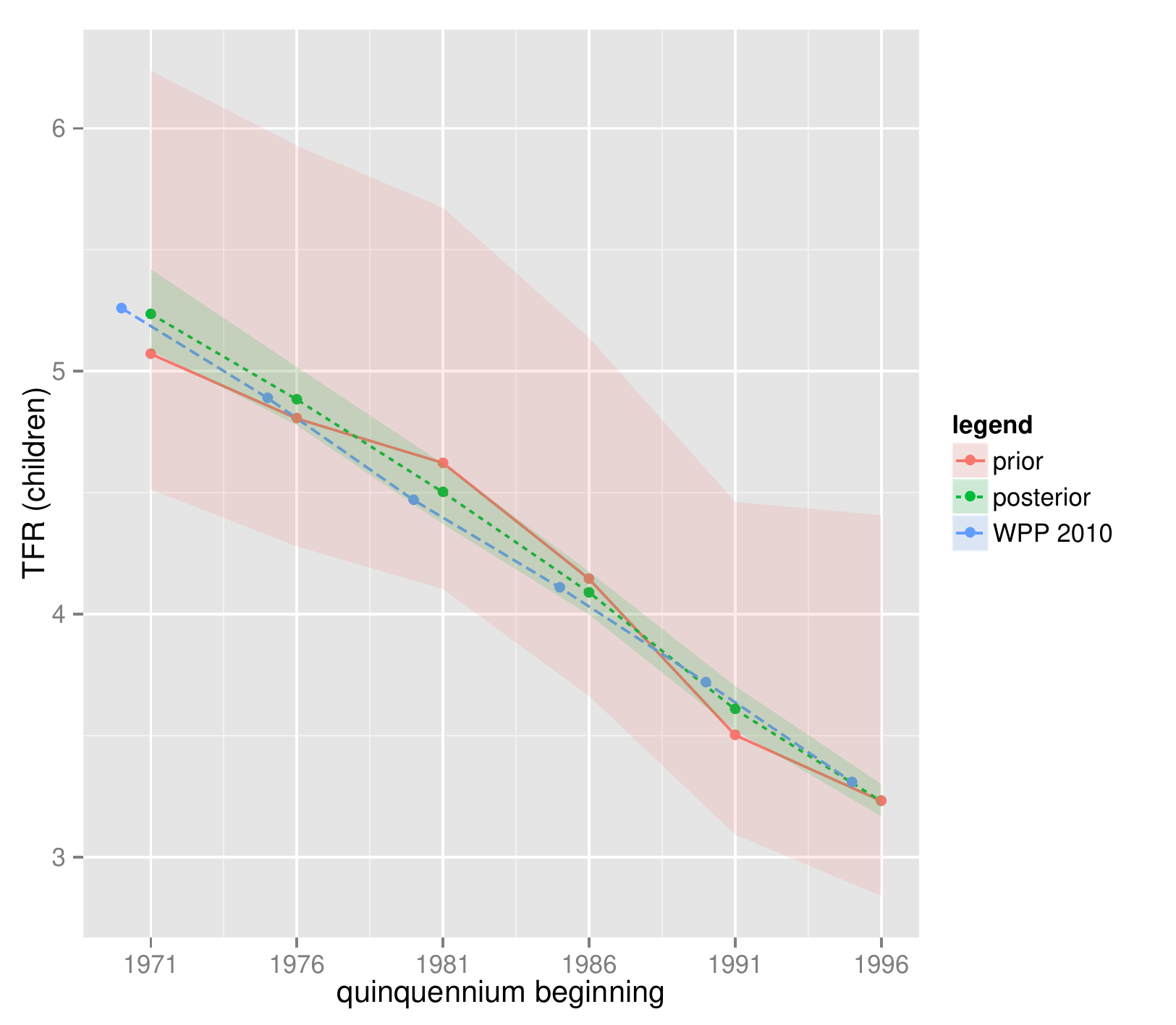}
\label{fig:indi-res-tfr}
}%
\subfloat[]{
\GinKARWidth{0.5\textwidth}
\includegraphics{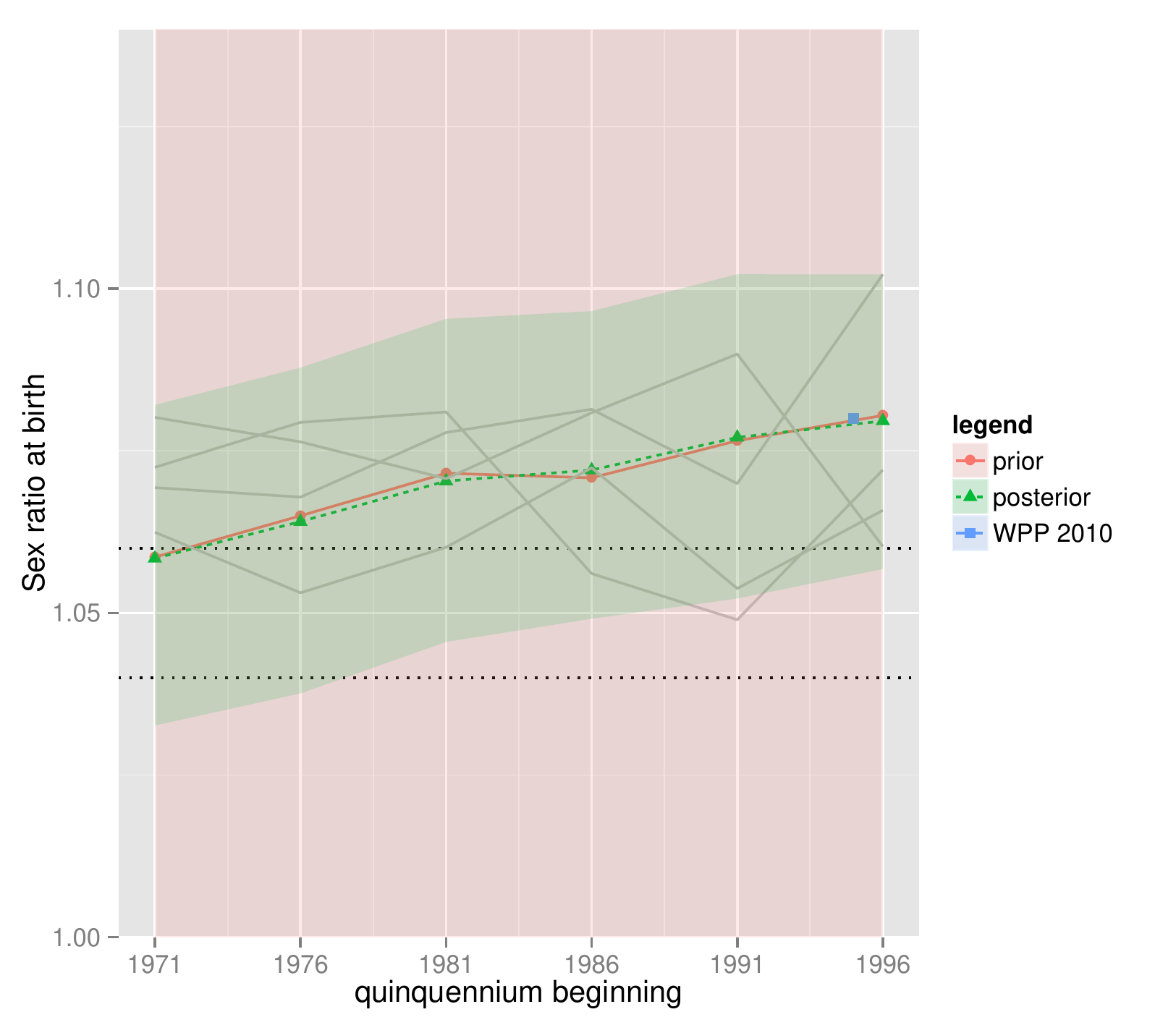}
\label{fig:indi-res-srb}
}
\caption{Prior and posterior medians and 95 percent credible intervals for the reconstructed population of India, 1971--2001: \protect\subref{fig:indi-res-tfr} \protect\acf{TFR}; \protect\subref{fig:indi-res-srb} \protect\acf{SRB} (four trajectories from the \protect\acf{MCMC} sample are also shown).}
\label{fig:indi-res-tfr-srb}
\end{figure}

Figures~\ref{fig:indi-res-tfr} and~\ref{fig:indi-res-srb} show posterior 95 percent intervals for \ac{TFR} and \ac{SRB} for India. Median \ac{TFR} decreased consistently and the posterior intervals have half-width %
0.11 %
children per woman. The marginal posterior for \ac{SRB} is centred above the range 1.04--1.06 from 1976--2001, which suggests that \acp{SRB} might have been atypically high over this period. There also appears to have been an increase in \ac{SRB} over the same period. Under Bayesian reconstruction, the posterior probabilities of these events can be estimated in a straightforward manner from the posterior sample. The posterior probabilities that \ac{SRB} exceeded 1.06 in each of the quinquennia are in Table~\ref{tab:indi-prob-var-quants-gt-106}a. Strong evidence for high \ac{SRB} was found for the period 1991--2001.

To investigate the trend further we looked at the posterior distributions of two measures of linear increase: 1) the difference between \acp{SRB} in the first and last quinquennia; and 2) the slope coefficient in the \ac{OLS} regression of \ac{SRB} on the start year of each quinquennium. Each quantity was calculated separately for each \ac{SRB} trajectory in the posterior sample. Some actual trajectories are shown in Figure~\ref{fig:indi-res-srb}. These measures summarize the posterior obtained from the reconstruction in simple ways; linear regression models were not used to obtain the sample from the posterior. The probabilities that the simple difference and slope coefficient were greater than zero are %
0.92 %
and %
0.93 %
respectively (Table~\ref{tab:indi-prob-dec-main-article}a).

\GinTextWidth

\begin{figure}[tbph]
  \centering
\subfloat[]{
\GinKARWidth{0.5\textwidth}
\includegraphics{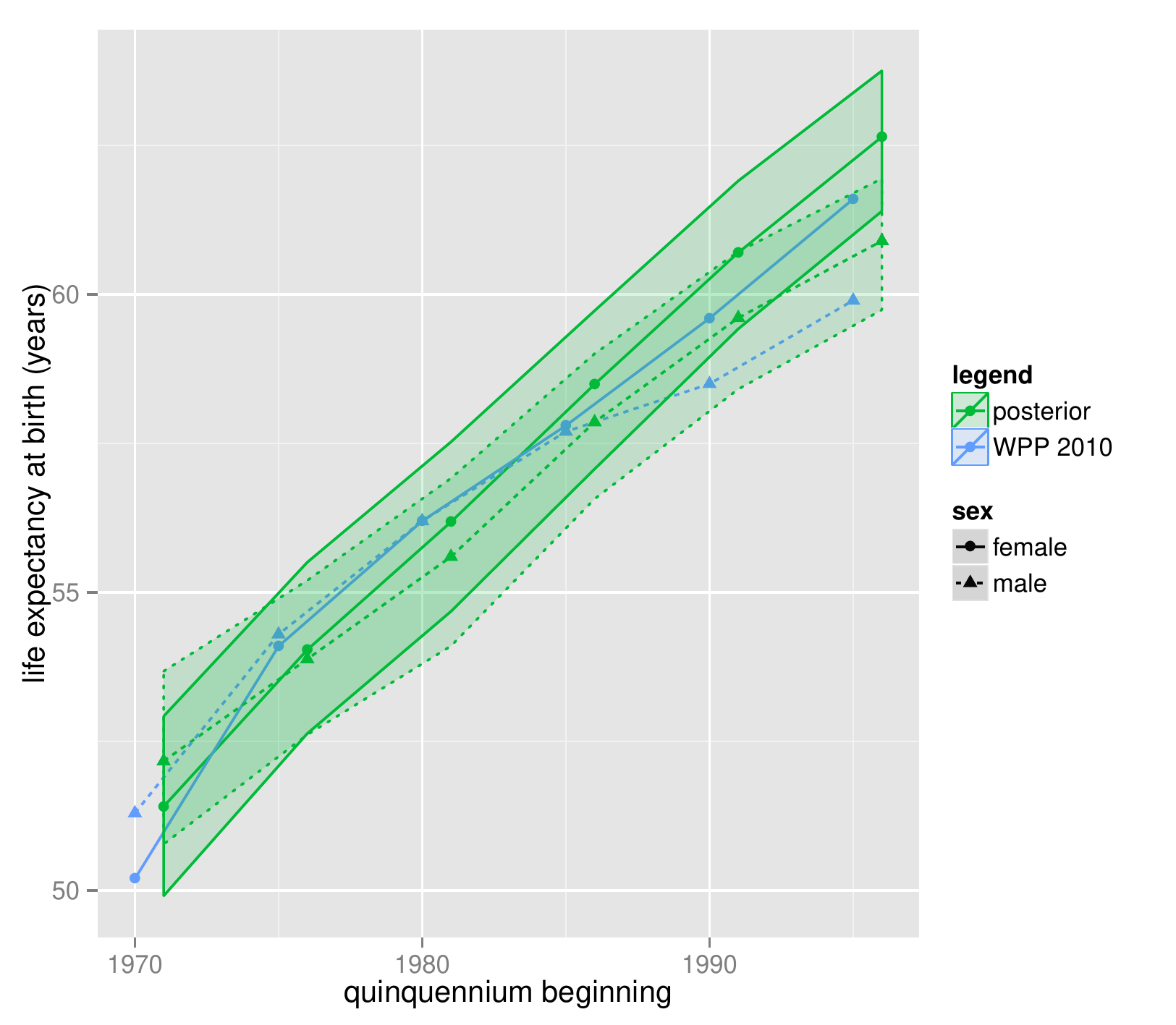}
\label{fig:indi-res-leb-femaleVSmale} 
} 
\subfloat[]{
\GinKARWidth{0.5\textwidth}
\includegraphics{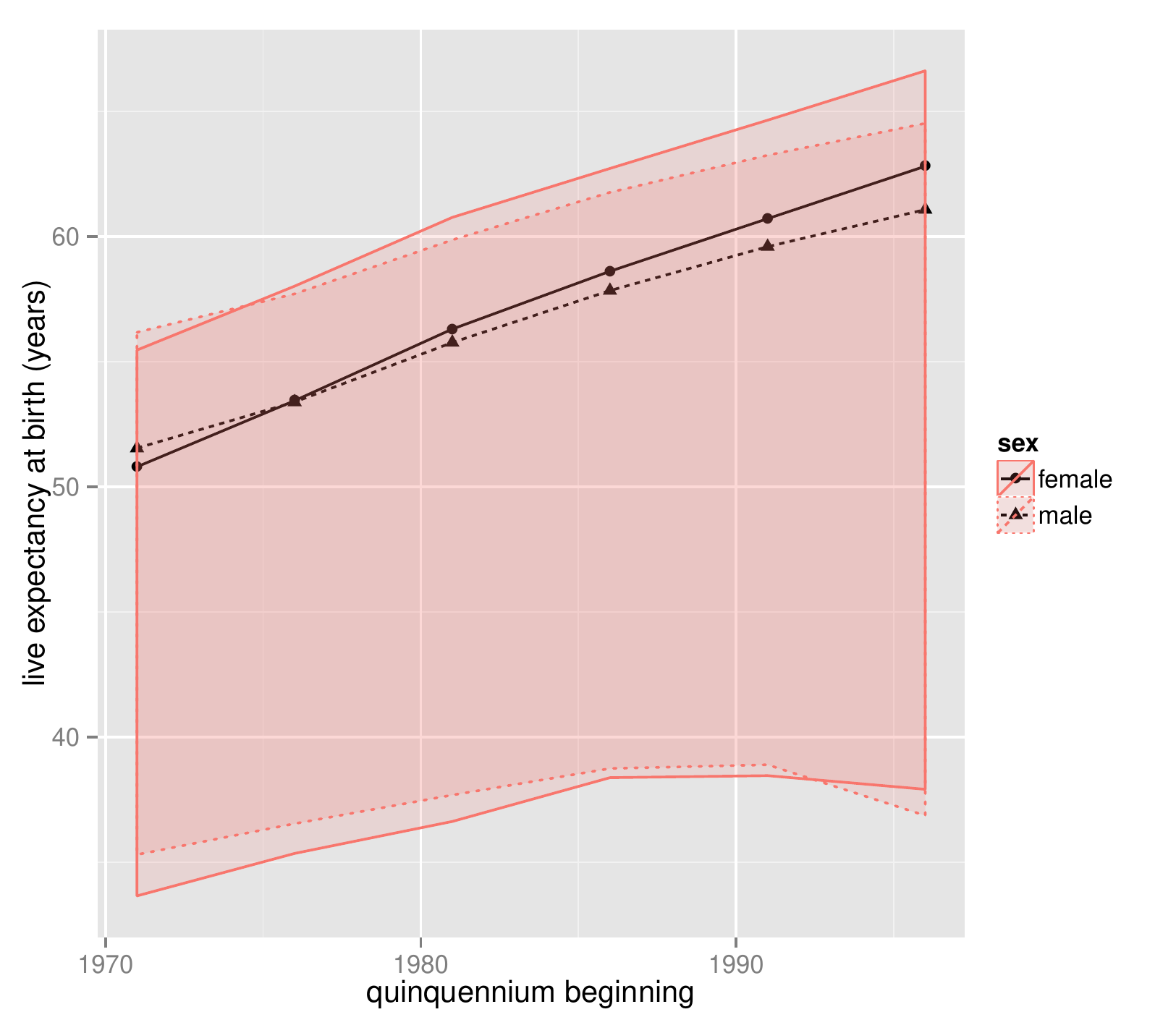}
\label{fig:indi-res-leb-femaleVSmale-prior} 
}\\ 
\subfloat[]{
\GinKARWidth{0.5\textwidth}
\includegraphics{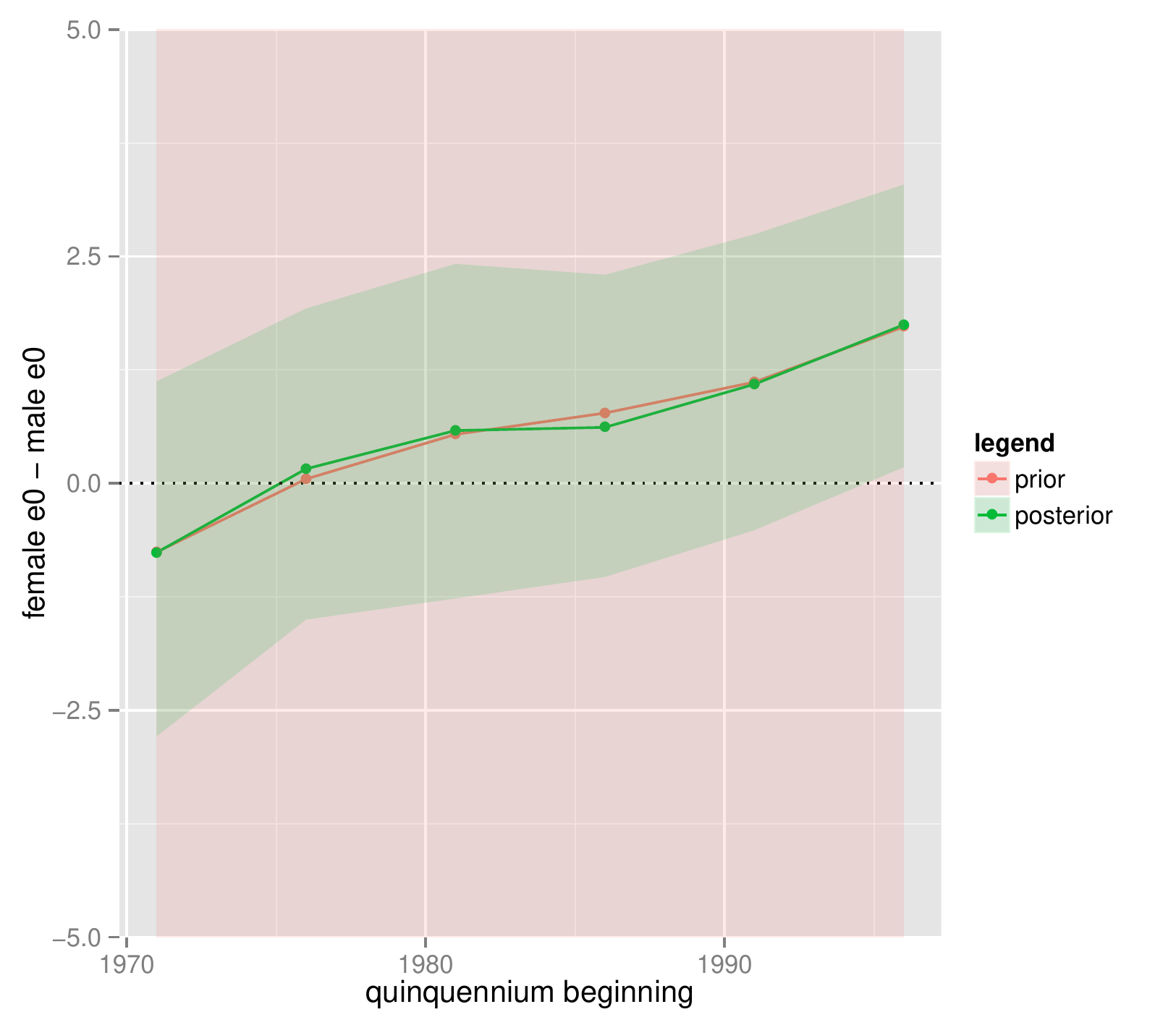}
\label{fig:indi-res-leb-female-male-difference} 
}
  \caption{Prior and posterior medians and 95 percent credible intervals for \protect\acf{e0} for the reconstructed population of India, 1971--2001. \protect\subref{fig:indi-res-leb-femaleVSmale}: Female and male posterior quantiles with \protect\acf{WPP} 2010 estimates; \protect\subref{fig:indi-res-leb-femaleVSmale-prior}: Female and male prior quantiles only; \protect\subref{fig:indi-res-leb-female-male-difference}: Sex difference (female$-$male).}
  \label{fig:indi-res-leb}
\end{figure}

Results for \ac{e0} are shown in Figure~\ref{fig:indi-res-leb}. \Acl{e0} increased for both sexes over the period of reconstruction (Figure~\ref{fig:indi-res-leb-femaleVSmale}) but the sex difference suggests that female \ac{e0} might have increased more rapidly than male \ac{e0} and even exceeded it in the period 1996--2001 (Figure~\ref{fig:indi-res-leb-female-male-difference}). The mean interval half-width for the sex difference in life expectancies is %
1.7 %
years. The posterior probabilities that female \ac{e0} exceeded male \ac{e0} support this (Table~\ref{tab:indi-prob-var-quants-gt-106}b). The possibility of an increase in the female$-$male difference in \ac{e0} was investigated using the same method applied to \ac{SRB}. The probability of an increase between 1971 and 2001 is %
0.98 %
and the probability that the slope is greater than zero is %
0.98 %
(Table~\ref{tab:indi-prob-dec-main-article}b); strong evidence of a positive time trend.

\GinTextWidth

\begin{figure}[tbph]
  \centering
\subfloat[]{
  \GinKARWidth{0.5\textwidth}
\includegraphics{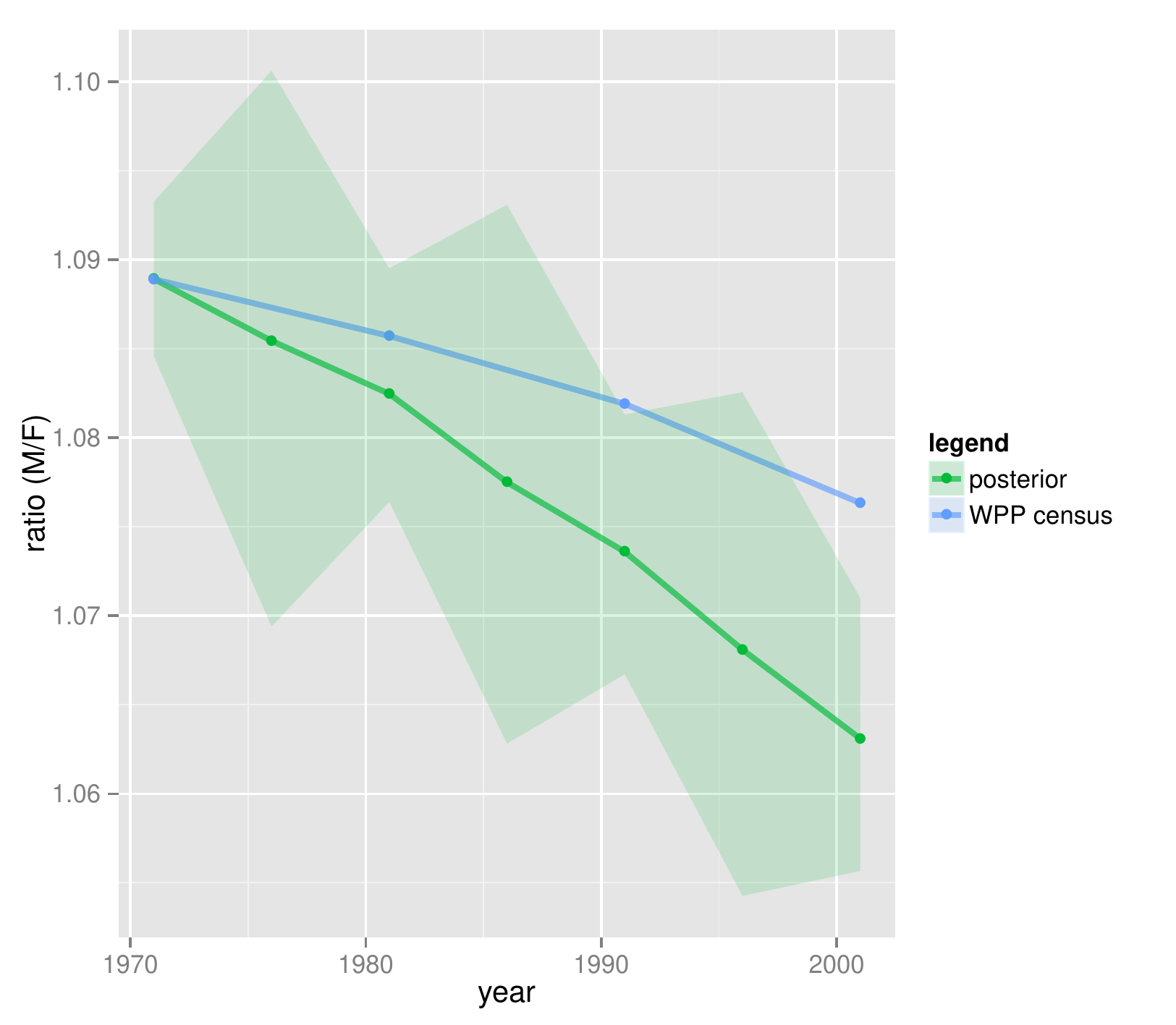}
\label{fig:indi-res-prtp}       
}
\subfloat[]{
  \GinKARWidth{0.5\textwidth}
\includegraphics{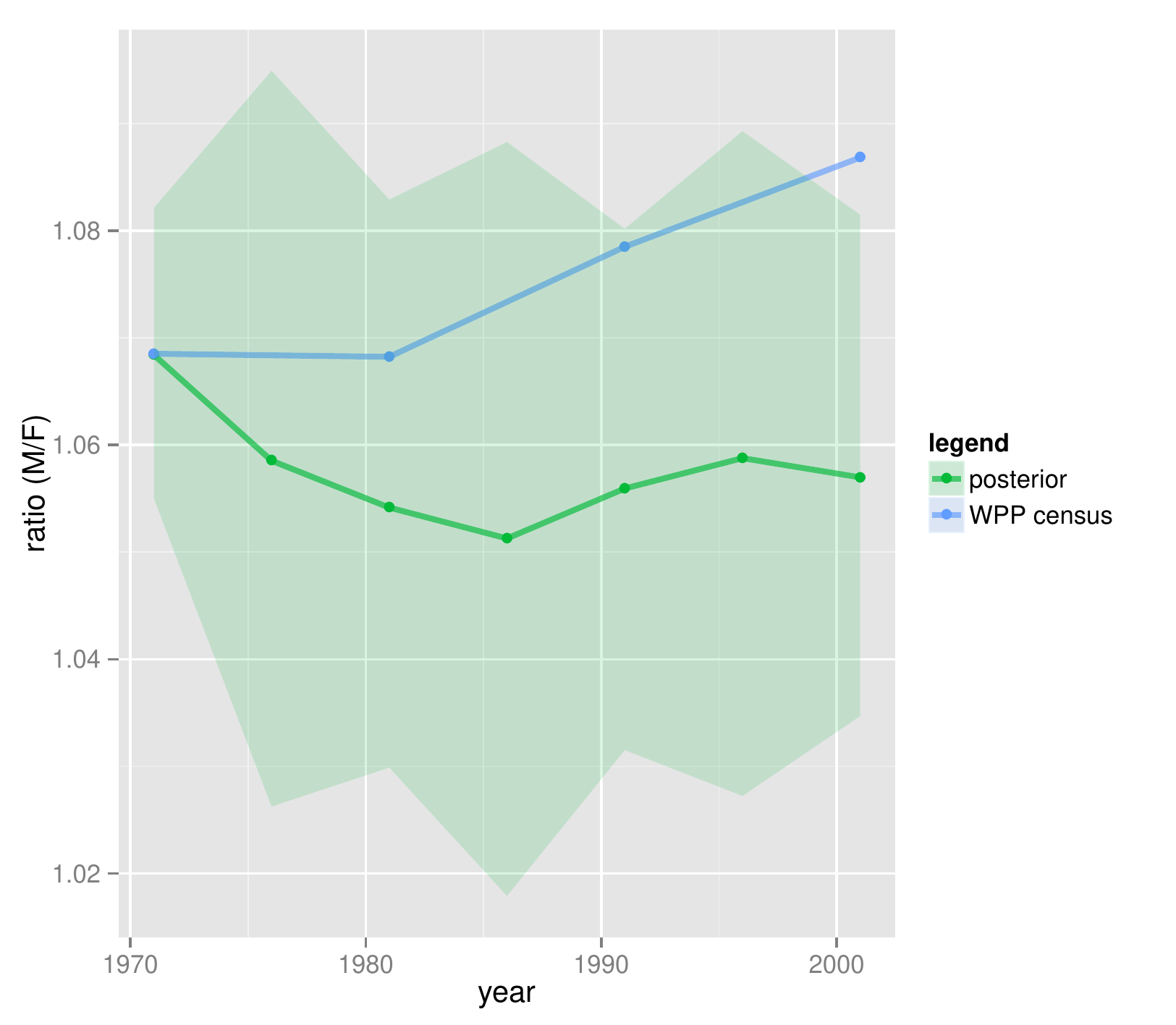}
\label{fig:indi-res-pr05} 
}

  \caption{Prior and posterior medians and 95 percent credible intervals for sex ratios in the reconstructed population of India, 1971--2001. \protect\subref{fig:indi-res-prtp}: Total population; \protect\subref{fig:indi-res-pr05}: Population aged 0--5
.}
  \label{fig:indi-res-prtp-pr05}
\end{figure}

Population sex ratios are shown in Figure~\ref{fig:indi-res-prtp}.
The probability of a decrease in \ac{SRTP} is %
1 %
and the probability that the \ac{OLS} slope was less than zero is %
1 %
(Table~\ref{tab:indi-prob-dec-main-article}c); very strong evidence for a decline over the period of reconstruction.
\Acp{SRU5} increased in the \ac{WPP} population counts, but our posterior median remained relatively constant after an initial decline.
The probability that the sex ratio declined is %
0.8;. %
the probability that the \ac{OLS} coefficient was negative is
0.71 %
(Table~\ref{tab:indi-prob-dec-main-article}d).
Mean half-widths of the intervals for \ac{SRTP} and the sex ratio in ages 0--5 are %
0.01 %
and %
0.027 %
respectively. Uncertainty is higher in years without a census.

\GinTextWidth

\begin{table}[bpth]
  \centering
  \caption{Probability that sex ratios and differences exceeded certain thresholds for the reconstructed population of India, 1971--2001, by quinquennium.}
  \label{tab:indi-prob-var-quants-gt-106}
\begin{tabular}{rrrrrr}
  \hline
  1971 & 1976 & 1981 & 1986 & 1991 & 1996 \\\hline\\[-1.5ex]
  \multicolumn{6}{c}{\textit{(a)} $\Pr(\text{\textit{SRB}} > 1.06)$}\\[1ex]
 0.44 & 0.66 & 0.83 & 0.86 & 0.93 & 0.96 \\
   [1ex]
  \multicolumn{6}{c}{\textit{(b)} $\Pr(\text{\textit{female} $e_0$} - \text{\textit{male} $e_0$} > 0)$}\\[1ex]
 0.21 & 0.58 & 0.74 & 0.78 & 0.91 & 0.99 \\
   [1ex]\hline
\end{tabular}
\end{table}

\begin{table}[tbph]
  \centering
  \caption{Probabilities of increasing linear trends for \protect\acf{SRB}, \protect\acf{SDe0}, 
    \protect\acf{SRTP}, and \protect\acf{SRU5}
    for the reconstructed population of India, 1971--2001. Two measures of trend are used: the difference over the period of reconstruction and the slope coefficient from the \protect\acf{OLS} regression on the start year of each quinquennium. Ninety-five percent credible intervals for the measures are also given.}
  \label{tab:indi-prob-dec-main-article}
  \begin{tabular}{lrr}
  \hline
  Measure of trend  &  95 percent CI  &  Prob $>$ 0 \\\hline\\[-1.5ex]

  \multicolumn{3}{c}{\textit{(a) \protect\Acf{SRB}}}\\[1ex]
 \protect\acs{SRB}$_{1996}$ $-$ \protect\acs{SRB}$_{1971}$ & [$-$0.011, 0.054] & 0.92 \\
  \protect\acs{OLS} slope (\protect\acs{SRB} $\sim$ year) &  [$-$0.00034, 0.0021] & 0.93 \\\mbox{}\\

  \multicolumn{3}{c}{\textit{(b) \protect\Acf{SDe0}}}\\[1ex]
  (\protect\acs{SDe0})$_{1996}$ $-$ (\protect\acs{SDe0})$_{1971}$ & [0.12, 5] & 0.98 \\
  OLS slope (\protect\acs{SDe0} $\sim$ year) &  [0.0066, 0.17] & 0.98 \\\mbox{}\\



  \multicolumn{3}{c}{\textit{(c) \protect\Acf{SRTP}}}\\[1ex]
  \protect\acs{SRTP}$_{1996}$ $-$ \protect\acs{SRTP}$_{1971}$ & [$-$0.034, $-$0.017] & 0.000027 \\
  \protect\acs{OLS} slope (\acs{SRTP} $\sim$ year) &  [$-$0.0013, $-$0.00045] & 0.00035\\\mbox{}\\

  \multicolumn{3}{c}{\textit{(d) \protect\Acf{SRU5}}}\\[1ex]
  \protect\acs{SRU5}$_{1996}$ $-$ \protect\acs{SRU5}$_{1971}$ & [$-$0.037, 0.017] & 0.2 \\
  \protect\acs{OLS} slope (\acs{SRU5} $\sim$ year) &  [$-$0.0011, 0.00062] & 0.29\\
  \hline
\end{tabular}
\end{table}

Using the sex ratio, the probability that female \ac{U5MR} exceeded that of males ranged from %
0.72 %
to %
0.89 %
over the period of reconstruction. Evidence of a linear trend was not found. Further results for \ac{U5MR} and net international migration, are given in the appendices.

\FloatBarrier

\subsubsection{Thailand, 1960--2000}
\label{sec:thailand}

\begin{figure}[tbph]
  \centering
\subfloat[]{
  \GinKARWidth{0.5\textwidth}
\includegraphics{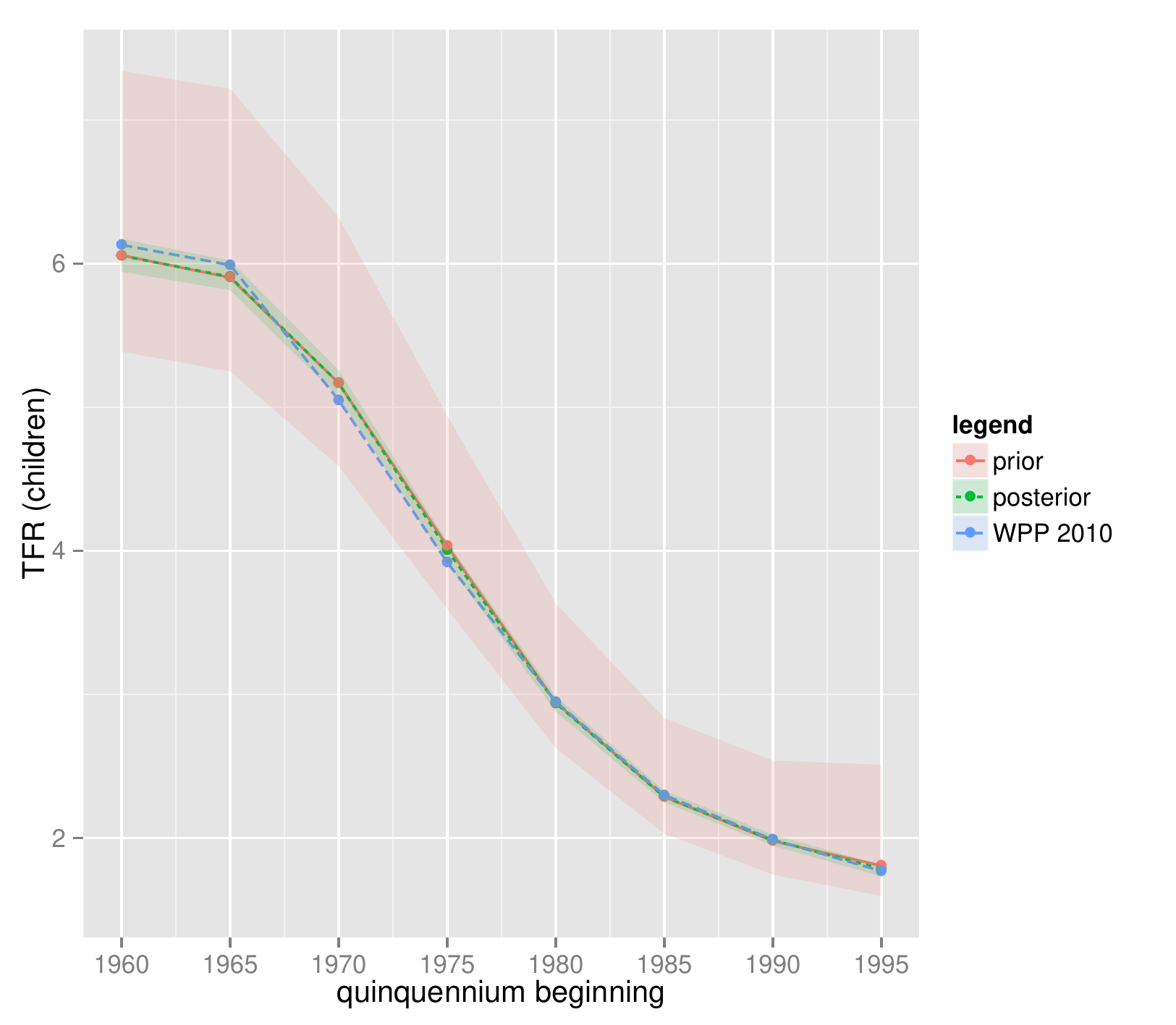}
\label{fig:thai-res-tfr}        
}
\subfloat[]{
  \GinKARWidth{0.5\textwidth}
\includegraphics{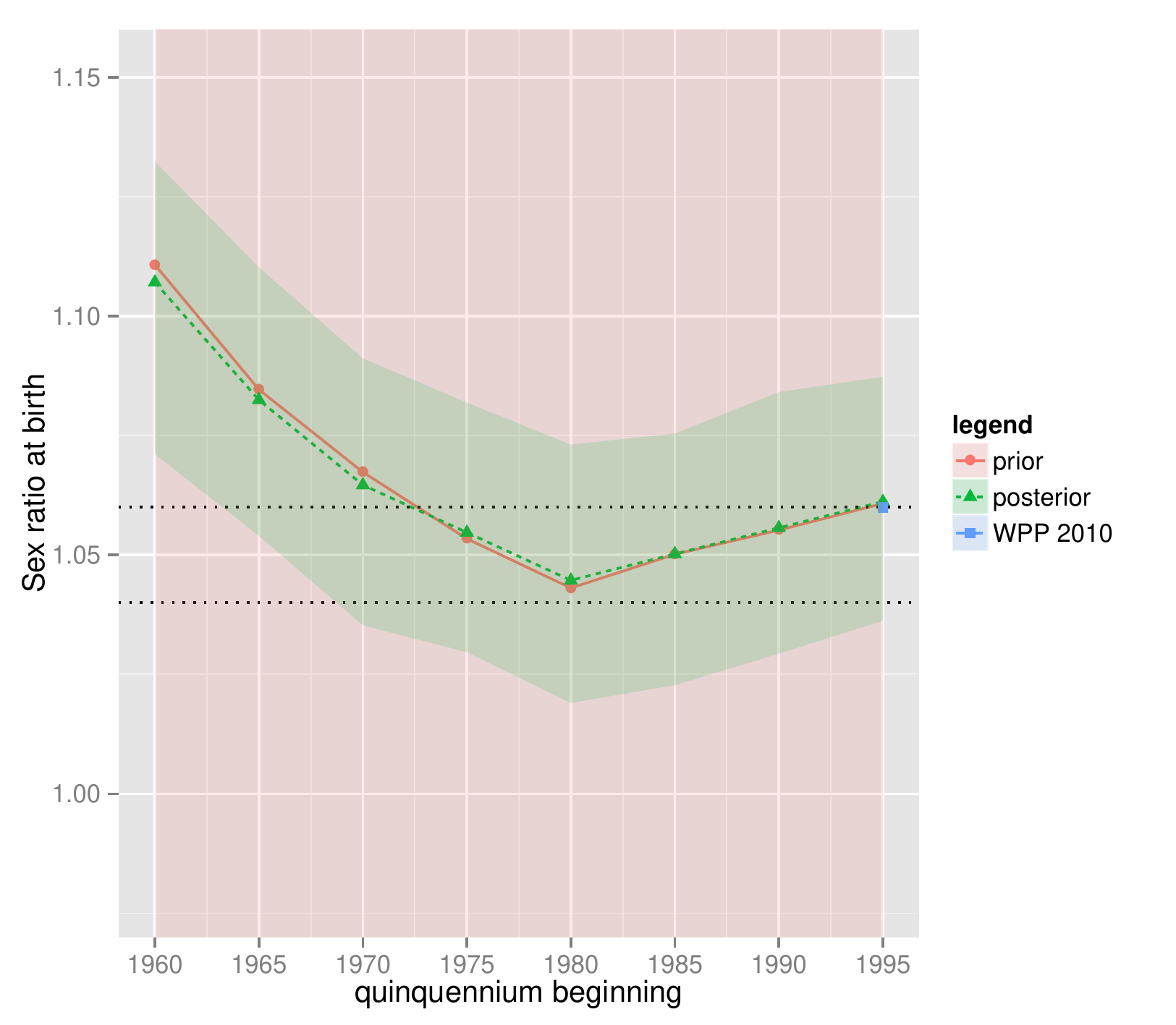}
\label{fig:thai-res-srb}        
}
  \caption{Prior and posterior medians and 95 percent credible intervals for the reconstructed population of Thailand, 1960--2000: \protect\subref{fig:thai-res-tfr} \protect\acf{TFR}; \protect\subref{fig:thai-res-srb} \protect\acf{SRB}.}
  \label{fig:thai-res-tfr-srb}
\end{figure}

\Acl{TFR} fell very steeply in Thailand from 1960--2000 (Figure~\ref{fig:thai-res-tfr}). Posterior uncertainty about this parameter is small; the mean half-width of the posterior intervals is %
0.07 %
children per woman.

Ninety five percent credible intervals for \ac{SRB} contain the typical range of 1.04--1.06 for all but the first two quinquennia (Figure~\ref{fig:thai-res-srb}). The probability that \ac{SRB} exceeded 1.06 in each period is given in Table~\ref{tab:thai-prob-srb-gt-106}. There is strong evidence that \acp{SRB} were atypically high in the period 1960--1969. The time trend for \ac{SRB} appears to be curvilinear, hence the simple linear summaries used to analyze the trend in Indian \ac{SRB} are not appropriate here. Piecewise linear regression models are available \citep[e.g.,][]{Hinkley1969_InfTwoPh_Biomka,Hinkley1971_InfTwoPh_JASA}, but each trajectory in the posterior sample consists of only eight values and they are quite volatile. These characteristics make identification of a change point difficult. Instead we partition the period of reconstruction into two sub-periods, 1960--1984 and 1985--2000, and summarize the time trend with the following two difference quantities: $\text{SRB}_{1960}-\text{SRB}_{1980}$ and $\text{SRB}_{1985}-\text{SRB}_{1995}$ (the subscripts indicate the start years of the quinquennia). The posterior joint probability of a decrease from 1960 to 1984 followed by an increase to 1995--1999 (i.e., $\Pr(\{\text{SRB}_{1960}-\text{SRB}_{1980} >0\} \cap \{\text{SRB}_{1985}-\text{SRB}_{1995} < 0\})$) is %
0.84.

\GinTextWidth

\begin{table}[bpth]
  \centering
  \caption{Probability that \protect\acf{SRB} was greater than 1.06 for the reconstructed population of Thailand, 1960--2000, by quinquennium.}
  \label{tab:thai-prob-srb-gt-106}
\begin{tabular}{rrrrrrrr}
  \hline
  1960 & 1965 & 1970 & 1975 & 1980 & 1985 & 1990 & 1995 \\\hline
 0.99 & 0.95 & 0.66 & 0.32 & 0.11 & 0.19 & 0.35 & 0.54 \\ \hline
\end{tabular}
\end{table}

\begin{figure}[tbph]
  \centering
\subfloat[]{
\GinKARWidth{0.5\textwidth}
\includegraphics{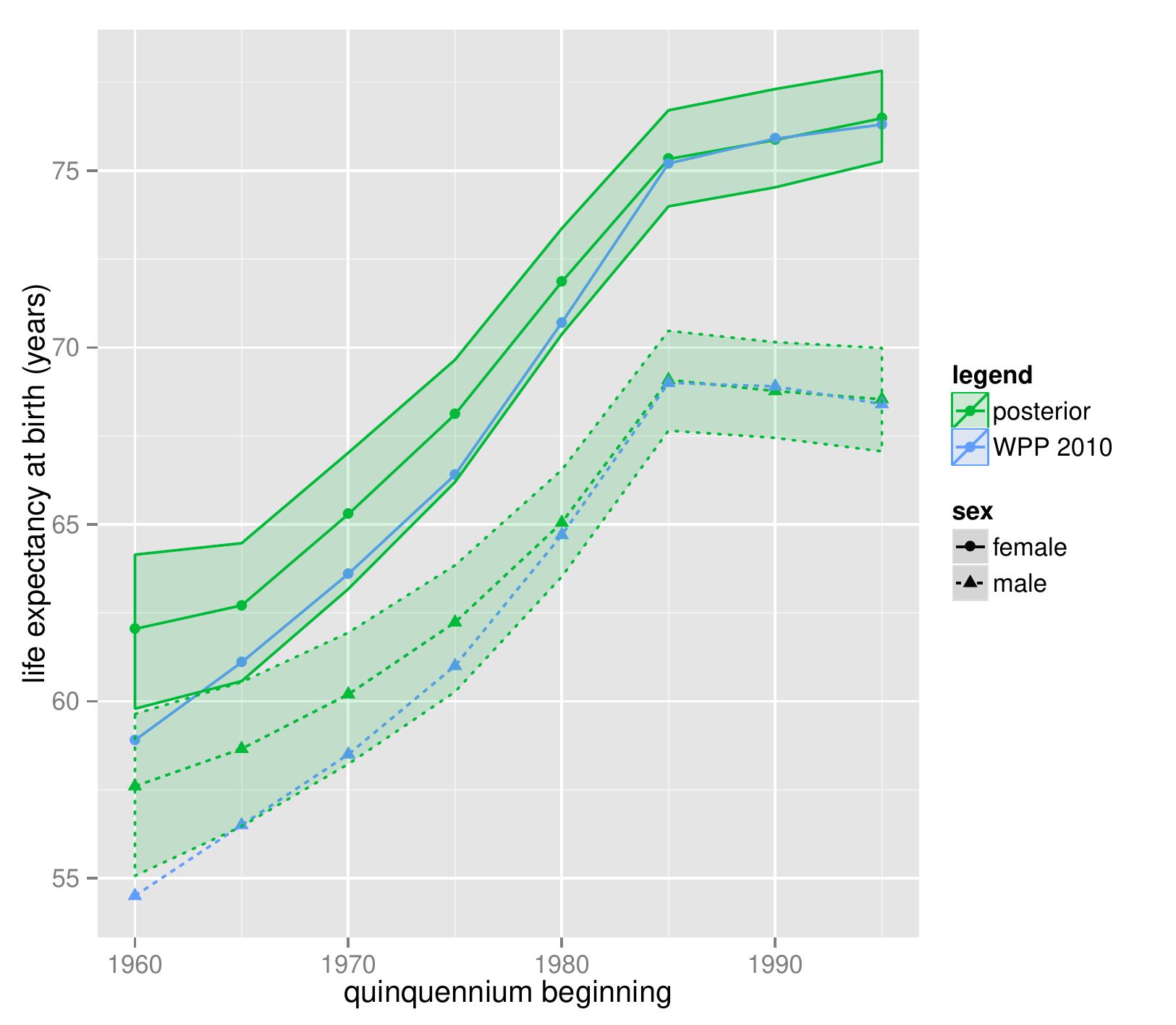}
\label{fig:thai-res-leb-femaleVSmale} 
} 
\subfloat[]{
\GinKARWidth{0.5\textwidth}
\includegraphics{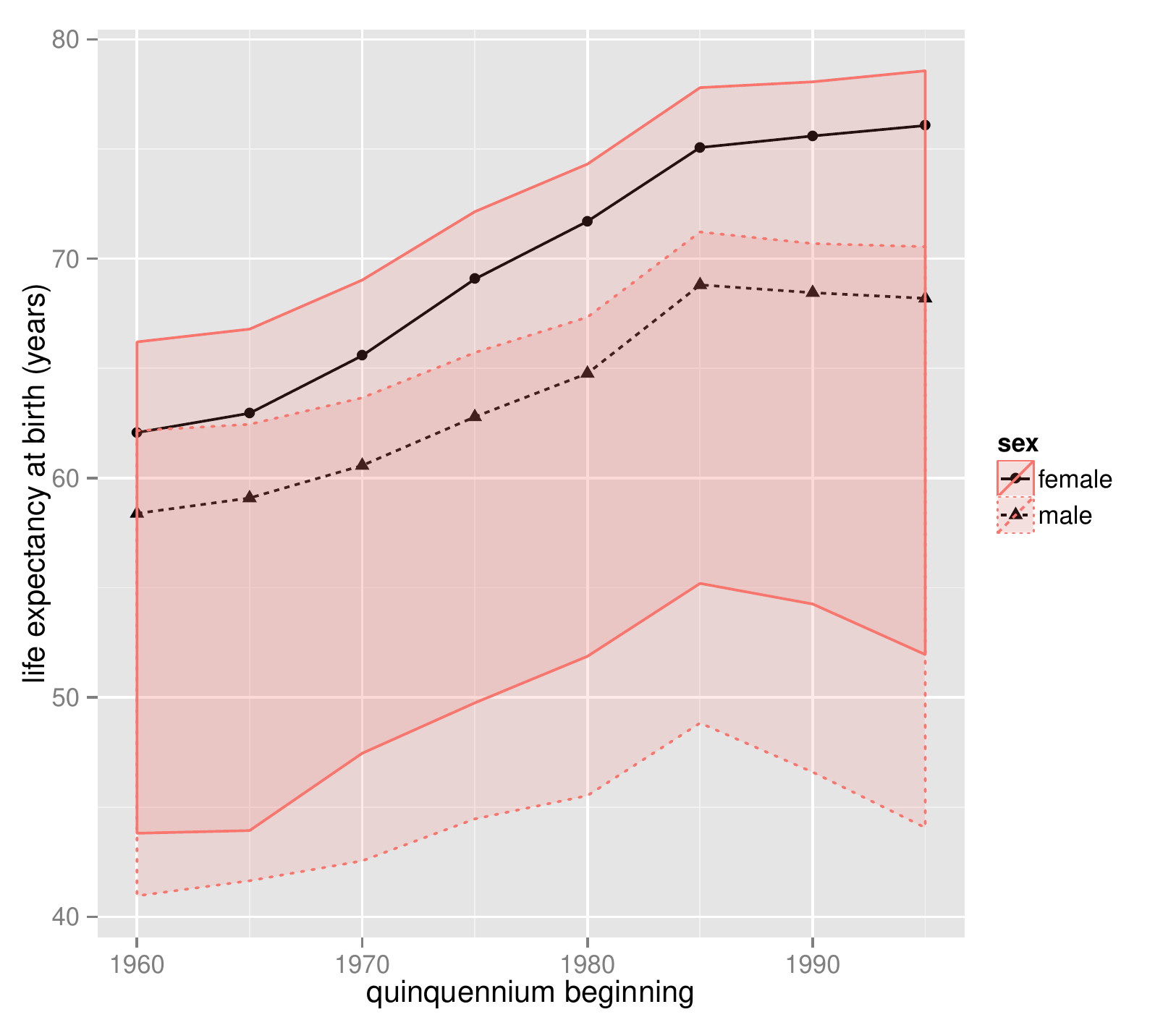}
\label{fig:thai-res-leb-femaleVSmale-prior} 
}\\ 
\subfloat[]{
\GinKARWidth{0.5\textwidth}
\includegraphics{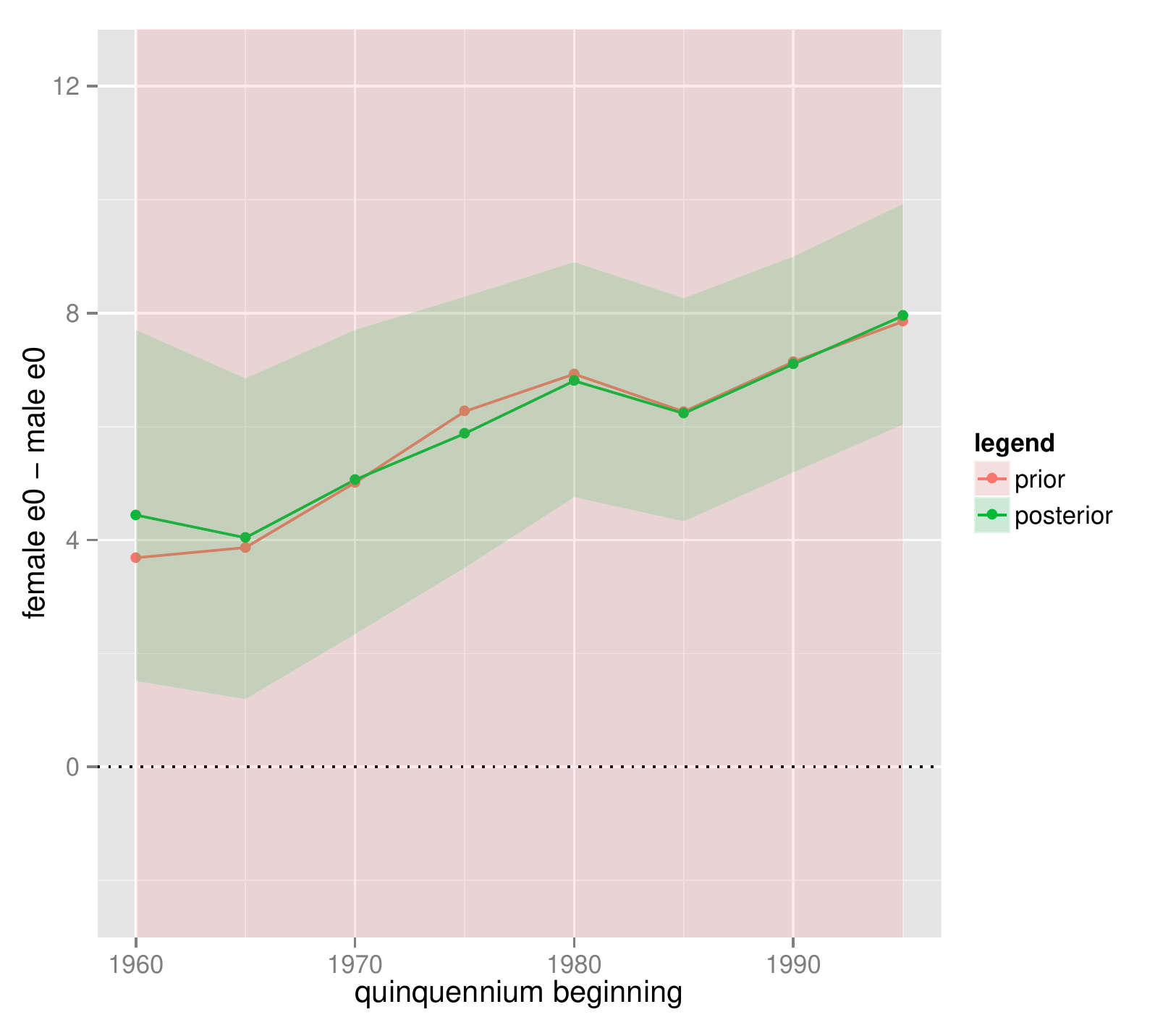}
\label{fig:thai-res-leb-female-male-difference} 
}
  \caption{Prior and posterior medians and 95 percent credible intervals for \protect\acf{e0} for the reconstructed population of Thailand, 1960--2000. \protect\subref{fig:thai-res-leb-femaleVSmale}: Female and male posterior quantiles with \protect\acf{WPP} 2010 estimates; \protect\subref{fig:thai-res-leb-femaleVSmale-prior}: Female and male prior quantiles only; \protect\subref{fig:thai-res-leb-female-male-difference}: Sex difference (female$-$male).}
  \label{fig:thai-res-leb}
\end{figure}

Results for the sex difference in \ac{e0} are shown in Figure~\ref{fig:thai-res-leb}. Our posterior intervals for the sex difference in \ac{e0} lie entirely above zero in each quinquennium, suggesting that female longevity was greater than that of males in Thailand from 1960--2000 (mean half-width of the difference: %
2.4 %
years). There is also strong evidence for a positive trend; the probability that the simple difference (1995 period minus 1960 period) in the sex differences in \ac{e0} was greater than 0 is %
0.96 %
and the probability that the \ac{OLS} slope is greater than 0 is %
1 %
(Table~\ref{tab:thai-prob-dec-lin-sde0}).

\GinTextWidth

The posterior for \ac{SRU5MR} suggests that mortality at ages 0--5 was similar for both sexes. In no quinquennium is there convincing evidence for a male-to-female ratio less than one. Similarly there is not strong evidence for a linear trend in this parameter over the period of reconstruction. Details are in the appendices along with results for net international migration and population sex ratios.

\begin{table}[tbph]
  \centering
  \caption{Probabilities of an increasing linear trend and 95 percent credible intervals for \protect\acf{SDe0} for the reconstructed population of Thailand, 1960--2000. Two measures of trend are used: the difference over the period of reconstruction and the slope coefficient from the \protect\acf{OLS} regression on the start year of each quinquennium.}
  \label{tab:thai-prob-dec-lin-sde0}
  \begin{tabular}{lrr}
  \hline
  Measure of trend  &  95 percent CI  &  Prob $>$ 0 \\\hline
  (\protect\acs{SDe0})$_{1996}$ $-$ (\protect\acs{SDe0})$_{1971}$ & [$-$0.42, 7.1] & 0.96 \\
  OLS slope (\protect\acs{SDe0} $\sim$ year) &  [0.027, 0.18] & 1 \\
  \hline
\end{tabular}
\end{table}

\subsubsection{Laos, 1985--2004}
\label{sec:laos}

\begin{figure}[tbph]
  \centering
\subfloat[]{
  \GinKARWidth{0.5\textwidth}
\includegraphics{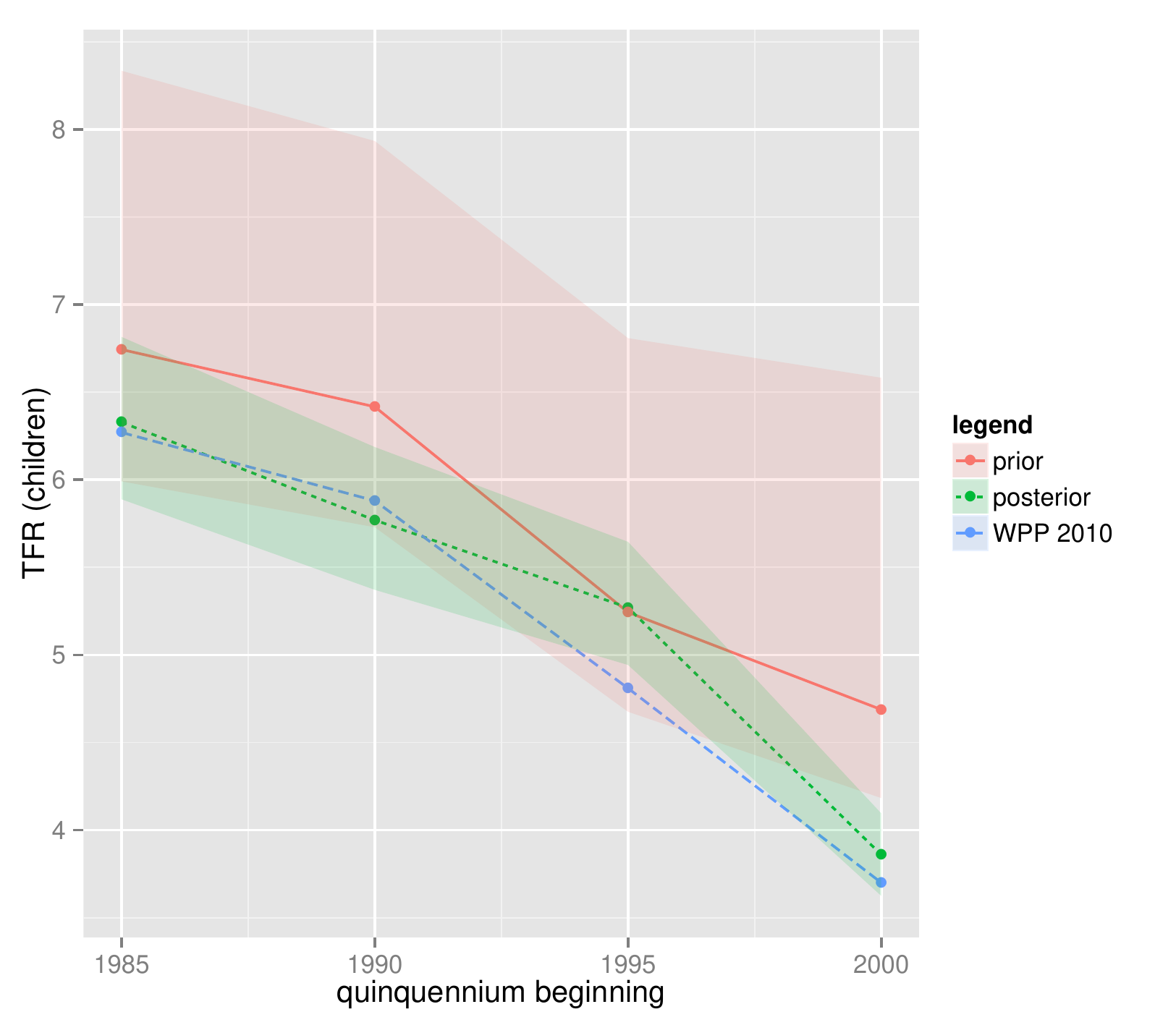}
\label{fig:laos-res-tfr}        
}
\subfloat[]{
  \GinKARWidth{0.5\textwidth}
\includegraphics{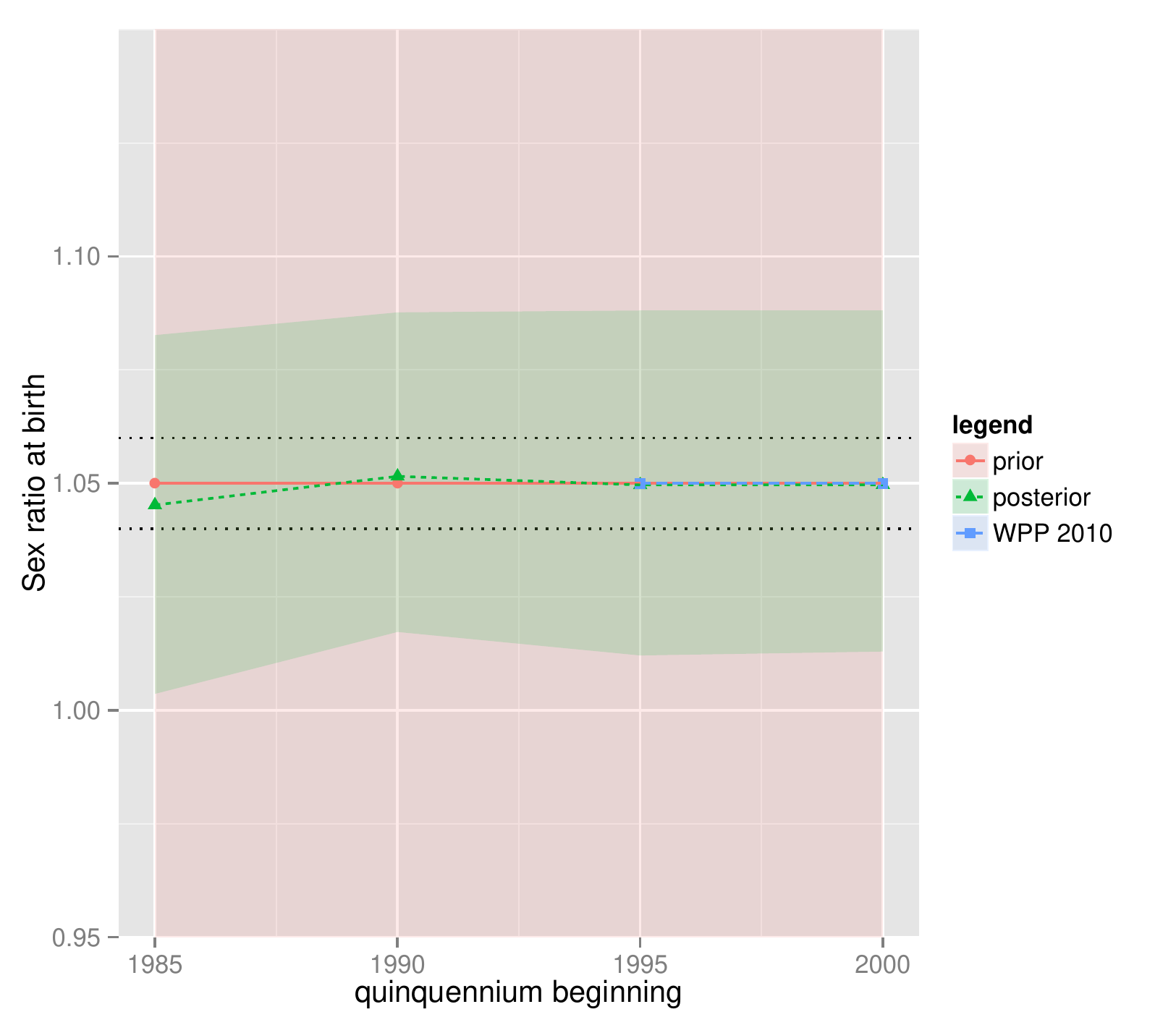}
\label{fig:laos-res-srb}        
}
  \caption{Prior and posterior medians and 95 percent credible intervals for the reconstructed population of Laos, 1985--2004: \protect\subref{fig:laos-res-tfr} \protect\acf{TFR}; \protect\subref{fig:laos-res-srb} sex ratio at birth (SRB).}
  \label{fig:laos-res-tfr-srb}
\end{figure}

Medians and prior and posterior credible intervals for \ac{TFR} and \ac{SRB} are shown in Figure~\ref{fig:laos-res-tfr-srb}. The posterior for \ac{TFR} obtained here (Figure~\ref{fig:laos-res-tfr}) is very similar that obtained by \citet{WheldonRafteryClarkEtAl_Bayesian__PAA2013} who reconstructed the female-only population and did not estimate \ac{SRB}; it was kept fixed at 1.05 throughout, a demographic convention \citep{PrestonHeuvelineGuillot2001}.

There was very little data on \ac{SRB} for Laos therefore, in this study, the initial estimate of \ac{SRB} was fixed at 1.05 in all quinquennia but a posterior distribution was estimated using the model. The posterior median \ac{SRB} deviates very little from the initial estimate, although the uncertainty has been considerably reduced; the mean of the half-widths of the 95 percent credible intervals is %
0.038 %
compared with %
0.39 %
for the prior (Figure~\ref{fig:laos-res-srb}). The probability that \ac{SRB} was above 1.06 in any of the quinquennia is low (Table~\ref{tab:laos-prob-var-quants-gt-106}) and the evidence for a trend over time is equivocal (Table~\ref{tab:laos-prob-dec-lin-all-params}).

\GinTextWidth

\begin{figure}[tbph]
  \centering
\subfloat[]{
\GinKARWidth{0.5\textwidth}
\includegraphics{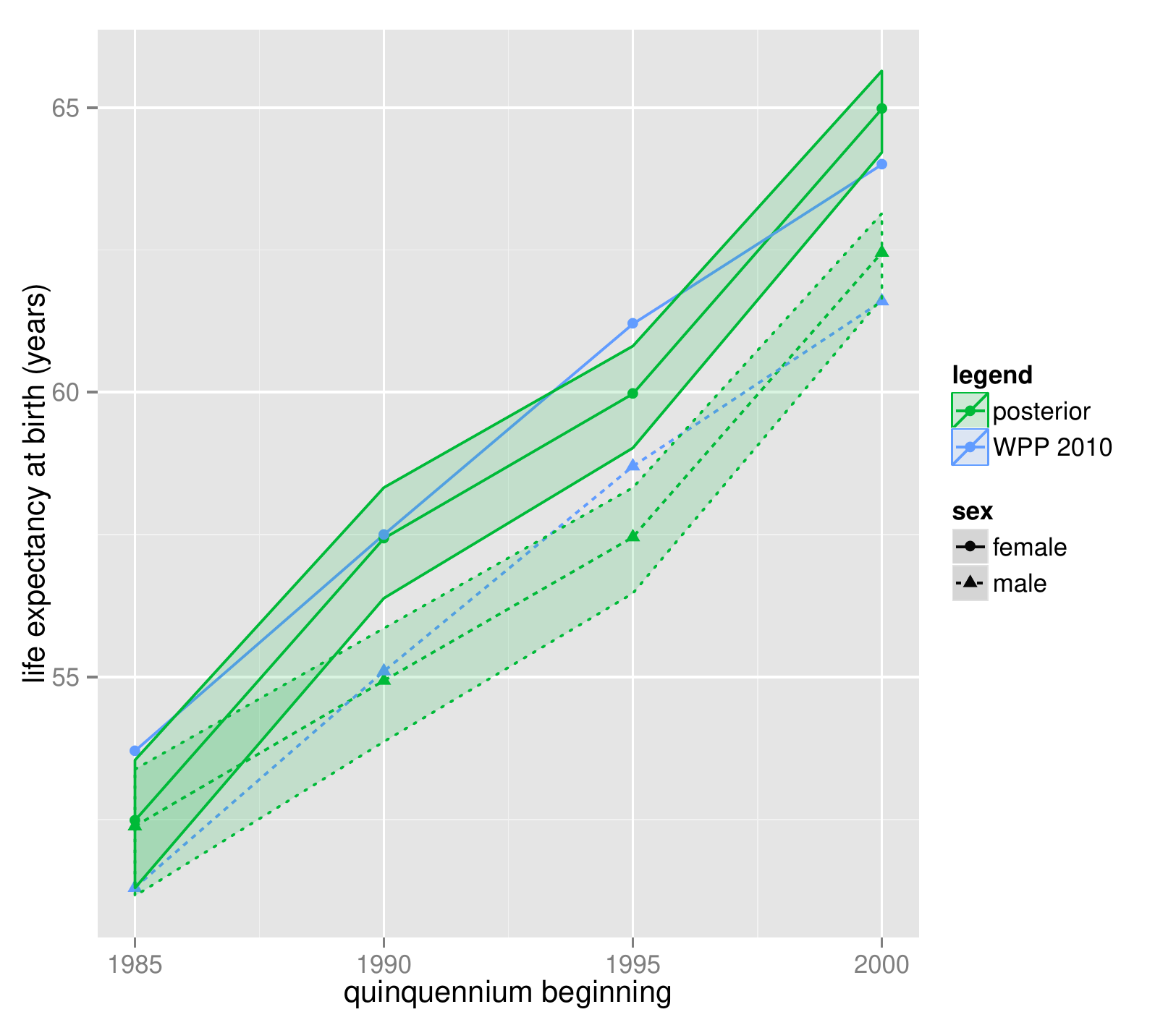}
\label{fig:laos-res-leb-femaleVSmale} 
} 
\subfloat[]{
\GinKARWidth{0.5\textwidth}
\includegraphics{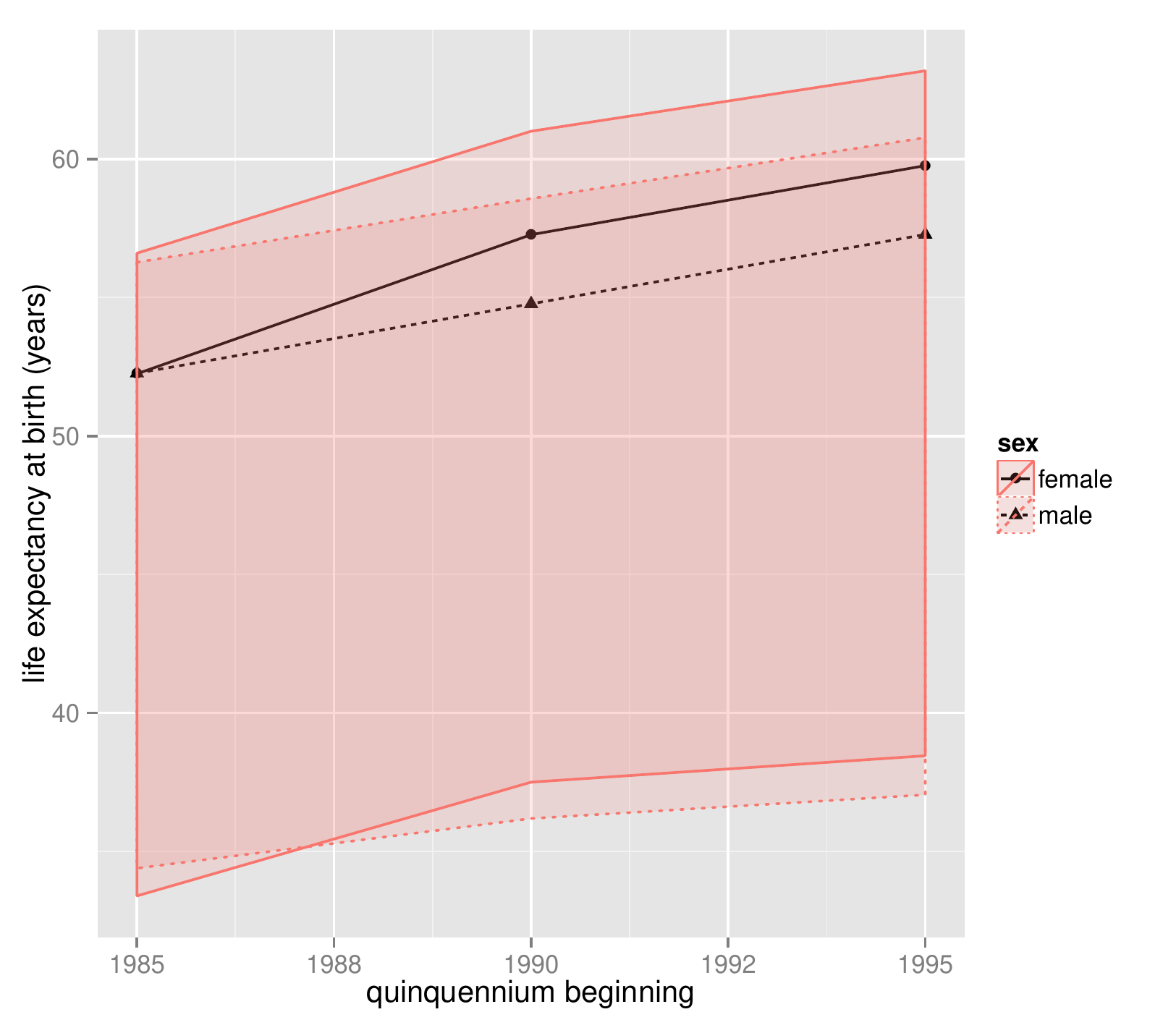}
\label{fig:laos-res-leb-femaleVSmale-prior} 
}\\ 
\subfloat[]{
\GinKARWidth{0.5\textwidth}
\includegraphics{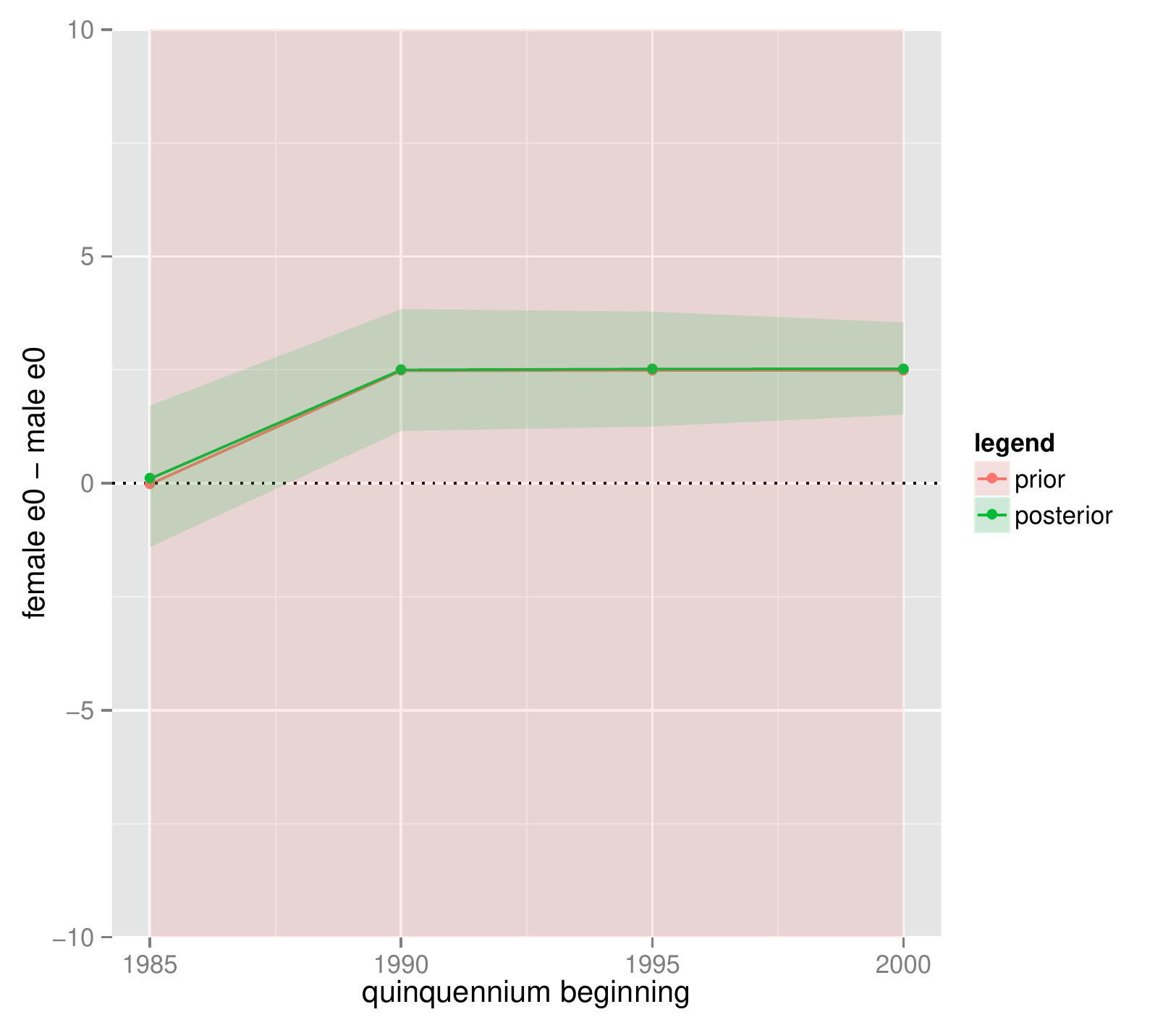}
\label{fig:laos-res-leb-female-male-difference} 
}
  \caption{Prior and posterior medians and 95 percent credible intervals for \protect\acf{e0} for the reconstructed population of Laos, 1985--2004. \protect\subref{fig:laos-res-leb-femaleVSmale}: Female and male posterior quantiles with \protect\acf{WPP} 2010 estimates; \protect\subref{fig:laos-res-leb-femaleVSmale-prior}: Female and male prior quantiles only; \protect\subref{fig:laos-res-leb-female-male-difference}: Sex difference (female$-$male).}
  \label{fig:laos-res-leb}
\end{figure}

There is strong evidence that female \ac{e0} was higher from 1990 through 2005 but there appears to be no evidence for a sex difference between 1985 and 1990 (Figure~\ref{fig:laos-res-leb}, Table~\ref{tab:laos-prob-var-quants-gt-106}b). The posterior distributions of both trend summaries provide strong evidence for an increase in the sex difference over the period of reconstruction (Table~\ref{tab:laos-prob-dec-lin-all-params}), although this is due primarily to the increase immediately following the 1985--1990 period.

Results for \ac{SRU5MR}, net international migration and population sex ratios are in the appendices.

\GinTextWidth

\begin{table}[bpth]
  \centering
  \caption{Probability that sex ratios and differences exceeded certain thresholds for the reconstructed population of Loas, 1985--2005, by quinquennium.}
  \label{tab:laos-prob-var-quants-gt-106}
\begin{tabular}{rrrr}
  \hline
  1985 & 1990 & 1995 & 2005 \\\hline\\[-1.5ex]
  \multicolumn{4}{c}{\textit{(a)} $\Pr(\text{\textit{SRB}} > 1.06)$}\\[1ex]
 0.20 & 0.30 & 0.27 & 0.27 \\
   [1ex]
  \multicolumn{4}{c}{\textit{(b)} $\Pr(\text{\textit{female} $e_0$} - \text{\textit{male} $e_0$} > 0)$}\\[1.25ex]
 0.56 & 1.00 & 1.00 & 1.00 \\
   [1ex]


\hline
\end{tabular}
\end{table}

\begin{table}[tbph]
  \centering
  \caption{Probabilities of an increasing linear trend and 95 percent credible intervals for \protect\acf{SDe0} 
    for the reconstructed population of Laos, 1985--2000. Two measures of trend are used: the difference over the period of reconstruction and the slope coefficient from the \protect\acf{OLS} regression on the start year of each quinquennium.}
  \label{tab:laos-prob-dec-lin-all-params}
  \begin{tabular}{lrr}
  \hline
  Measure of trend  &  95 percent CI  &  Prob $>$ 0 \\\hline\\[-1.5ex]

  \multicolumn{3}{c}{\textit{(a) \protect\Acf{SRB}}}\\[1ex]
 \protect\acs{SRB}$_{2000}$ $-$ \protect\acs{SRB}$_{1985}$ & [$-$0.046, 0.061] & 0.58 \\
  \protect\acs{OLS} slope (\protect\acs{SRB} $\sim$ year) &  [$-$0.0034, 0.0042] & 0.56 \\\mbox{}\\

  \multicolumn{3}{c}{\textit{(b) \protect\Acf{SDe0}}}\\[1ex]
  (\protect\acs{SDe0})$_{2000}$ $-$ (\protect\acs{SDe0})$_{1985}$ & [0.54, 4.2] & 0.99 \\
  OLS slope (\protect\acs{SDe0} $\sim$ year) &  [0.025, 0.26] & 0.99 \\

  \hline
\end{tabular}
\end{table}

\section{Discussion}
\label{sec:discussion}

We have described Bayesian reconstruction for two-sex populations, a method of reconstructing human populations of the recent past which yields probabilistic estimates of uncertainty \citep{WheldonRafteryClarkEtAl_Reconstructing_JotASA,WheldonRafteryClarkEtAl_Bayesian__PAA2013}. We reconstructed the populations of Laos from 1985--2005, Thailand from 1960--2000 and India from 1971--2001, paying particular attention to sex ratios of fertility and mortality indicators. We estimate that the posterior probability that \ac{SRB} was above 1.06 is greater than 0.9 in India between 1991 and 2001 and the probability it increased over this period is about %
0.92. The \ac{SRB} was also above 1.06 with high posterior probability in Thailand from 1960--1970. We estimate that the probability it decreased between 1960 and 1980, then increased from 1985 to 2000, is %
0.84. We found no evidence for atypically high \acp{SRB}, or a trend over the period of reconstruction, for Laos, a country with much less available data than Thailand and India. In both Thailand and Laos, we found strong evidence that \ac{e0} was greater for females and, in Thailand, that the difference increased over the period of reconstruction. In India, the probability that female \ac{e0} was lower during 1971--1976 was %
0.79 but there was strong evidence for a narrowing of the gap through to 2001.

In its original formulation, Bayesian reconstruction was for female-only populations; here we show how two-sex populations can be reconstructed using the same framework. The method takes a set of data-derived, bias-reduced initial estimates of age-specific fertility rates, and age-sex-specific survival proportions, migration proportions and population counts from censuses, together with expert opinion on the measurement error informed by data if available. Bayesian reconstruction updates initial estimates using adjusted census counts via a Bayesian hierarchical model. The periods of reconstruction used in our applications begin in the earliest census year for which non-census \ac{vitalrate} data were available, and end with the year of the most recent census. Reconstruction can be done further ahead, but without a census the results are based purely on the initial estimates.

The initial estimates, $f_{a,t}^*$, $s_{a,t,l}^*$, $g_{a,t,l}^*$, $n_{a,t_0,l}^*$, $\mathsrb_t^*$, enter the model as fixed prior medians (Section~\ref{sec:modell-uncert}), and so the posterior will be sensitive to changes in these inputs. This is actually desirable, because the initial estimates are based heavily on data specific to that estimate; the posterior should be sensitive to changes in these data.

Information about measurement error variance is needed to set the model hyper parameters (\ref{eq:prior-sigmasq-params}). In our case studies, this took the form of expert opinion, elicited as 90percent probability intervals centred at the data-derived initial estimates. If available, additional data on measurement error could be used to inform the elicitations, or to replace them altogether if they are comprehensive enough. This is the case in many developed countries (e.g., New Zealand, \citet{WheldonRafteryClarkEtAl_Bayesian__PAA2013}), where post-census enumeration surveys and vital registration coverage studies are undertaken. However, we expect Bayesian reconstruction to be most useful for less developed and developing countries where fragmentary and unreliable data lead to a substantial amount of uncertainty. For these countries, expert opinion is a main source of information. We do expect the posterior to be somewhat sensitive to the elicited intervals since the initial estimates themselves do not contain much information about measurement error.

Previous methods of population reconstruction were purely deterministic, were not designed to work with the type of data commonly available for many countries over the last sixty years, or did not account for measurement error \citep[e.g.,][]{Lee_Econometric_1971,Lee_Estimating_PS1974,WrigleySchofield_Population_1981,BertinoSonnino_Stochastic_IPTOaNA2004}. \citet{DaponteKadaneWolfson_Bayesian_JotASA1997} used a Bayesian approach to construct a counterfactual population, but age patterns of fertility were held fixed and mortality varied only through infant mortality. Bayesian reconstruction does not impose fixed age-patterns and mortality can vary at each age. Moreover, international migration is estimated in the same way as fertility and mortality.


We considered sex ratios of births and mortality because these are of interest to demographers and policy-makers, especially since they determine the \ac{SRTP} \citep{GriMatHi2000UnderstandingD,Guillot_Dynamics_PS_2002}. It is conventional to compare sex-specific \acp{U5MR} with a ratio but sex-specific \acp{e0} with a difference. We have not studied the associations among \ac{U5MR} ratios, \ac{e0} differences and population sex ratios. The \ac{SRTP} is a function of life-time cohort mortality but the \ac{U5MR} and \ac{e0} presented here are period measures for which the relationship used by \citet{Guillot_Dynamics_PS_2002} does not hold. Our results add to previous work on \acp{SRM}, especially that of \citet{Sawyer_Child_PM_2012} who studied sex ratios of \ac{U5MR} and called for further work to quantify its uncertainty. \citet{Sawyer_Child_PM_2012} decomposed \ac{U5MR} into mortality between ages 0 and 1 (infant mortality) and mortality between ages 1 and 5 (child mortality). We reconstructed populations in age- time-intervals of width five because these are the intervals for which data is most widely available across all countries.

Many methods of adjusting vital registration using census data have been proposed \citep[e.g.,][]{BennettHoriuchi_Estimating_PI1981,Hill_Estimating__1987,LutherRetherford_Consistent_MPS1988}, but these deal with intercensal migrations by essentially truncating the age groups most affected by migration \citep{MurrayRajaratnamMarcusEtAl_What_PM_2010}, or ignore them altogether. \citet{LutherDhanasakdiArnold_Consistent__1986} and \citet{HillVapattanawongPrasartkulEtAl_Epidemiologic_IJoE_2007} applied these methods to Thailand to estimate undercounts. The aim of these methods is to produce improved point estimates of \acp{vitalrate}. We have not used any census data to derive our initial estimates. For example, initial estimates of survival were not based on inter-censal cohort survival and initial estimates of migration were not based on ``residual'' counts. Doing so would have amounted to using the census data twice and would have underestimated the uncertainty. The outputs of Bayesian reconstruction are interval estimates which quantify uncertainty probabilistically.

\subsection{Sex Ratios in Asia}
\label{sec:sex-ratios-asia}

The \ac{SRTP} is a crude measure of the balance of sexes in a population since it is not age-specific; sex ratios may be quite different among age groups, for example. Nevertheless, there is a large literature devoted to estimating \acp{SRTP} and exploring the causes and consequences, especially in Asia where \acp{SRTP} are the highest in the world \citep{Nations2011World}. A persistently high sex ratio among younger cohorts could lead to a ``marriage squeeze'' in which many young males will struggle to marry due to a lack of eligible females \citep{Guilmoto_Skewed_D_2012}. Formally, high \acp{SRTP} are the result of high \acp{SRB} and life-time sex ratios of cohort mortality \citep{Guillot_Dynamics_PS_2002}. The relative effects of these two factors may vary by time and country.

A normal range for \ac{SRB} is believed to be 1.04--1.06 \citep{Fund2010UNFPA}; on average, slightly more than half of all newborns are male. Concerns about quality, or a complete lack of data, have made it difficult to accurately estimate \acp{SRB} in many Asian countries. Bayesian reconstruction of these populations quantifies the accuracy of \ac{SRB} estimates probabilistically.

With very few exceptions, country-level \acp{e0} for females exceed those of males. This pattern is consistent with a female survival advantage; in virtually all contemporary human populations, females age slower and live longer than males  \citep{SoliLuchGeneticDemog2006,VallinMortality_Demog2006}. Behavioural factors appear to play an important role, especially in developed countries where the prevalence of harmful activities, such as smoking, consumption of alcohol, and risky behaviour leading to accidental death, tends to be lower among women than among men \citep{Waldron_What_PBotUN_1985,Waldron_Gender__,LalicRaftery_Joint__PAA2012}. The persistence of lower male \ac{e0} across the development spectrum suggests there are other factors as well. In less developed countries, \ac{e0} is determined more by infant and child mortality. \citet{Pongou2012_WhyDiff_Demog} studied infant mortality in Africa and found an association with parental characteristics extant prior to conception. Biological, as well as behavioural, determinants have also been proposed but there is little consensus on the specific mechanism \citep{Austad_Sex__2011}. The only countries in which female \ac{e0} is not higher are countries in Africa and some in South and Central Asia. The African countries are those with generalized HIV epidemics where mortality due to AIDS changes the typical sex difference in \ac{e0} \citep{Nations2011World}. Those in Asia include, most notably, India where cultural preference for males is thought to be the major cause.

\subsubsection{India}
\label{sec:india}

The \ac{SRTP} in India has received considerable attention, particularly as an indicator of discrimination against women \citep{Fund2010UNFPA}. \citet{GriMatHi2000UnderstandingD} showed that only slightly elevated \acp{SRB} and \acp{SRM} at young ages are sufficient to produce the observed \ac{SRTP} in India, if they persist for a long period of time. The relative contribution of these two factors may have varied over time. \citet{Bhat_Trail_EaPW_2002a} and \citet{GuilmotoCharacteristicsIndia2007,Guilmoto_Sex_PaDR_2009} argue that India experienced a transition in the 1970s whereby high \ac{SRB} replaced low \ac{SRM} as the cause of the high \acp{SRTP} observed throughout the period (recall that a low \ac{SRM} is a result of lower male mortality). Evidence suggests that low \acp{SRM} were due to female neglect and infanticide. In the 1970s, these practices gave way to sex selective abortion which raised the \ac{SRB} instead.

The transition hypothesis is based on several pieces of evidence. Data suggest a possible rise in \ac{SRB} in India above the typical range of 1.04--1.06 in the mid 1980s \citep{GuilmotoCharacteristicsIndia2007}. In the 1970s, amniocentesis started to became available as a method for determining the sex of a foetus and abortion was legalized. Ultrasonography, a less invasive way of determining foetal sex, started to became available in many parts of India in the 1980s. In certain regions, such as the northern states and highly urbanized areas, there is a long standing tradition of preference for sons over daughters \citep{Mayer_Indias_PaDR_1999,Bongaarts_Fertility_PaDR_2001}. The arrival of these new technologies in this context appears to have led to an increase in sex selective abortions in India \citep{Bhat_Trail_EaPW_2002a,GuilmotoCharacteristicsIndia2007_ES,JhaKumarVasaEtAl_Low_TL_2006,Fund2010UNFPA,JKKRRABC2011TrendsTL}. The steep decline in Indian \ac{TFR}, which began in the early 1970s, could have increased the prevalence of such procedures. Several studies have found evidence that \ac{SRB} is higher at higher parities (birth orders), both in India \citep{GuptaBhat_Fertility_PS_1997,Bhat_Trail_EaPW_2002a,JhaKumarDhingra_Sex_TL_2006,JKKRRABC2011TrendsTL} and other Asian countries \citep{Gupta_Explaining_PaDR_2005}. The increase appears to be greater if none of the earlier births were male. As \ac{TFR} decreases so does the average family size, so the risk of having no sons increases \citep{Guilmoto_Sex_PaDR_2009}. Therefore, in cultures where sons play important economic and social functions, or where families benefit materially much more from the marriage of a son than of a daughter, the incentive to use sex selective abortion increases \citep{Mayer_Indias_PaDR_1999,Guilmoto_Sex_PaDR_2009}.

After combining all available data and including uncertainty, we estimate that the probability there was an increase in \ac{SRB} between 1971 and 2001 is above %
0.92. %
The probability that female \ac{U5MR} was higher over this period is estimated to be between %
0.72 %
and %
0.89, %
but there was no evidence of a trend. Therefore, our results provide support for the \ac{SRB} part of the transition hypothesis but not the \ac{U5MR} part.

Overall mortality decreased rapidly in India from about 1950 as infectious diseases were brought under control, food security increased and health services became more widely available \citep{Bhat_Trail_EaPW_2002a}. Our results suggest that, after taking account of uncertainty, there was an increase in \ac{e0}  and a continual decrease in \ac{U5MR} between 1971 and 2001 for both sexes. Using Sample Registration System data, \citep{Bhat_Trail_EaPW_2002a} noted that the decrease was greater for females than males and our analysis of the change in the sex difference of \ac{e0} supports this; we found that the probability that the difference increased over the period of reconstruction is about %
0.98. %
The probability of a decline in the \ac{SRTP} was found to be similarly strong. However, as with \ac{SRU5MR}, there was little evidence for a trend in \ac{SRU5}.

India's large population makes it a very important case for the study of sex ratios in Asia \citep{GuilmotoCharacteristicsIndia2007} and, like other authors \citep[e.g.,][]{Guillot_Dynamics_PS_2002,Nations2011World}, we have focused on country level estimates only. However, where available, data suggest that there are large regional variations in \acp{SRB} and population sex ratios, with estimates for urban areas and northern states being much higher than other areas \citep{Bhat_Trail_EaPW_2002a,Guilmoto_Sex_PaDR_2009,JKKRRABC2011TrendsTL}. Currently, Bayesian reconstruction is not able to produce sub-national estimates, but could be extended to do so in the future.

\subsubsection{Thailand}
\label{sec:thailand-5}

Like other parts of Asia, Thailand experienced rapid economic growth and a rapid fall in \ac{TFR} beginning in 1960. The \ac{TFR} decline was accompanied by an increase in the widespread use of modern contraceptive methods made available through government supported, voluntary family planning programs \citep{KamnuansilpaChamratrithirongKnodel_Thailands_IFPP_1982,Knodel_Thailands_SS_1987,KnodelRuffoloRatanalangkarnEtAl_Reproductive_SiFP_1996}. Unlike India, Thailand is not considered to have had a high \ac{SRB} \citep{Guilmoto_Sex_PaDR_2009}. Vital registration data, which formed the basis of our initial estimates, indicate that \ac{SRB} was high during the period 1960 to 1970, but remained within the typical range thereafter. This is consistent with studies conducted after the early 1970s which found that small families of two or three children consisting of at least one boy and one girl was the most commonly desired configuration in Thailand. \Ac{TFR} decline in Thailand may have intensified this preference \citep{KnodelPrachuabmoh_Preferences_SiFP_1976,KnodelRuffoloRatanalangkarnEtAl_Reproductive_SiFP_1996}.

Posterior intervals for sex ratios of mortality in Thailand reflect the typical pattern which is one of higher female life expectancy. There is no evidence that this pattern was also true for \ac{U5MR} (on-line resources).

\subsubsection{Laos}
\label{sec:laos-5}

Fertility in Laos remained high relative to its neighbours. For example, the estimated \ac{TFR} for 1985 in Laos is comparable to the 1960 estimate for Thailand \citep{Frisen1991PopulationA}. Posterior uncertainty about \ac{SRB} is high and our results provide no evidence to suggest levels were atypical between 1985 and 2005. All-age mortality as summarized by \ac{e0} does appear to have been higher for females from 1990 onward but, as with Thailand, there is no evidence that this advantage held for \ac{U5MR} (on-line resources).

\subsection{Further Work and Extensions}
\label{sec:extens-furth-work}

Our prior distributions were constructed from initial point estimates of the \ac{CCMPP} input parameters, together with information about measurement error. In the examples given here, this was elicited from \ac{UN} analysts who are very familiar with the data sources and the demography of each country. In cases where good data on measurement error are available, they can be used. For example, \citet{WheldonRafteryClarkEtAl_Bayesian__PAA2013} used information from post-enumeration surveys to estimate the accuracy of New Zealand censuses. Data of this kind are rarely available in developing countries for either census or \ac{vitalrate} data. Nevertheless, surveys come with a substantial amount of meta-data which can be used to model accuracy. This approach was taken by \citet{AlkemaRafteryGerlandEtAl_Estimating_DR_2012} who developed a method for estimating the quality of survey-based \ac{TFR} estimates in West Africa by modelling bias and measurement error variance as a function of data quality covariates. \citet{WheldonRafteryClarkEtAl_Reconstructing_JotASA} used these results in their reconstruction of the female population of Burkina Faso. Extending this work to more countries could provide a source of initial estimates together with uncertainty that could be used as inputs to Bayesian reconstruction.

Posterior uncertainty in estimates of \ac{U5MR} was found to be substantial and we did not find strong evidence for a skewed sex ratio in any of our examples. \ac{U5MR} is under-identified in the model because the only census count it affects, that for ages 0--5, is also dependent on \ac{SRB} and \ac{TFR}. Recent work by the \ac{UN} \ac{IGME} has focused on producing probabilistic estimates of \ac{U5MR} \citep{HillYouInoueEtAl_Child_PM_2012,AlkemaAnn_2011_EstU5_PLoSO}. Further work could look at ways of using these estimates in Bayesian reconstruction to improve estimation of \ac{U5MR}.

The census counts we used were not raw census output but adjusted counts published in \ac{WPP}. In some cases, these counts are adjusted by the \ac{UNPD} to reduce bias due to factors such differential undercount by age. This may have led to an underestimate of uncertainty. The use of raw census outputs instead is worth investigating, but the effects of undercount would still need to be addressed through modifications to the method.

Our prior distributions for international migration were centred at zero with large variances. This is a sensible default when accurate data are not available. Further work could investigate the possibility of using stocks of foreign-born, often collected in censuses, to provide more accurate initial estimates. Data on refugee movements is another source that could be investigated where available.

We have reconstructed national level populations only, principally because this is the level at which the \ac{UN} operate. We have already mentioned that sub-national reconstructions might be of interest. Subnational reconstructions could be done without any modifications to the method if the requisite data are available; national level initial estimates and population counts would just be replaced with their subnational equivalents and the method applied as above. Reconstructing adjacent regions, or regions between which there is likely to be a lot of migration would require special care, however, as there is no way of accounting for dependence among migratory flows under the current approach.

\clearpage
\appendix

\section{Derivation of Demographic Indicators}
\label{sec:defin-age-summ}

\subsection{Total Fertility Rate}
\label{sec:total-fertility-rate}

The \ac{TFR} for the period $[t,t+5)$ is the average number of children born to women of a hypothetical cohort which survives through ages $a_L^{\text{[fert]}}$ to $a_U^{\text{[fert]}}$, all the way experiencing the age-specific fertility rates, $f_{a,t}$. Its definition is:
\begin{equation*}
  \mathit{TFR}_t \equiv 5 \cdot \sum_{a_L^{\text{[fert]}}}^{a_U^{\text{[fert]}}} f_{a,t}.
\end{equation*}

\subsection{Life Expectancy at Birth}
\label{sec:life-expectancy-at}

\Acl{e0} is the average age at death for members of a hypothetical cohort which, at each age, experience age-specific survival $s_{a,t,l}$. Its definition is:
\begin{equation*}
  e_{0,t,l} = 5 \sum_{a=0}^{A} \prod_{i=0}^a s_{i,t,l} + 5\left(\prod_{i=0}^{A} s_{i,t,l}\right) \bigl(s_{A+5,t,l}/(1-s_{A+5,t,l})\bigr).
\end{equation*}
The derivation is straightforward and can be found in \citet{WheldonRafteryClarkEtAl_Reconstructing_JotASA}.

\subsection{Under 5 Mortality Rate}
\label{sec:under-5-mortality}

The \ac{U5MR} is constructed in the same way as the standard ``Infant Mortality Rate'' \citep[e.g.,][Ch.~2]{PrestonHeuvelineGuillot2001}, except for the age interval [0,5). It is defined as follows:
\begin{equation*}
  \mathit{U5MR}_t \equiv \frac{\text{No.\ deaths to those aged [0,5) in the interval [$t$,$t$+5)}}{\text{No.\ births in the interval [$t$,$t$+5)}}.
\end{equation*}
It is neither a true demographic rate nor a probability but, nevertheless, is in common use.

\section{Application: Further Details About Data Sources}
\label{sec:data-sources}

\subsection{India}
\label{sec:india}

\subsubsection{Population Counts}
\label{sec:population-counts}

Censuses were conducted in 1971, 1981, 1991 and 2001. We used the counts in \ac{WPP} 2010 which were adjusted for slight undercount in some age groups.

\subsubsection{Sex Ratio at Birth}
\label{sec:sex-ratio-at-1}

Data on sex ratio at birth came from the same sources as data on age-specific fertility (Section~\ref{sec:fertility-2}). A weighted cubic spline was used to smooth the available data and initial estimates were derived by evaluating the spline at the mid-points of the quinquennia 1971--1976, \ldots, 1996--2001. Elicited relative error for this parameter was set at 10 percent.

\subsubsection{Fertility}
\label{sec:fertility-2}

Initial estimates of age-specific fertility were based on  data from the Indian \ac{SRS} \citep{The_Vital__2011} and \acp{NFHS} conducted between 1992 and 2006 \citep{Population_National__2009}. These were weighted and smoothed using the same procedure as for Thailand. Elicited relative error for this parameter was set at 10 percent.

\subsubsection{Mortality}
\label{sec:mortality-2}

Initial estimates of survival proportions were calculated from abridged life tables based on data from the \ac{SRS} from 1968-1969 through 2008. The average number of person-years lived in each interval (${}_na_x$s) were computed using \poscite{Greville_Short_TRotAIoA_1943} formula from age 15 and above. Values for ages under 5 are based on the formulae of \citet{CoaleDemeny_Regional_1983} using the West pattern. All other values were set to 2.5 \citep[see also][Ch.~6]{United_Model_1982}.

Additional sources were used for infant and child mortality: data on births and deaths under-five were calculated from maternity-history data from the \acp{NFHS} conducted between 1992 and 2006 and data on children ever-born and surviving classified by age of mother from these surveys and earlier ones as well as from the 2002--04 \ac{RCHS}. Weighted cubic splines were used to smooth estimates of ${}_1q_0$ and ${}_4q_1$ from these data sources. The weights were determined by \ac{UN} analysts based on their expert judgement about the relative reliability of each source. Elicited relative error was set to 10 percent.

\subsubsection{Migration}
\label{sec:migration-2}

We used the same initial estimates for international migration as for Laos.

\subsection{Thailand}
\label{sec:thailand-1}

\subsubsection{Fertility}
\label{sec:fertility-1}

Initial estimates of age-specific fertility were based on direct and indirect estimates of current fertility and \ac{CEB} taken from the Surveys of Population Change conducted between 1964 and 1996 \citetext{\citealp{Thailand_Report__1970, Thailand_Report__1977, Thailand_Report__1992,Thailand_Report__1997}; see also \citealp{Vallin_La_PE_1976}}, the \acp{WFS} \citep{InternationalSocial_Fertility__1993}, the Thai Longitudinal Study of Economic, Social and Demographic Change \citep{KnodelPitaktepsombati_Fertility_SiFP_1975}, the 1968--1972 Longitudinal Study of Economic, Social and Demographic Change \citep{National_Fertility__1980}, the 1978 and 1996 National Contraceptive Prevalence Surveys \citep{SuvanajataKamnuansilpa_Thailand__1979,ChamratrithirongMahidonSangkhomEtAl_National__1997}, the 1987 \ac{DHS}, and vital registration.

Age patterns and levels of fertility were estimated separately from the available data. The available data were smoothed within age and within year using weighted splines. The weights were determined by \ac{UN} analysts based on their expert judgement about the relative reliability of each source. Estimates of \ac{TFR}, based on summed age-specific estimates, were obtained similarly. The final set of initial estimates was obtained by multiplying the smoothed \acp{TFR} by the smoothed age patterns. Elicited relative error for this parameter was set at 10 percent.

\subsubsection{Mortality}
\label{sec:mortality-1}

Initial estimates of mortality for both sexes were based on life tables calculated from vital registration. Vital registration is thought to have underestimated \ac{U5MR} \citep{HillVapattanawongPrasartkulEtAl_Epidemiologic_IJoE_2007} so the following additional sources were used to derive initial estimates of under five mortality: (a) the 1974--1976, 1984--86, 1989, 1991, 1995--1996, Surveys of Population Change \citetext{\citealp{Thailand_Report__1970, Thailand_Report__1977, Thailand_Report__1992,Thailand_Report__1997}}, (b) maternity-history data from the 1975 \ac{WFS} and 1987 \ac{DHS}, (c) data on children ever born and surviving from these surveys, the 1981--1986 \acp{CPS}. All available non-census data on ${}_5q_0$ from these sources were weighted by \ac{UN} analysts who used their expert judgement about potential biases and smoothed over time using cubic splines. The splines were evaluated at 1962.5, 1968.5, \ldots 1998.5 and the values substituted for the ${}_5q_0$s in the life tables based solely on vital registration. Survival proportions were derived from these ``spliced'' life tables.

\subsection{Laos}
\label{sec:laos-1}

Initial estimates for females were the same as those used by \citet{WheldonRafteryClarkEtAl_Bayesian__PAA2013}. Initial estimates for males were derived using the same procedures.

\subsubsection{Population Counts}
\label{sec:sex-ratio-at}

National censuses were conducted in 1985, 1995 and 2005. Following \citet{WheldonRafteryClarkEtAl_Bayesian__PAA2013}, we used the census year counts in \ac{WPP} 2010; there were no post-enumeration surveys, but these counts were adjusted to compensate for undercount in certain age groups.

\subsubsection{Fertility}
\label{sec:fertility}

Direct and indirect estimates were based on \ac{CEB} and recent births (preceding 12 and 24 months), all by age of mother, collected by the 1993 Laos Social Indicator Survey, the 1995 and 2005 censuses, the 1994 Fertility and Birth Spacing Survey, the 2000 and 2005 Lao Reproductive Health Surveys, the 1986--1988 multi-round survey and the 2006 MICS3 survey. \poscite{Zaba_Use__1981} Relational Gompertz Model, \poscite{Arriaga_Estimating_1983} method, the $P/F$ method \citep{Brassothers_Demography_1968}, \poscite{Arriaga_Estimating_1983} modified $P/F$ method and the Brass fertility polynomial \citep{Brass_Graduation_PS_1960,BrassCELADE_Methods__1975} were used to derive indirect estimates.

Age patterns of fertility were estimated by taking medians across all available data points within the quinquennia 1985--1990, \ldots, 2000--2005. \ac{TFR} was estimated separately by converting each age-specific data series into series of \ac{TFR} by summing. Medians within quinquennia were then taken. The final initial estimates of the age-specific rates was obtained by multiplying the median age-patterns by the median \acp{TFR}. Data points were not weighted.

The elicited relative error was set to 10 percent. Hence, the medians of the initial estimate distributions for age-specific fertility rates were the initial point estimates and the upper 95 percent and lower five percent quantiles were equal to the initial point estimates plus and minus 10 percent, respectively.

\subsubsection{Mortality}
\label{sec:mortality}

The available data only provide estimates of mortality under age five. These are maternity histories and \ac{CEB} and surviving data from the 1993 Laos Social Indicator Survey, the 1994 Fertility and Birth Spacing Survey, the 1995 and 2000 censuses and the 2000 and 2005 Lao Reproductive Health Surveys (intercensal survival estimates were not used). Weighted cubic smoothing splines were used to smooth these to produce single initial point estimates of the average $\ltp{1}{q}{0}$ and $\ltp{5}{q}{0}$ over each quinquennium 1985--1990, \ldots, 2000--2005. For each quinquennium, two \ac{CDWest} life tables were found; one with $\ltp{1}{q}{0}$ closest to that produced by the smoothing procedure and one with $\ltp{5}{q}{0}$ closest. Within each pair, the \acp{e0} in these two tables were averaged. Age-specific survival proportions calculated from the \ac{CDWest} table with \ac{e0} closest to this average were then taken as the initial estimates of mortality at all ages. The elicited relative error of these initial point estimates was set to 10 percent.

\subsubsection{Migration}
\label{sec:migration}

There is very little quantitative information about international migration in Laos between 1985 and 2005. To reflect this, initial estimates for females and males were set to zero and the elicited relative error was high, at 20 percent.

\section{Application: Further Results}
\label{sec:further-results}

\subsection{India}
\label{sec:india-2}

\begin{figure}[tbph]
  \centering
\subfloat[]{
\GinKARWidth{0.5\textwidth}
\includegraphics{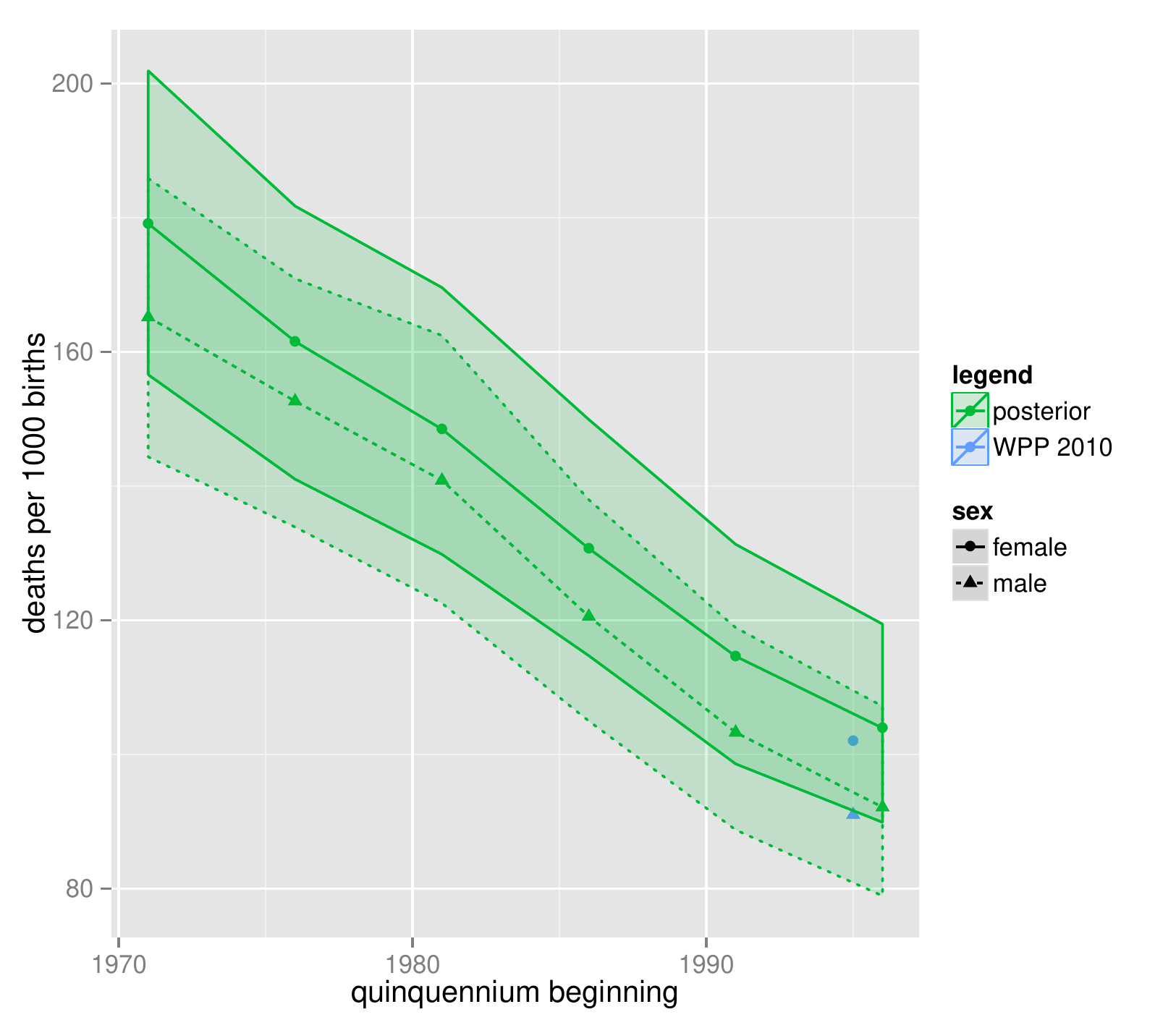}
\label{fig:indi-res-IMR-femaleVSmale} 
}
\subfloat[]{
\GinKARWidth{0.5\textwidth}
\includegraphics{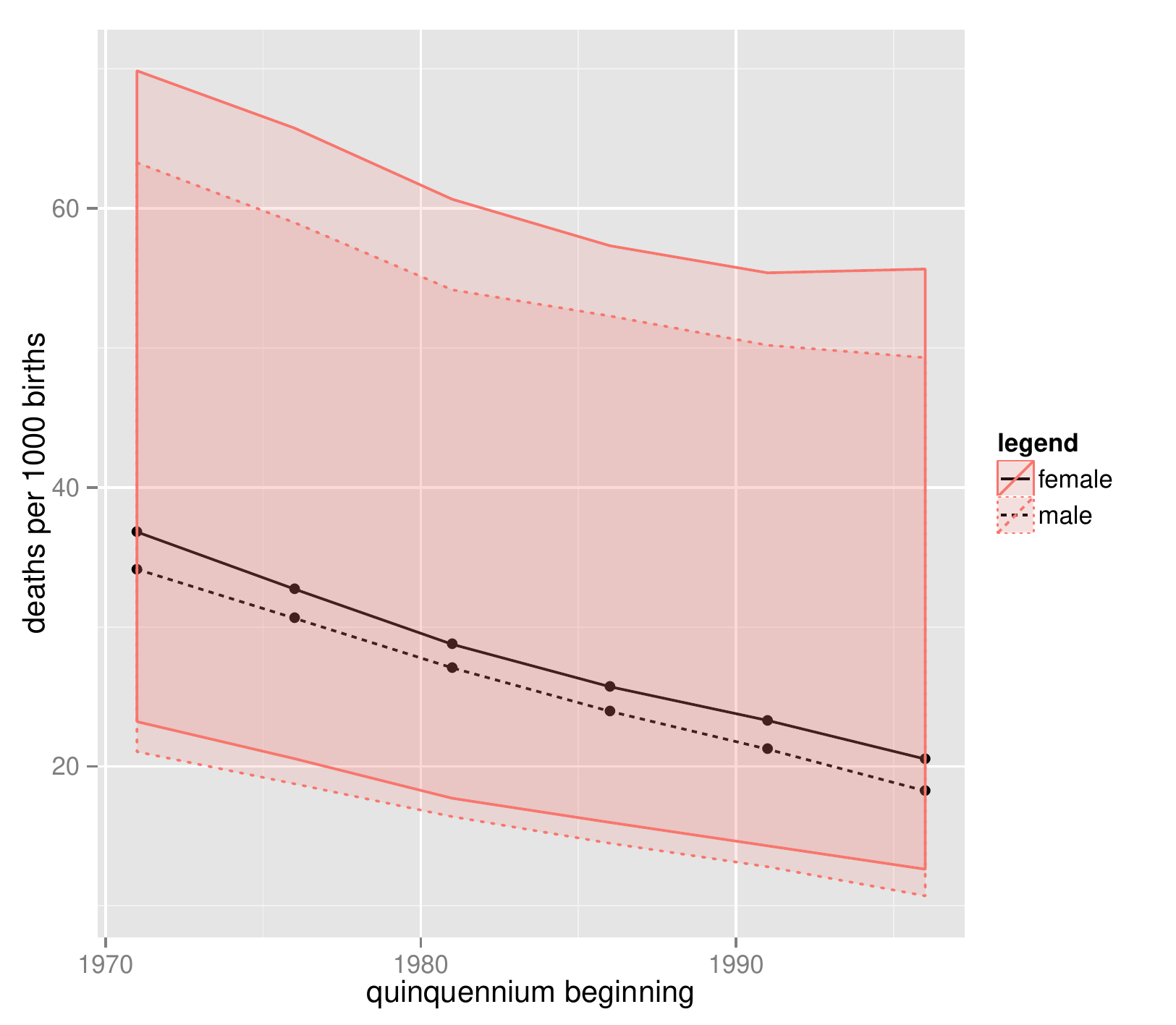}
\label{fig:indi-res-IMR-femaleVSmale-prior} 
}\\ 
\subfloat[]{
\GinKARWidth{0.5\textwidth}
\includegraphics{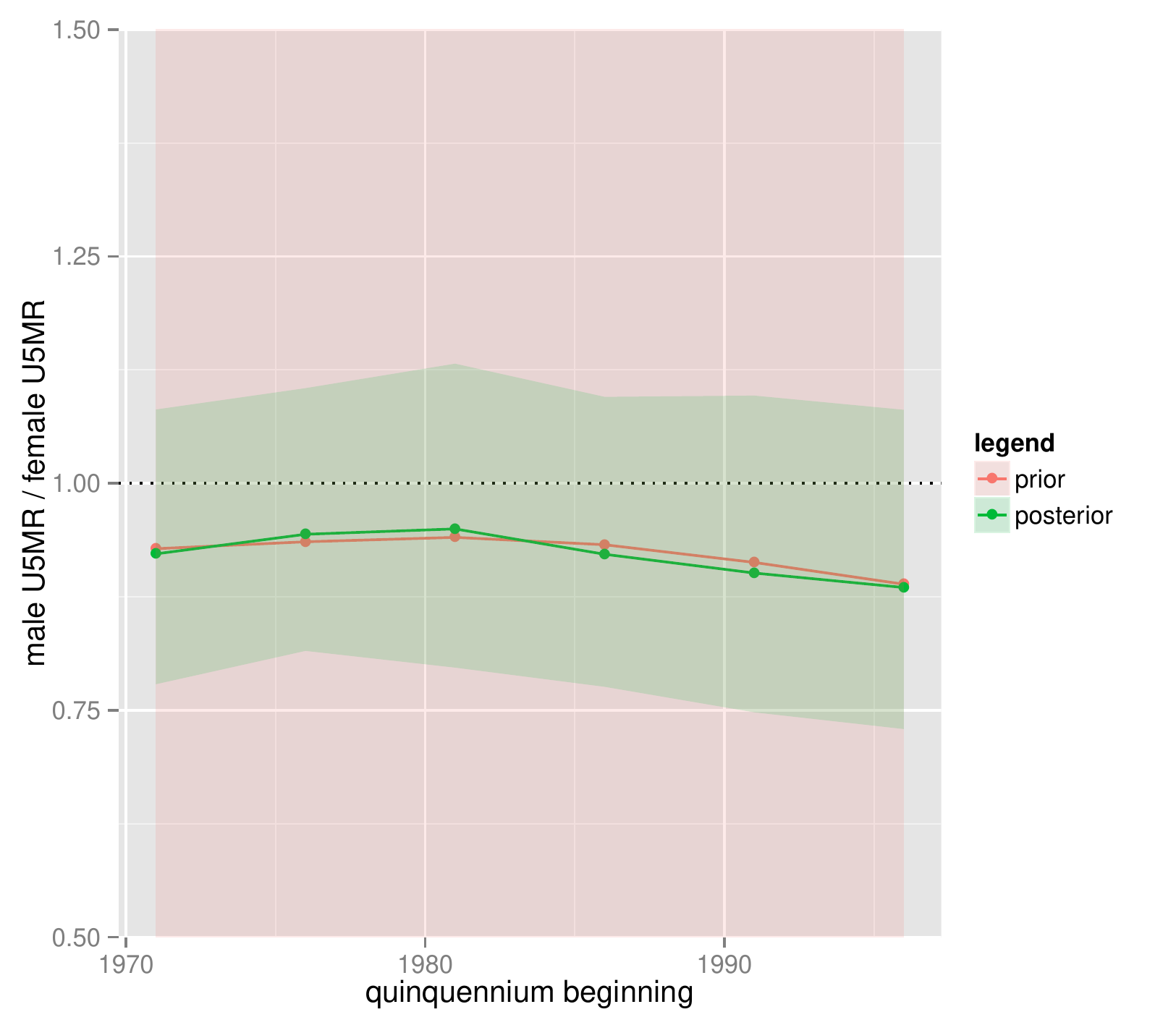}
\label{fig:indi-res-IMR-female-male-ratio} 
}
  \caption{Prior and posterior medians and 95 percent credible intervals for \protect\acf{U5MR} (deaths per 1,000 live births) for the reconstructed population of India, 1971--2001 \protect\subref{fig:indi-res-IMR-femaleVSmale}: Female and male posterior quantiles and \protect\acf{WPP} 2010 estimates; \protect\subref{fig:indi-res-IMR-femaleVSmale-prior}: Female and male prior quantiles only; \protect\subref{fig:indi-res-IMR-female-male-ratio}: Male-to-female ratio.}
  \label{fig:indi-res-IMR}
\end{figure}

Results for \ac{U5MR} are in Figure~\ref{fig:indi-res-IMR}. Posterior medians of Indian \ac{U5MR} for males are lower than those for females. However, although the sex ratio in \ac{U5MR} is centred below one, the posterior intervals are relatively wide and contain it (mean half-width %
0.16 %
deaths per live births; Figure~\ref{fig:indi-res-IMR}). The probability that female \ac{U5MR} exceeded that of males in each quinquennium is given in Table~\ref{tab:indi-prob-sex-ratio-U5MR-gt-1}; by the period 1996--2001 this had peaked at %
0.89. %
The evidence for a linear decrease over the period of reconstruction is weak (Table~\ref{tab:indi-prob-dec-online-supp}a). There is a slight decline in the posterior median after 1981 but the evidence for a trend after this point is similar: the probability of a simple decrease is %
0.7 %
and the probability that the \ac{OLS} coefficient was less than zero is %
0.68 %
(Table~\ref{tab:indi-prob-dec-online-supp}b).

\GinTextWidth

\begin{table}[bpth]
  \centering
  \caption{Probability that \protect\acf{SRU5MR} was less than one for the reconstructed population of India, 1971--2001, by quinquennium.}
  \label{tab:indi-prob-sex-ratio-U5MR-gt-1}
\begin{tabular}{rrrrrr}
  \hline
  1971 & 1976 & 1981 & 1986 & 1991 & 1996 \\\hline
 0.85 & 0.77 & 0.72 & 0.83 & 0.86 & 0.89 \\
   [1ex]\hline
\end{tabular}
\end{table}

\begin{table}[tbph]
  \centering
  \caption{Probabilities of an increasing linear trend and 95 percent credible intervals for 
    \protect\acf{SRU5MR} 
    for the reconstructed population of India, 1971--2001. Two measures of trend are used: the difference over the period of reconstruction and the slope coefficient from the \protect\acf{OLS} regression on the start year of each quinquennium.}
  \label{tab:indi-prob-dec-online-supp}
  \begin{tabular}{lrr}
  \hline
  Measure of trend  &  95 percent CI  &  Prob $>$ 0 \\\hline\\[-1.5ex]



  \multicolumn{3}{c}{\textit{(a) \protect\Acf{SRU5MR}, 1971--2001}}\\[1ex]
  \protect\acs{SRU5MR}$_{1996}$ $-$ \protect\acs{SRU5MR}$_{1971}$ & [$-$0.26, 0.2] & 0.38 \\
  \protect\acs{OLS} slope (\acs{SRU5MR} $\sim$ year) &  [$-$0.0097, 0.0061] & 0.32\\\mbox{}\\

  \multicolumn{3}{c}{\textit{(b) \protect\Acf{SRU5MR}, 1981--2001}}\\[1ex]
  \protect\acs{SRU5MR}$_{1996}$ $-$ \protect\acs{SRU5MR}$_{1971}$ & [$-$0.3, 0.17] & 0.3 \\
  \protect\acs{OLS} slope (\acs{SRU5MR} $\sim$ year) &  [$-$0.0097, 0.0061] & 0.32\\


  \hline
\end{tabular}
\end{table}

\begin{figure}[tbph]
  \centering
\subfloat[]{
  \GinKARWidth{0.5\textwidth}
\includegraphics{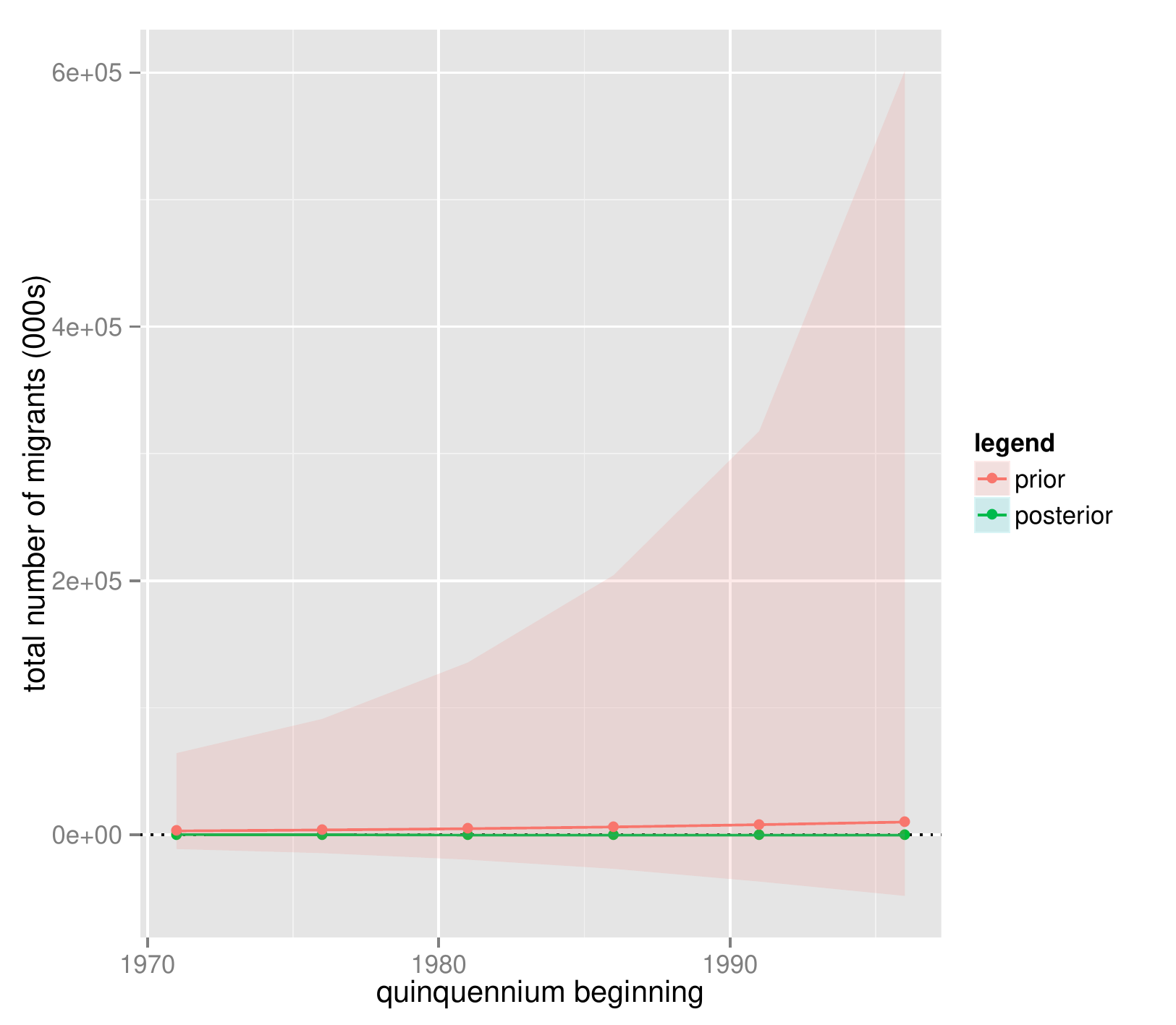}
\label{fig:indi-res-totmig-female} 
} 
\subfloat[]{
\GinKARWidth{0.5\textwidth}
\includegraphics{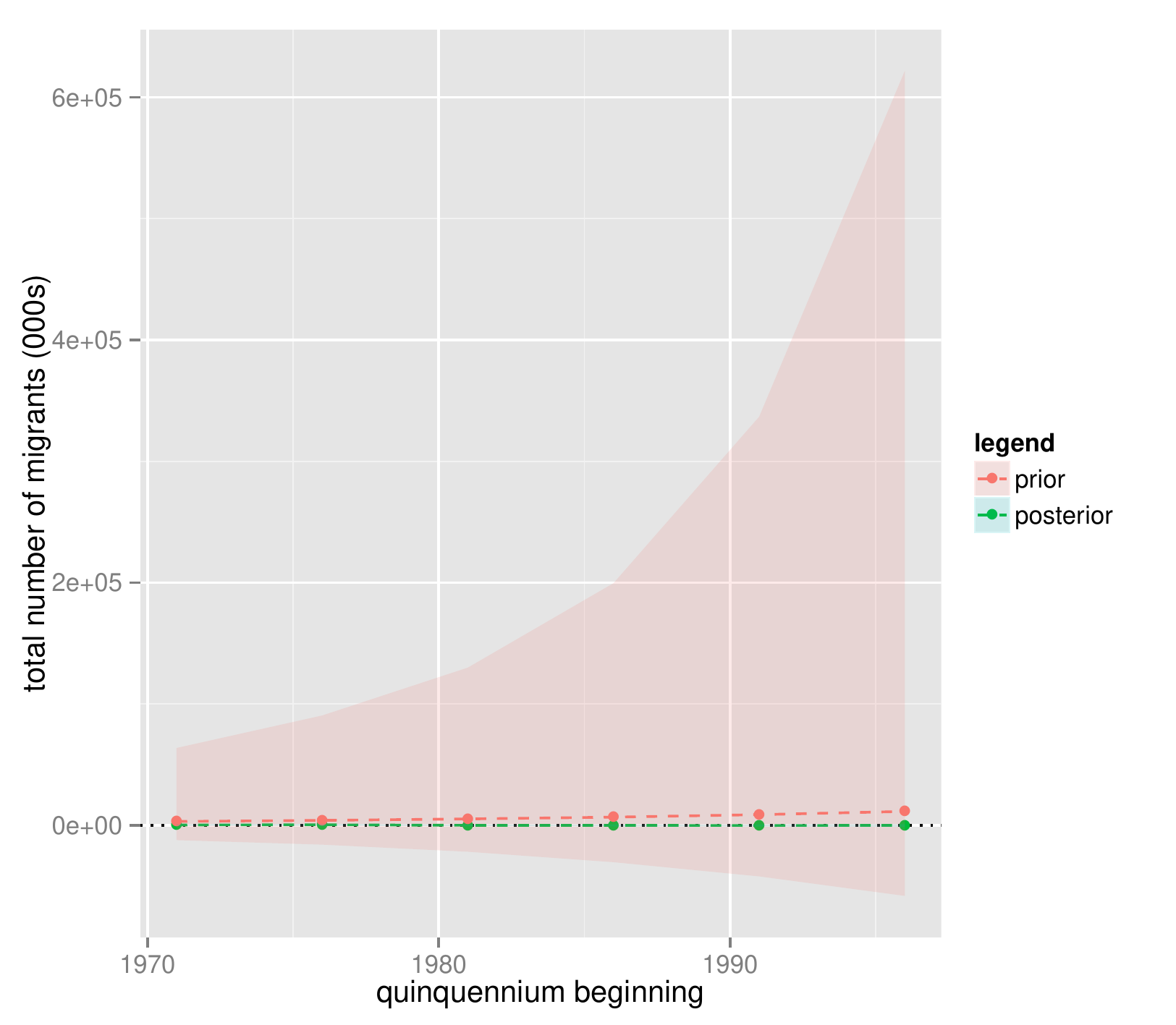}
\label{fig:indi-res-totmig-male} 
}\\ 
\subfloat[]{
\GinKARWidth{0.5\textwidth}
\includegraphics{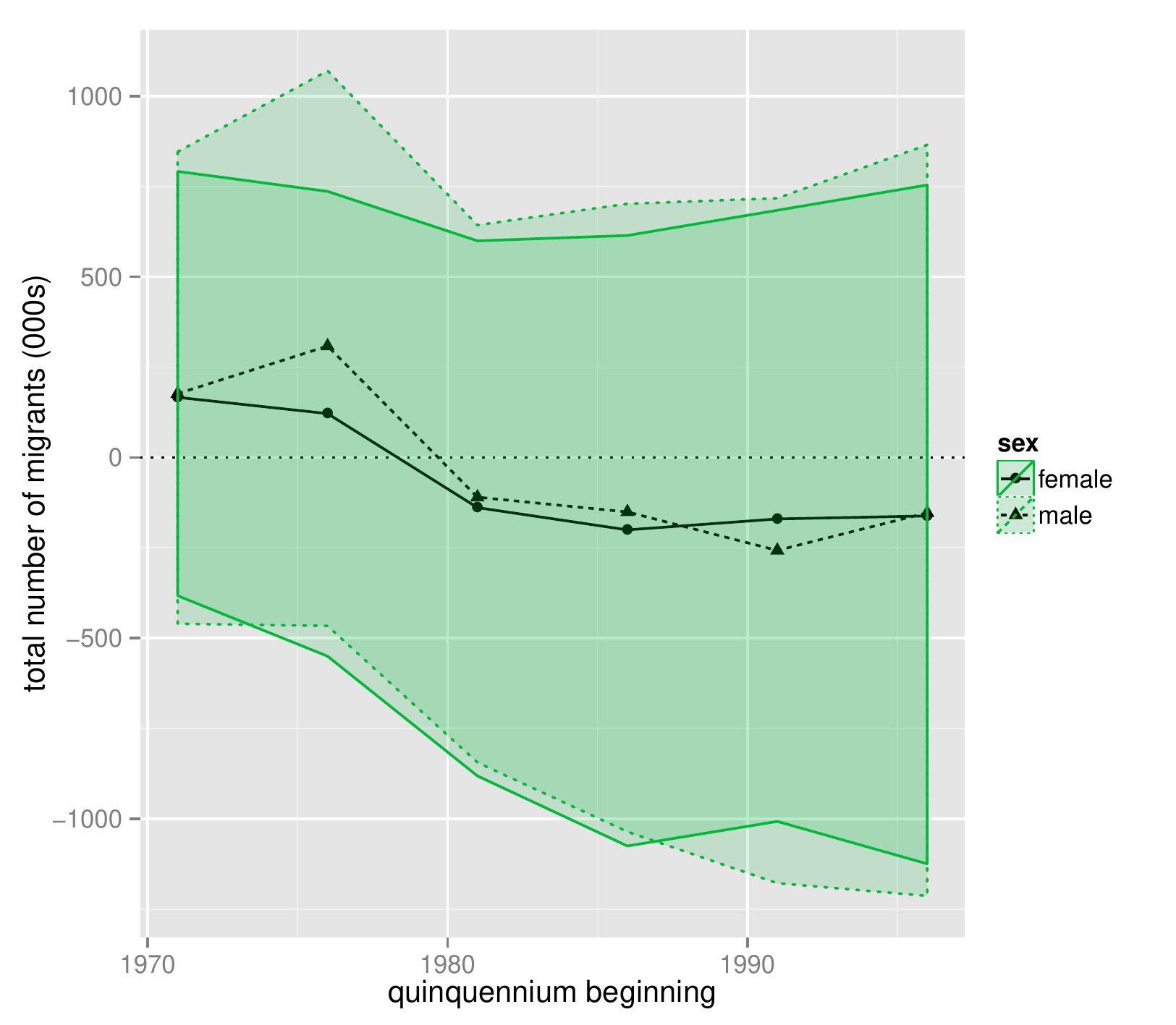}
\label{fig:indi-res-totmig-femaleVSmale} 
}
\subfloat[]{
\GinKARWidth{0.5\textwidth}
\includegraphics{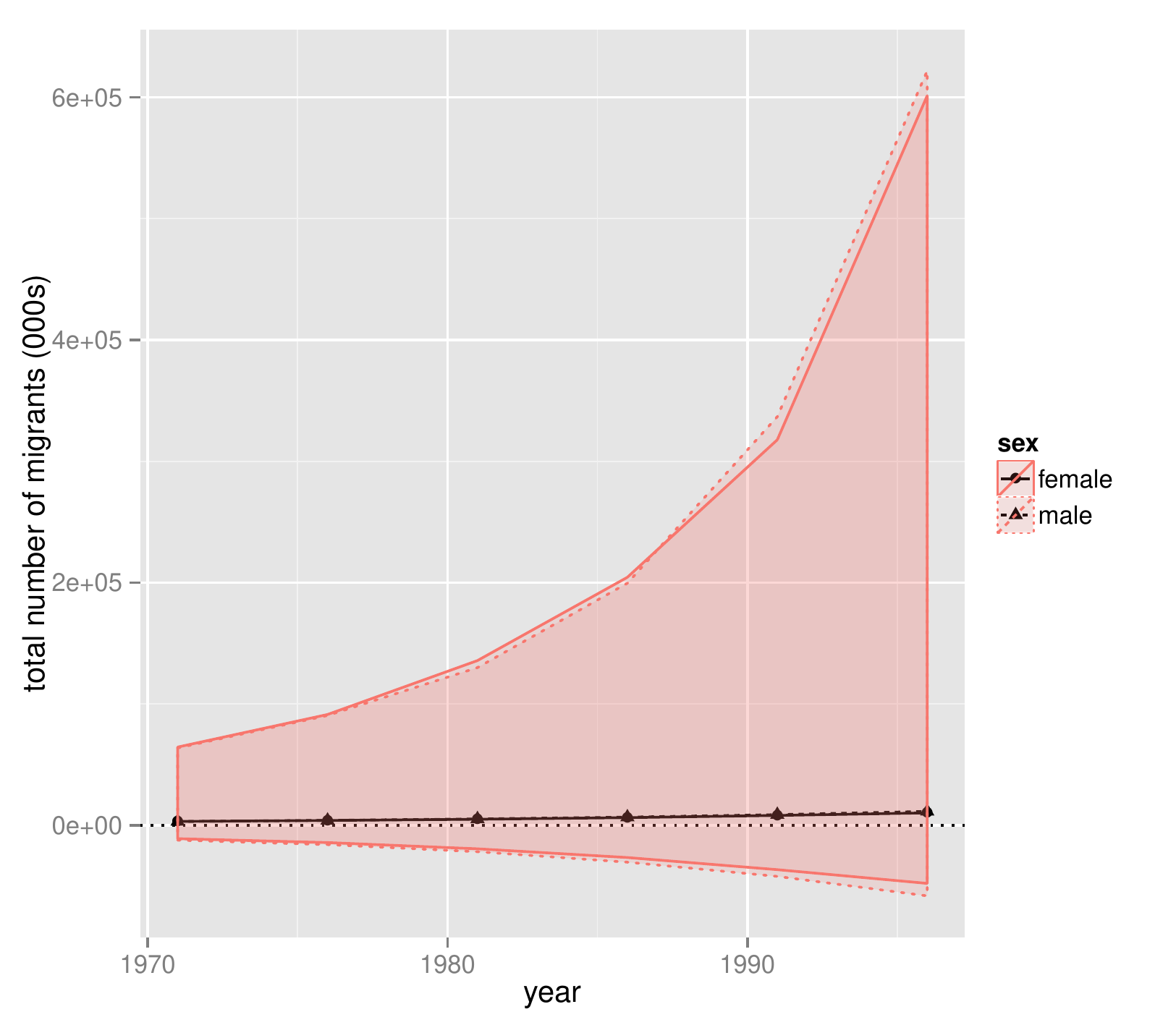}

\label{fig:indi-res-totmig-femaleVSmale-prior} 
}
  \caption{Prior and posterior medians and 95 percent credible intervals for the average annual net number of migrants for the reconstructed population of India, 1985--2004. \protect\subref{fig:indi-res-totmig-female}: Females; \protect\subref{fig:indi-res-totmig-male}: Males; \protect\subref{fig:indi-res-totmig-femaleVSmale}: Female and male posterior quantiles only; \protect\subref{fig:indi-res-totmig-femaleVSmale-prior}: Female and male prior quantiles only.}
  \label{fig:indi-res-totmig}
\end{figure}

Results for the average annual net number of migrants are in Figure~\ref{fig:indi-res-totmig}. The mean posterior half-width is 801,000.

\GinTextWidth

\FloatBarrier
\subsection{Thailand}
\label{sec:thailand-3}

\begin{figure}[tbph]
  \centering
\subfloat[]{
\GinKARWidth{0.5\textwidth}
\includegraphics{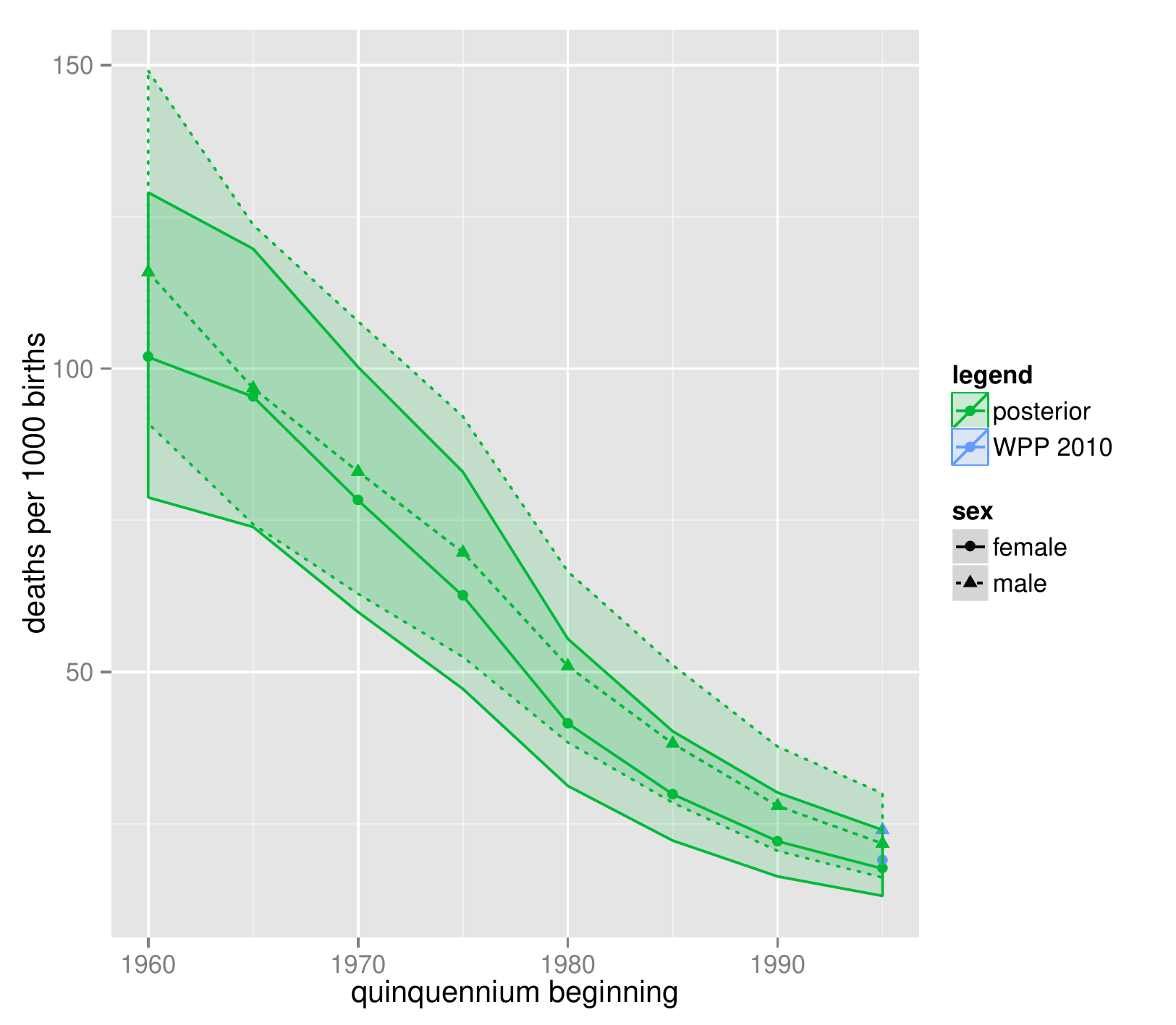}
\label{fig:thai-res-IMR-femaleVSmale} 
}
\subfloat[]{
\GinKARWidth{0.5\textwidth}
\includegraphics{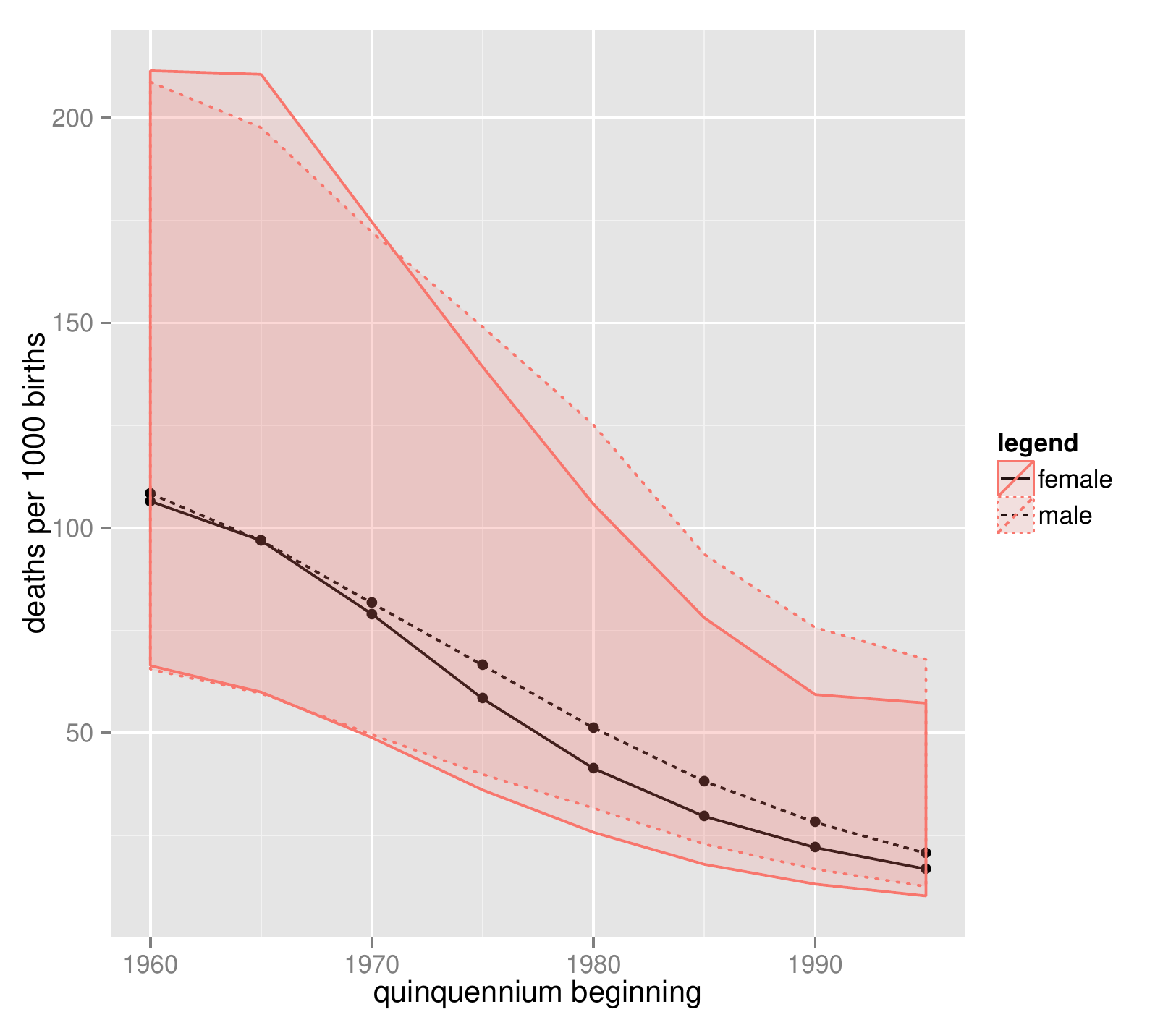}
\label{fig:thai-res-IMR-femaleVSmale-prior} 
}\\ 
\subfloat[]{
\GinKARWidth{0.5\textwidth}
\includegraphics{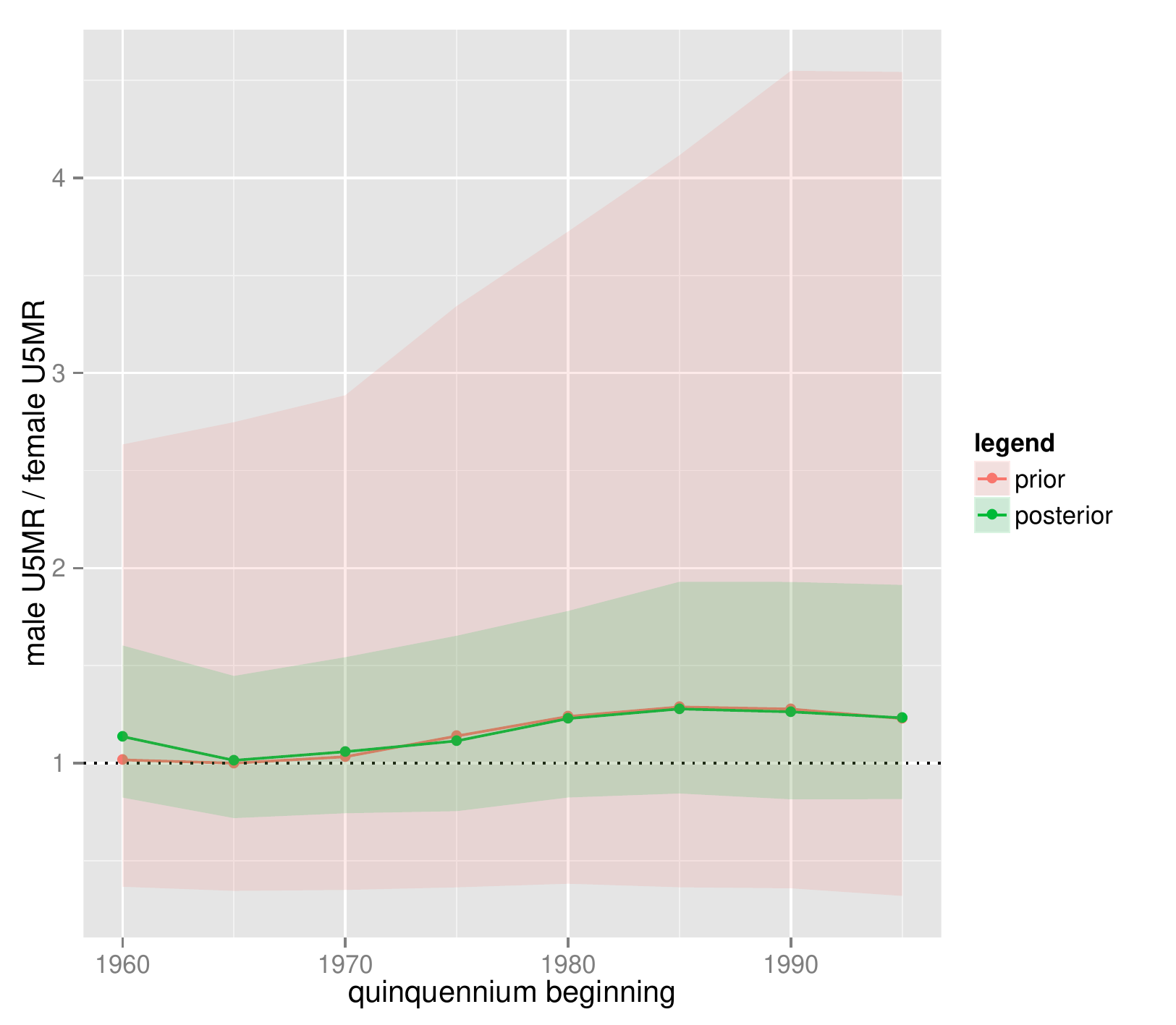}
\label{fig:thai-res-IMR-female-male-ratio} 
}
  \caption{Prior and posterior medians and 95 percent credible intervals for \protect\acf{U5MR} (deaths per 1,000 live births) for the reconstructed population of Thailand, 1960--2000 \protect\subref{fig:thai-res-IMR-femaleVSmale}: Female and male posterior quantiles and \protect\acf{WPP} 2010 estimates; \protect\subref{fig:thai-res-IMR-femaleVSmale-prior}: Female and male prior quantiles only; \protect\subref{fig:thai-res-IMR-female-male-ratio}: Male-to-female ratio.}
  \label{fig:thai-res-IMR}
\end{figure}

Results for \ac{SRU5MR} are shown in Figure~\ref{fig:thai-res-IMR}. The posterior suggests mortality at ages 0--5 was similar for both sexes. The 95 percent credible interval is centred above one for the period of reconstruction (mean half-width %
0.47) %
and the probabilities that the male-to-female ratios of \ac{U5MR} were less than one are small (Table~\ref{tab:thai-prob-sex-ratio-U5MR-gt-1}). Based on the measures of change over time used above, there is no evidence for a strong trend in this parameter over the period of reconstruction (Table~\ref{tab:thai-prob-dec-lin-sru5mr}).

\GinTextWidth

\begin{table}[tpbh]
  \centering
  \caption{Probability that \protect\acf{SRU5MR} was less than one for the reconstructed population of Thailand, 1971--2001, by quinquennium.}
  \label{tab:thai-prob-sex-ratio-U5MR-gt-1}
\begin{tabular}{rrrrrrrr}
  \hline
  1960 & 1965 & 1970 & 1975 & 1980 & 1985 & 1990 & 1995 \\\hline
 0.23 & 0.46 & 0.38 & 0.29 & 0.15 & 0.12 & 0.13 & 0.16 \\
   [1ex]\hline
\end{tabular}
\end{table}

\begin{table}[tbph]
  \centering
  \caption{Probabilities of an increasing linear trend and 95 percent credible intervals for \protect\acf{SRU5MR} for the reconstructed population of Thailand, 1960--2000. Two measures of trend are used: the difference over the period of reconstruction and the slope coefficient from the \protect\acf{OLS} regression on the start year of each quinquennium.}
  \label{tab:thai-prob-dec-lin-sru5mr}
  \begin{tabular}{lrr}
  \hline
  Measure of trend  &  95 percent CI  &  Prob $>$ 0 \\\hline
  \protect\acs{SRU5MR}$_{1996}$ $-$ \protect\acs{SRU5MR}$_{1971}$ & [$-$0.54, 0.84] & 0.61 \\
  \protect\acs{OLS} slope (\acs{SRU5MR} $\sim$ year) &  [$-$0.0075, 0.022] & 0.81\\
  \hline
\end{tabular}
\end{table}

\begin{figure}[tbph]
  \centering
\subfloat[]{
  \GinKARWidth{0.5\textwidth}
\includegraphics{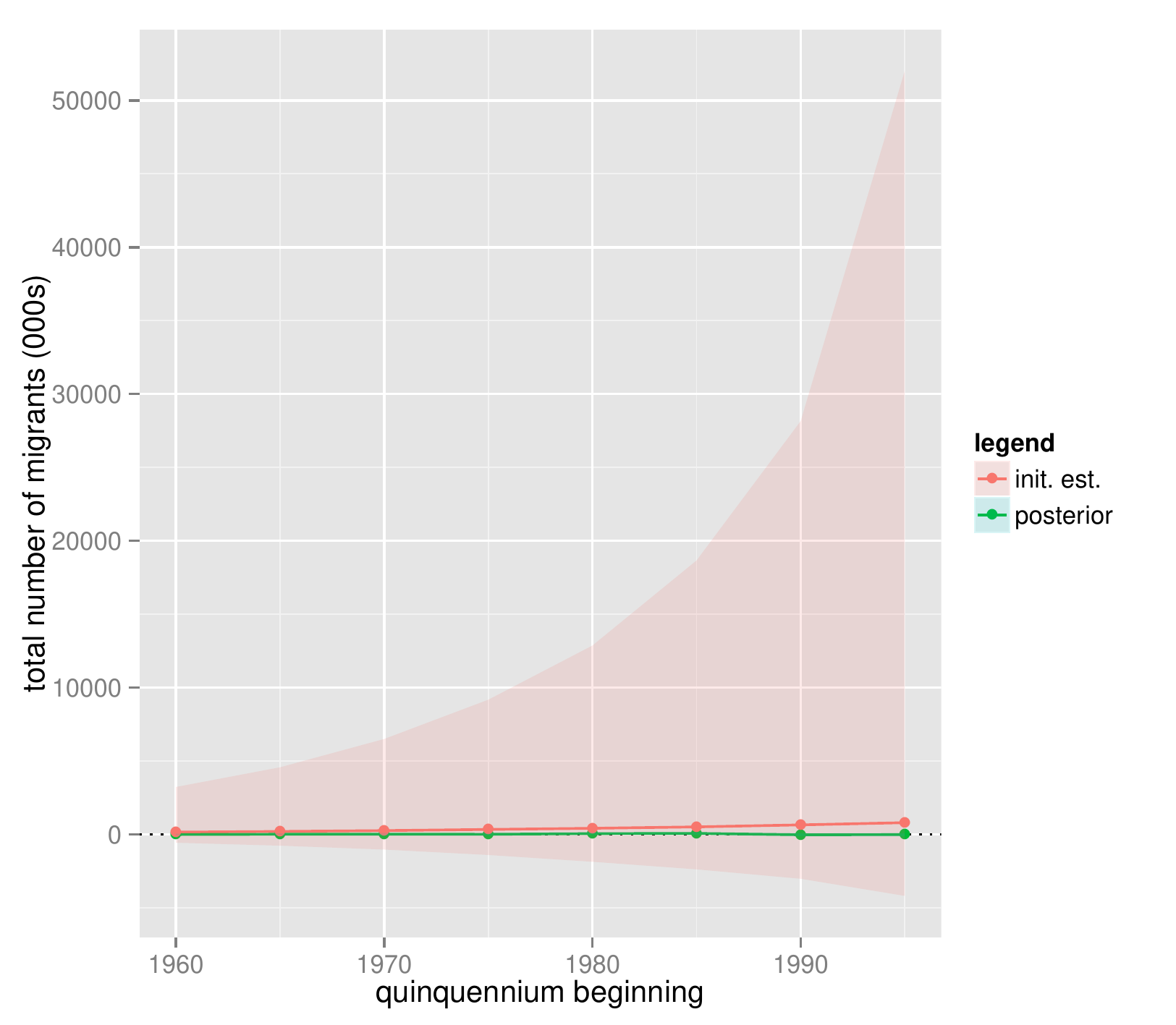}
\label{fig:thaiF2-res-totmig-female} 
} 
\subfloat[]{
\GinKARWidth{0.5\textwidth}
\includegraphics{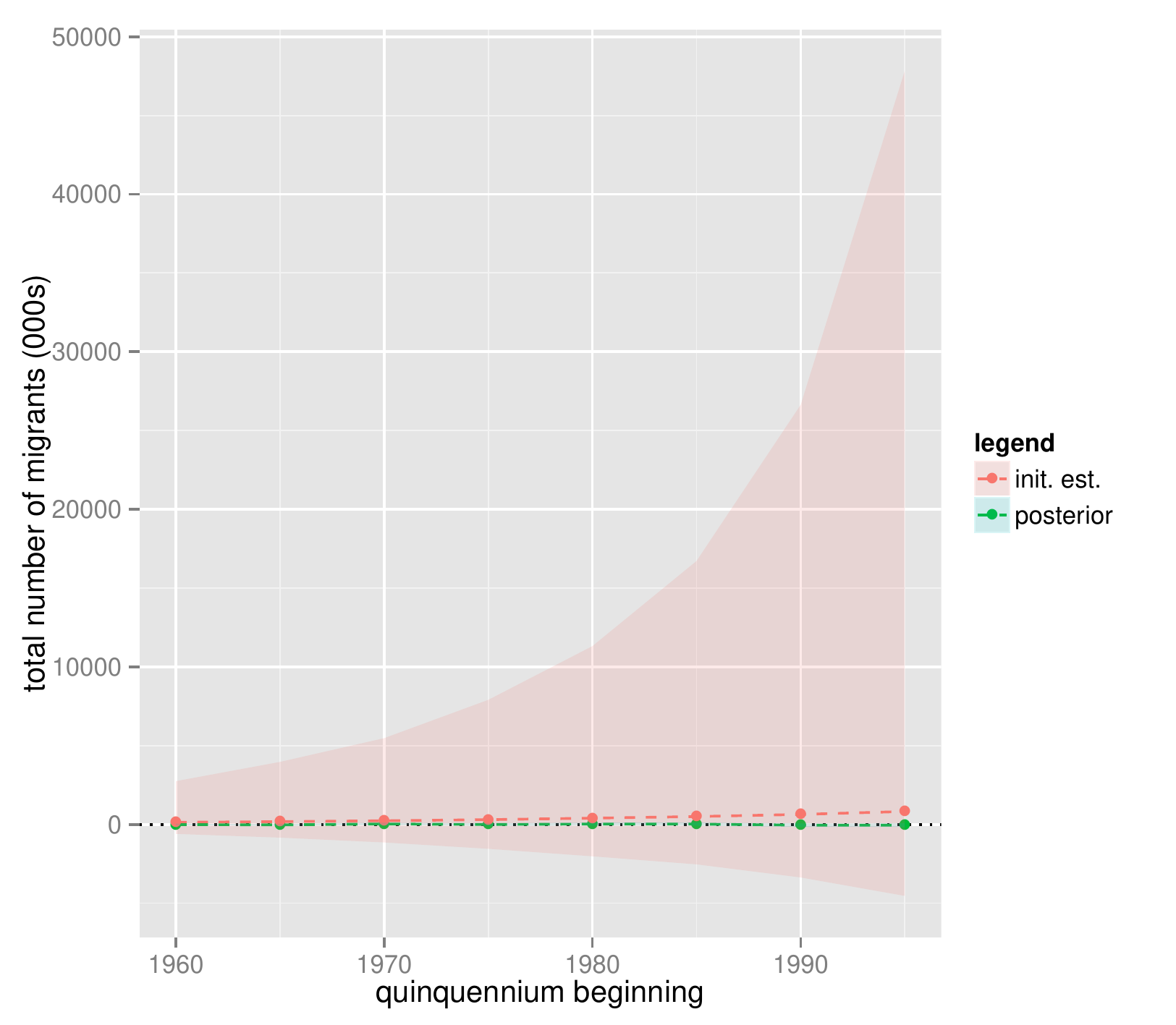}
\label{fig:thaiF2-res-totmig-male} 
}\\ 
\subfloat[]{
\GinKARWidth{0.5\textwidth}
\includegraphics{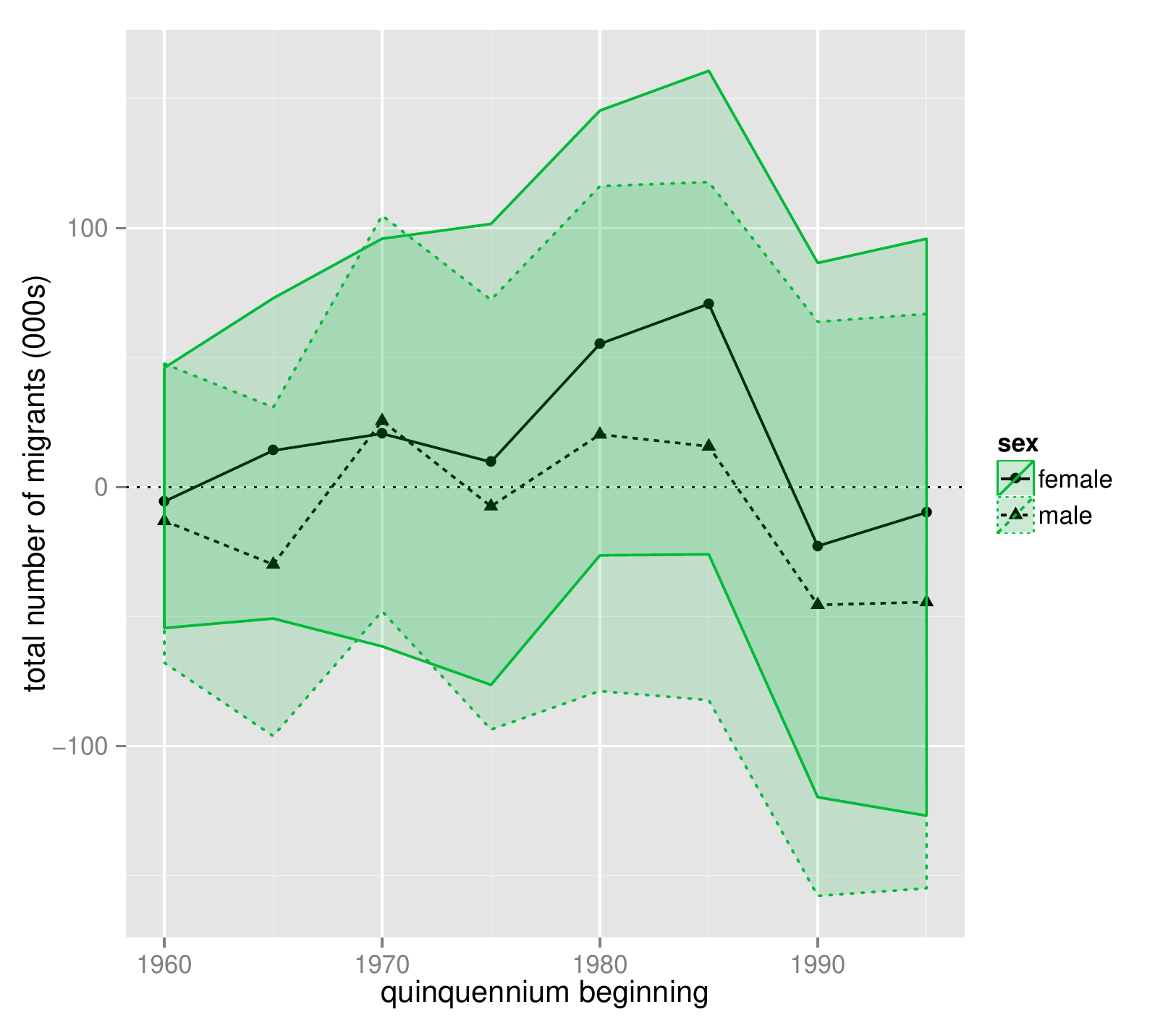}
\label{fig:thaiF2-res-totmig-femaleVSmale} 
}
\subfloat[]{
\GinKARWidth{0.5\textwidth}
\includegraphics{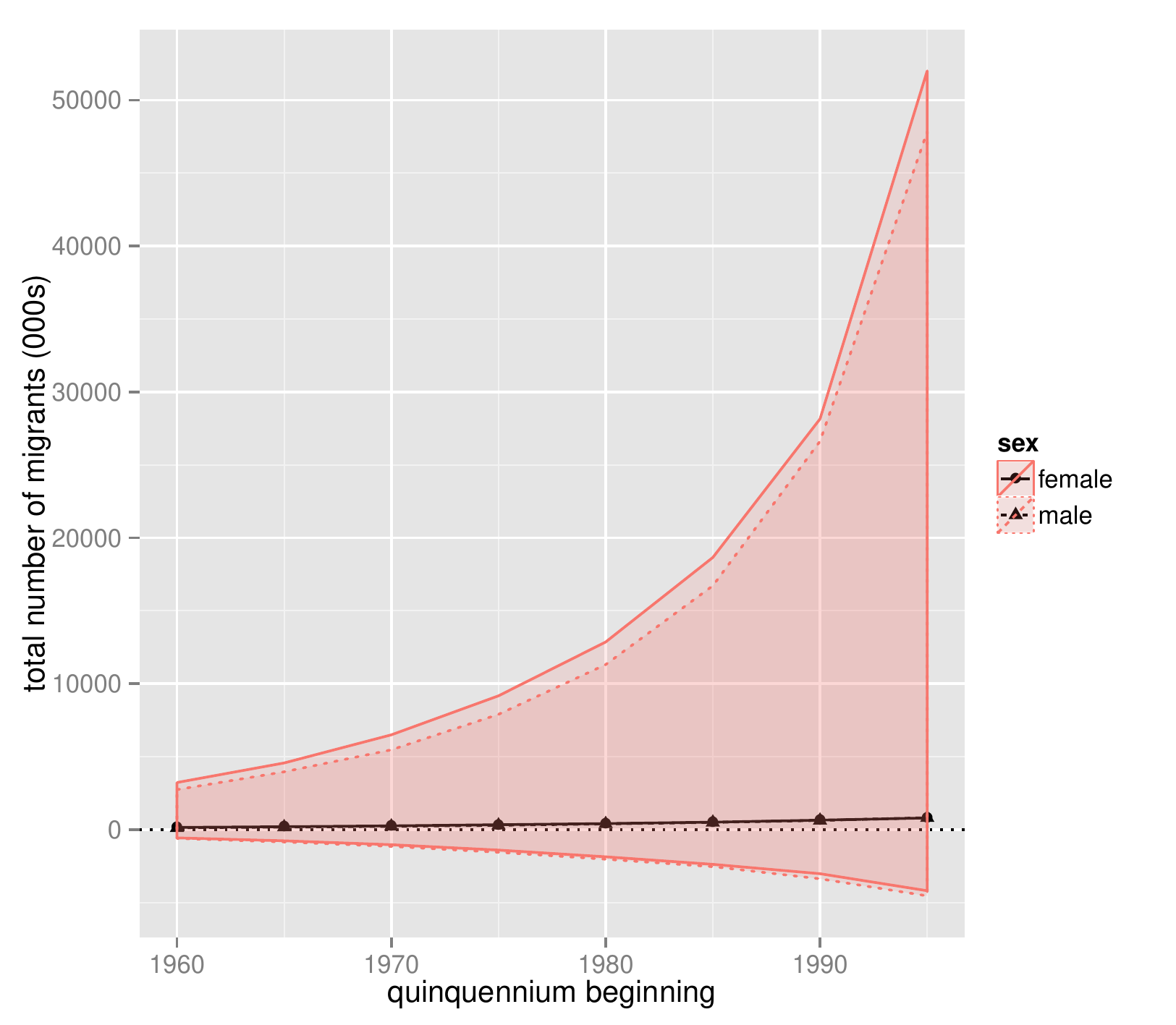}
\label{fig:thai-res-totmig-femaleVSmale-prior} 
}
  \caption{Prior and posterior medians and 95 percent credible intervals for the average annual net number of migrants for the reconstructed population of Thailand, 1985--2004. \protect\subref{fig:thaiF2-res-totmig-female}: Females; \protect\subref{fig:thaiF2-res-totmig-male}: Males; \protect\subref{fig:thaiF2-res-totmig-femaleVSmale}: Female and male posterior quantiles only; \protect\subref{fig:thai-res-totmig-femaleVSmale-prior}: Female and male prior quantiles only.}
  \label{fig:thaiF2-res-totmig}
\end{figure}

Results for the average annual net number of migrants are in Figure~\ref{fig:thaiF2-res-totmig}. The mean posterior half-width is %
85,800.

\GinTextWidth

\begin{figure}[tbph]
  \centering
\subfloat[]{
  \GinKARWidth{0.5\textwidth}
\includegraphics{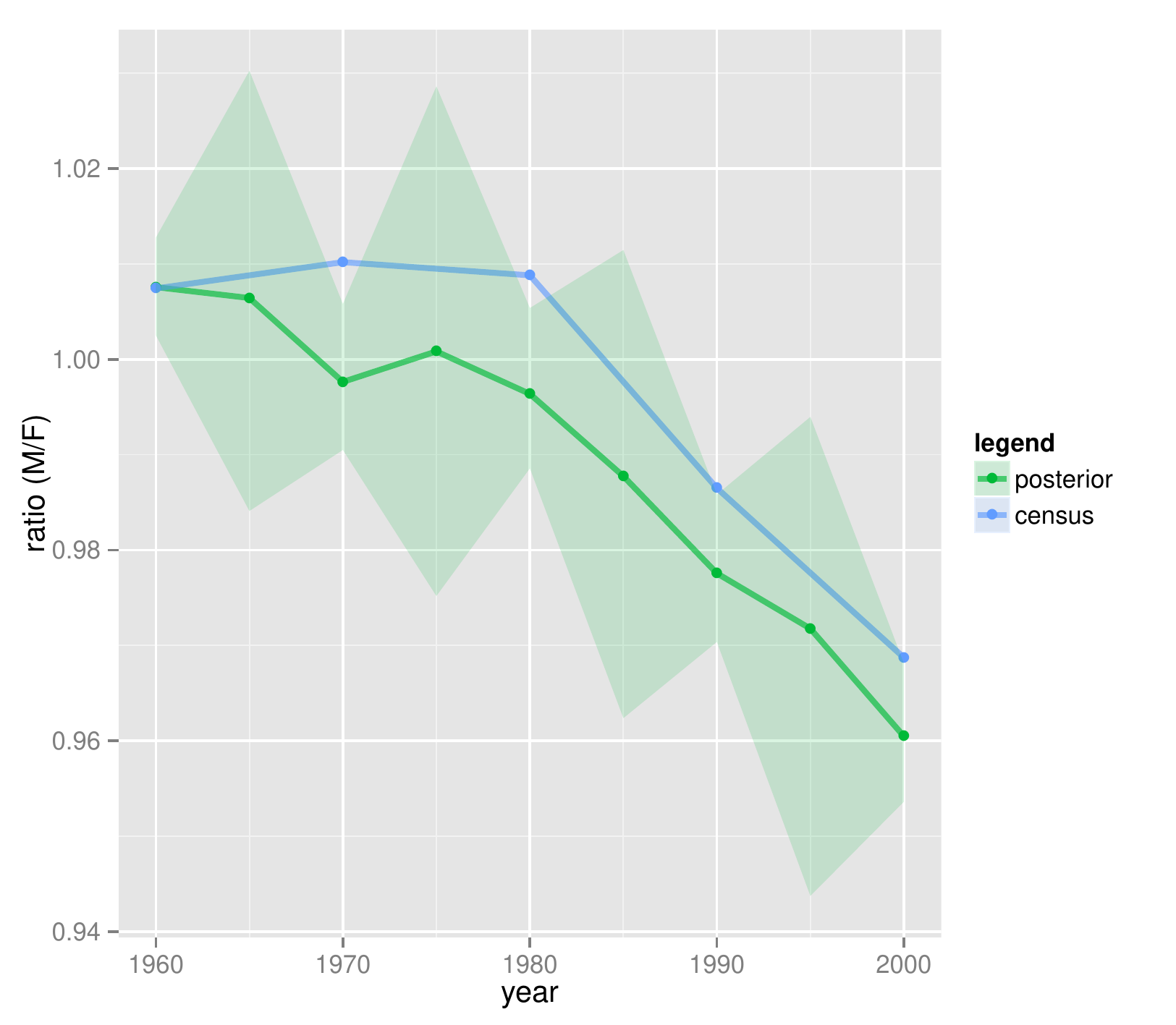}
\label{fig:thai-res-prtp}       
}
\subfloat[]{
  \GinKARWidth{0.5\textwidth}
\includegraphics{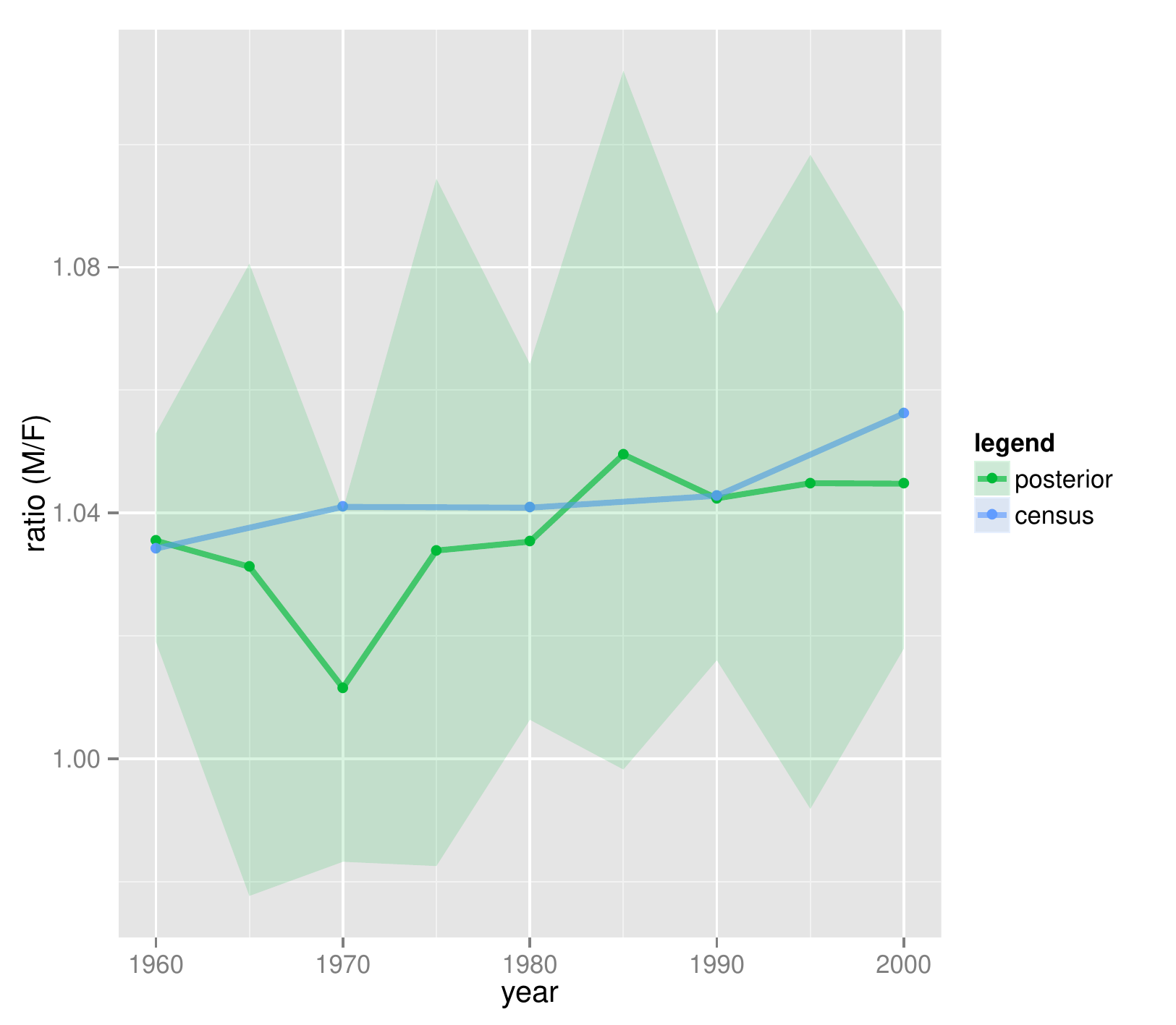}
\label{fig:thai-res-pr05} 
}

  \caption{Prior and posterior medians and 95 percent credible intervals for the reconstructed population of Thailand, 1960--2000. Sex ratios (male/female) for \protect\subref{fig:thai-res-prtp}: total population; \protect\subref{fig:thai-res-pr05}: population aged 0--5; 
.}
  \label{fig:thai-res-prtp-pr05}
\end{figure}

Population sex ratios are shown in Figure~\ref{fig:thai-res-prtp-pr05}. Posterior medians follow the ratios in the \ac{WPP} counts relatively closely.

\GinTextWidth

\FloatBarrier
\subsection{Laos}
\label{sec:laos-3}

\begin{figure}[tbph]
  \centering
\subfloat[]{
\GinKARWidth{0.5\textwidth}
\includegraphics{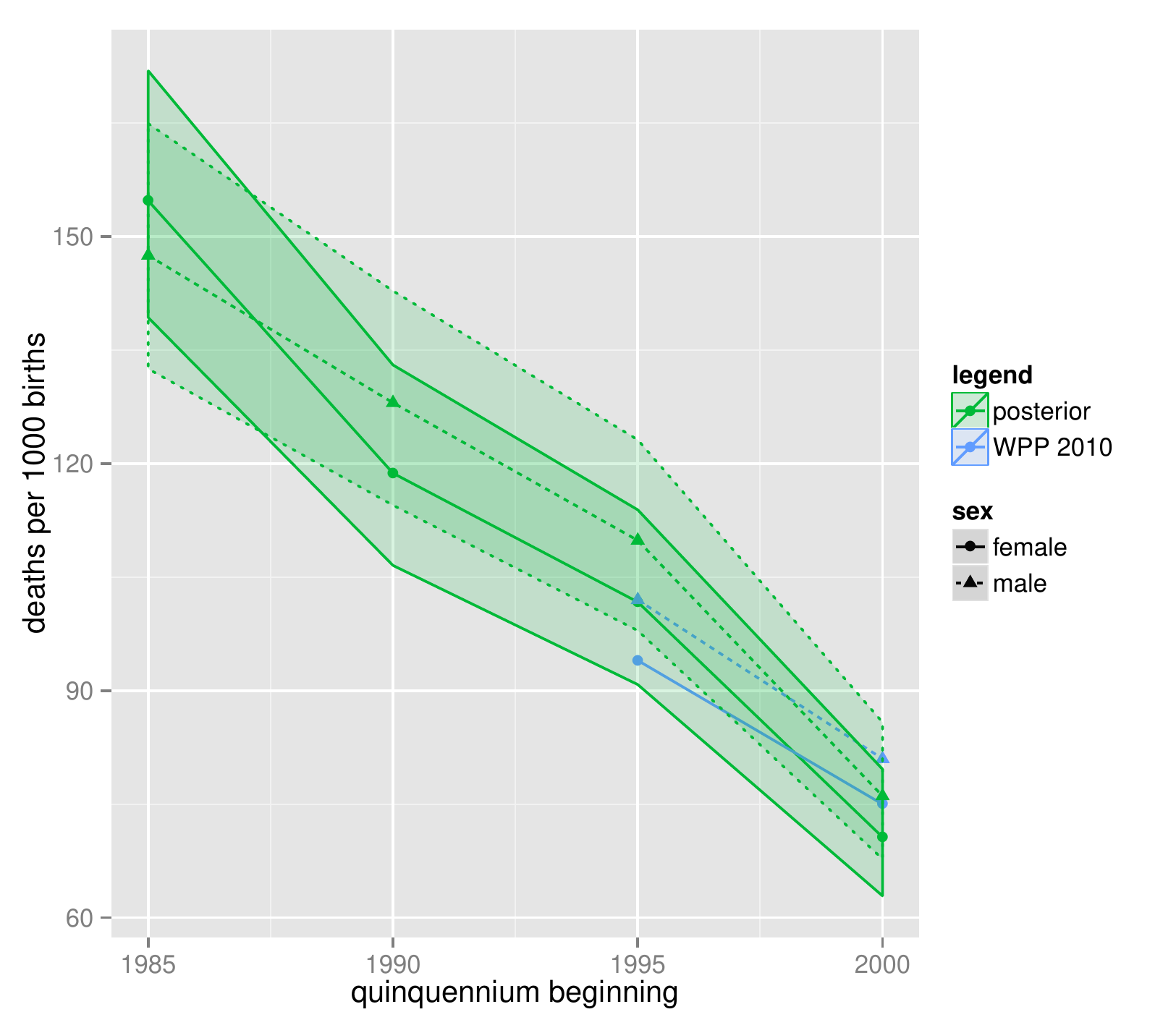}
\label{fig:laos-res-IMR-femaleVSmale} 
}
\subfloat[]{
\GinKARWidth{0.5\textwidth}
\includegraphics{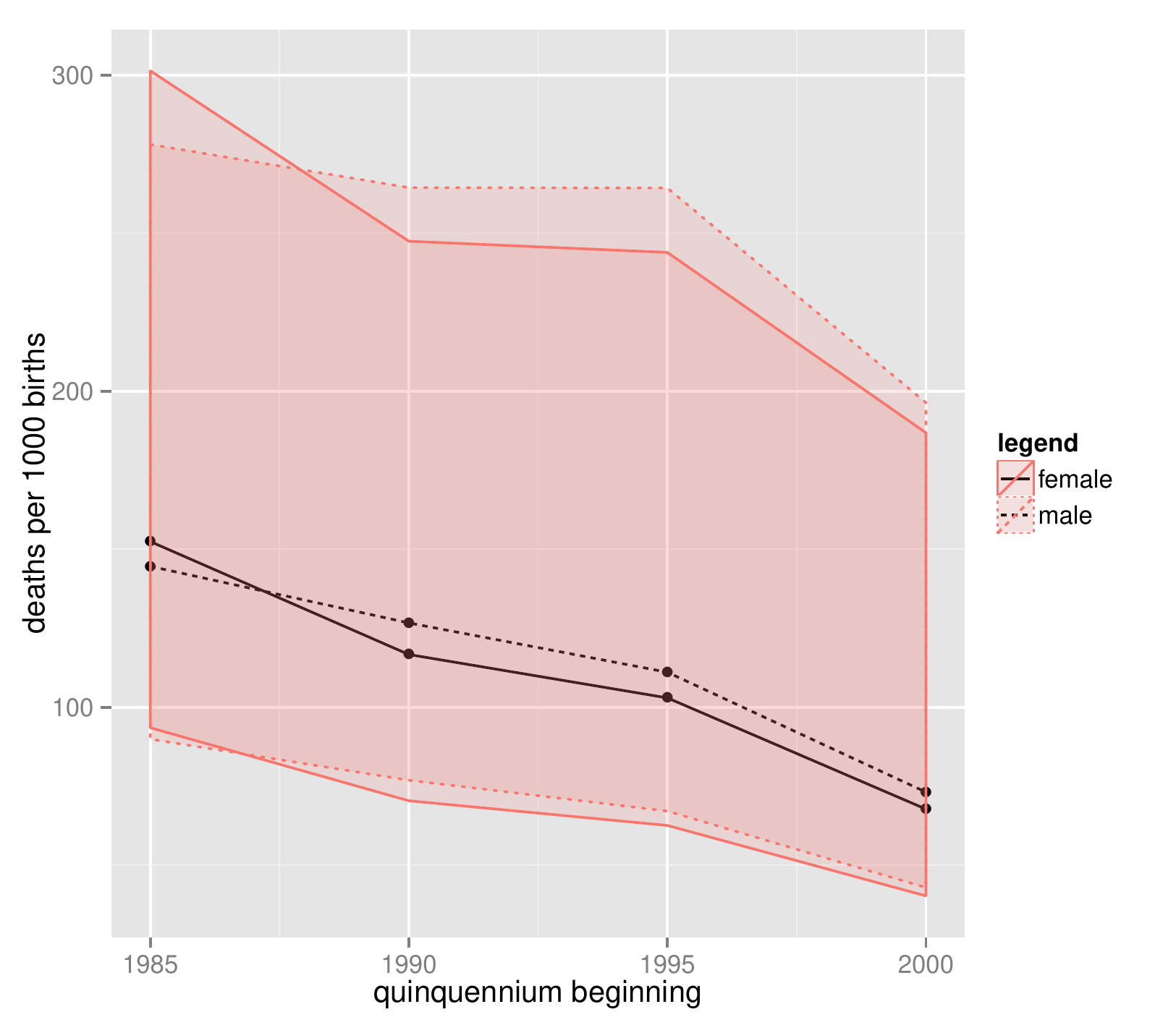}
\label{fig:laos-res-IMR-femaleVSmale-prior} 
}\\ 
\subfloat[]{
\GinKARWidth{0.5\textwidth}
\includegraphics{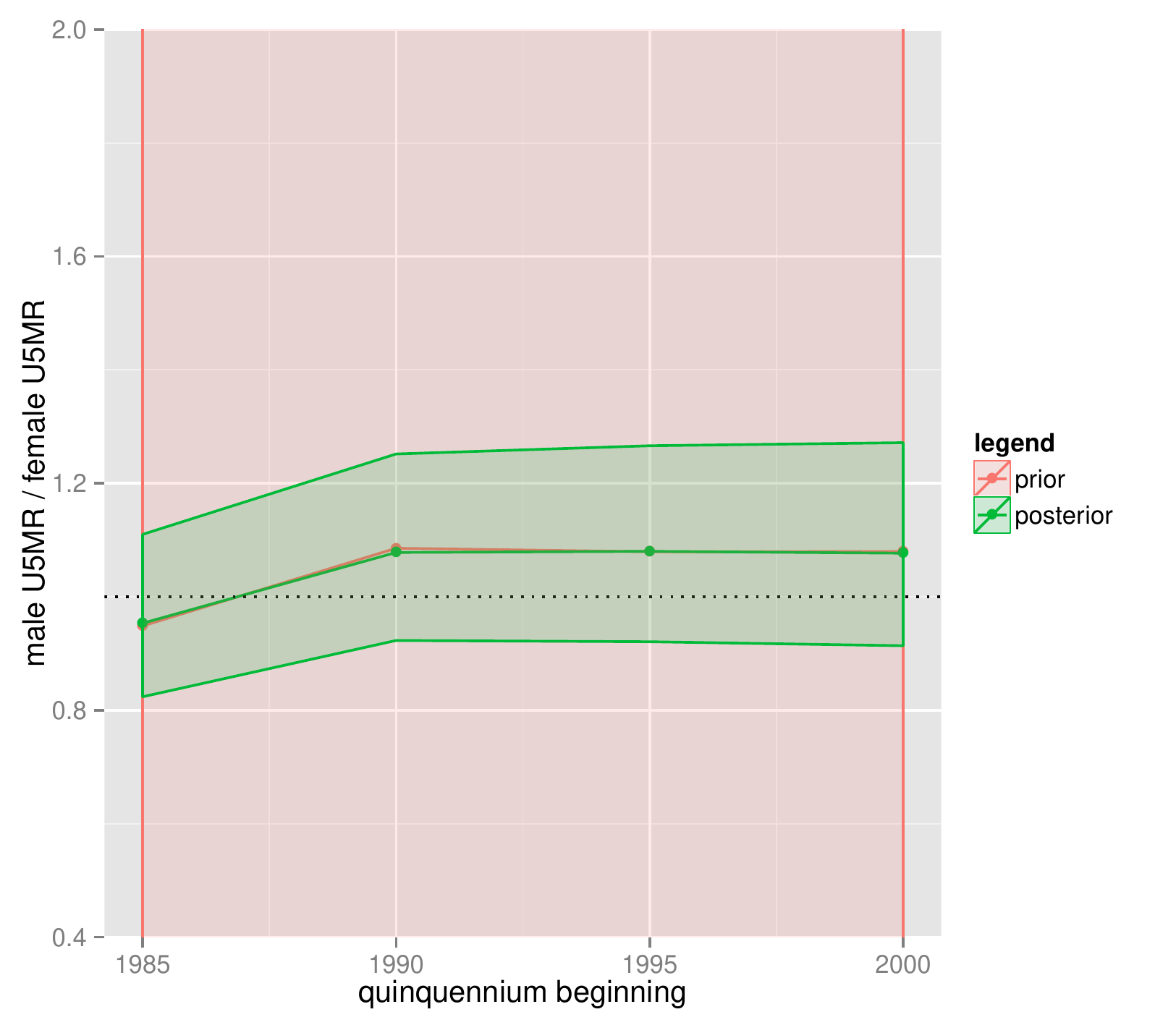}
\label{fig:laos-res-IMR-female-male-ratio} 
}
  \caption{Prior and posterior medians and 95 percent credible intervals for \protect\acf{U5MR} (deaths per 1,000 live births) for the reconstructed population of Laos, 1985--2004. \protect\subref{fig:laos-res-IMR-femaleVSmale}: Female and male posterior quantiles and \protect\acf{WPP} 2010 estimates; \protect\subref{fig:laos-res-IMR-femaleVSmale-prior}: Female and male prior quantiles only; \protect\subref{fig:laos-res-IMR-female-male-ratio}: Male-to-female ratio.}
  \label{fig:laos-res-IMR}
\end{figure}

The results for \ac{U5MR} are shown in Figure~\ref{fig:laos-res-IMR}. Posterior median estimates of \ac{U5MR} decreased. The posterior intervals for \ac{U5MR} straddle one for the entire period of reconstruction (Figure~\ref{fig:laos-res-IMR-female-male-ratio}; mean half-width of the ratio %
0.16). %
The probability that the male-to-female ratio of \ac{U5MR} was less than one was %
0.77 %
in 1985 but less than %
0.2 %
in all other periods (Table~\ref{tab:laos-prob-sex-ratio-U5MR-gt-1}). The probabilities of an increasing linear trend are given in Table~\ref{tab:laos-prob-dec-lin-sru5mr}.

\GinTextWidth

\begin{table}[bpth]
  \centering
  \caption{Probability that \protect\acf{SRU5MR} was less than one for the reconstructed population of Loas, 1985--2005, by quinquennium.}
  \label{tab:laos-prob-sex-ratio-U5MR-gt-1}
\begin{tabular}{rrrr}
  \hline
  1985 & 1990 & 1995 & 2005 \\\hline






 0.77 & 0.14 & 0.14 & 0.16 \\
   [1ex]\hline
\end{tabular}
\end{table}

\begin{table}[tbph]
  \centering
  \caption{Probabilities of an increasing linear trend and 95 percent credible intervals for 
    \protect\acf{SRU5MR}
    for the reconstructed population of Laos, 1985--2000. Two measures of trend are used: the difference over the period of reconstruction and the slope coefficient from the \protect\acf{OLS} regression on the start year of each quinquennium.}
  \label{tab:laos-prob-dec-lin-sru5mr}
  \begin{tabular}{lrr}
  \hline
  Measure of trend  &  95 percent CI  &  Prob $>$ 0 \\\hline\\[-1.5ex]



  \protect\acs{SRU5MR}$_{2000}$ $-$ \protect\acs{SRU5MR}$_{1985}$ & [$-$0.1, 0.35] & 0.88 \\
  \protect\acs{OLS} slope (\acs{SRU5MR} $\sim$ year) &  [$-$0.0071, 0.022] & 0.87\\
  \hline
\end{tabular}
\end{table}

\begin{figure}[tbph]
  \centering
\subfloat[]{
  \GinKARWidth{0.5\textwidth}
\includegraphics{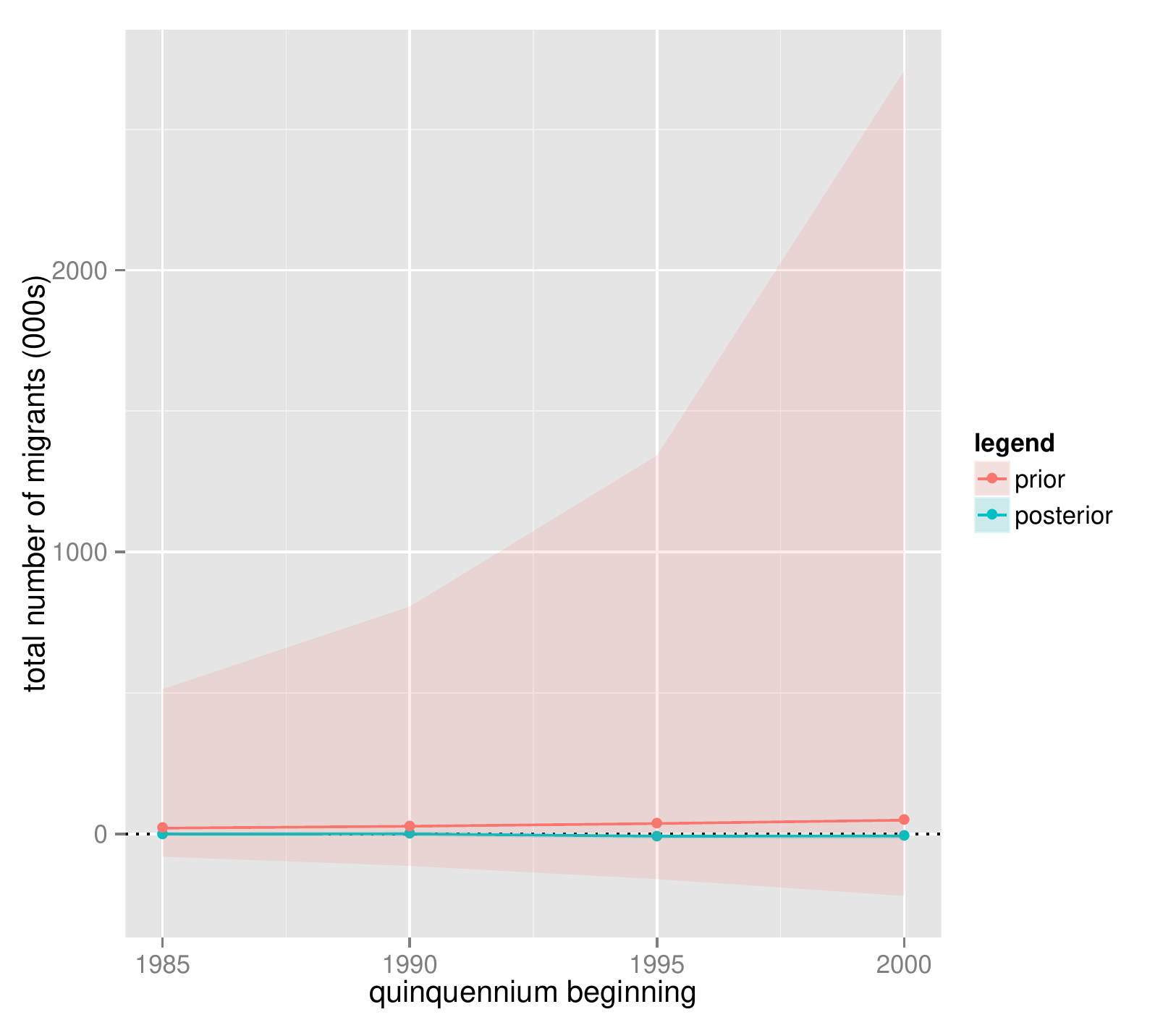}
\label{fig:laos-res-totmig-female} 
} 
\subfloat[]{
\GinKARWidth{0.5\textwidth}
\includegraphics{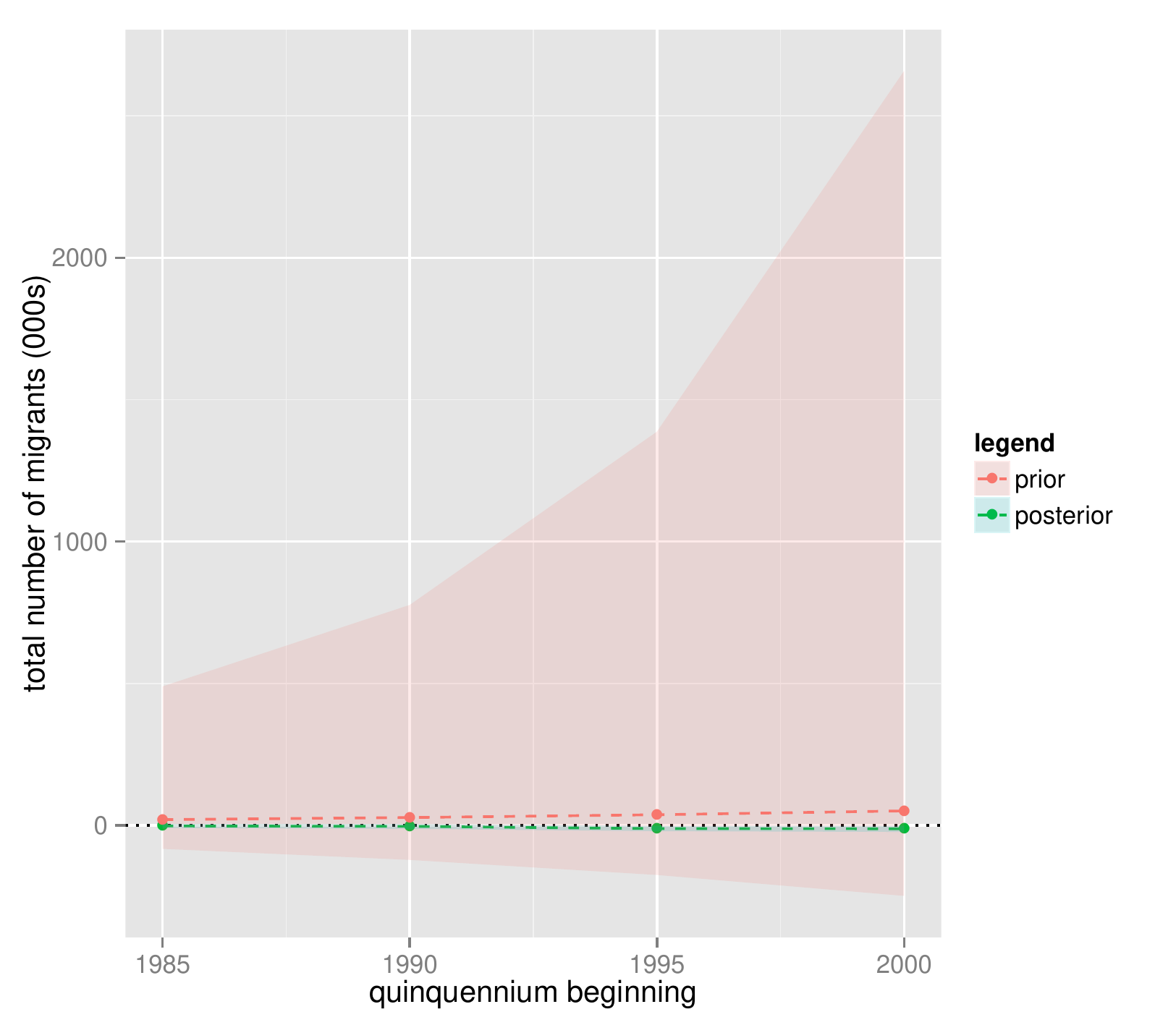}
\label{fig:laos-res-totmig-male} 
}\\ 
\subfloat[]{
\GinKARWidth{0.5\textwidth}
\includegraphics{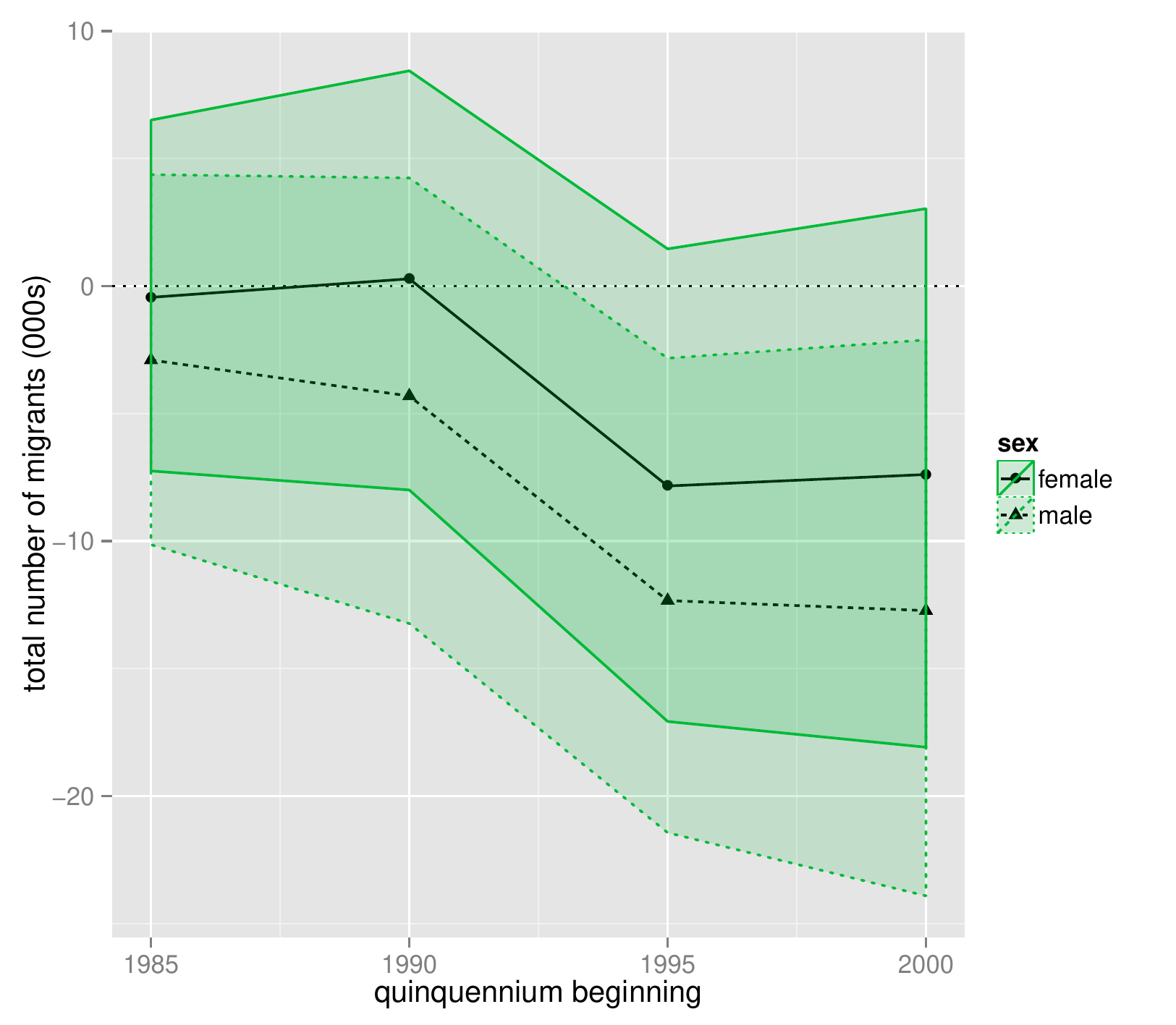}
\label{fig:laos-res-totmig-femaleVSmale} 
}
\subfloat[]{
\GinKARWidth{0.5\textwidth}
\includegraphics{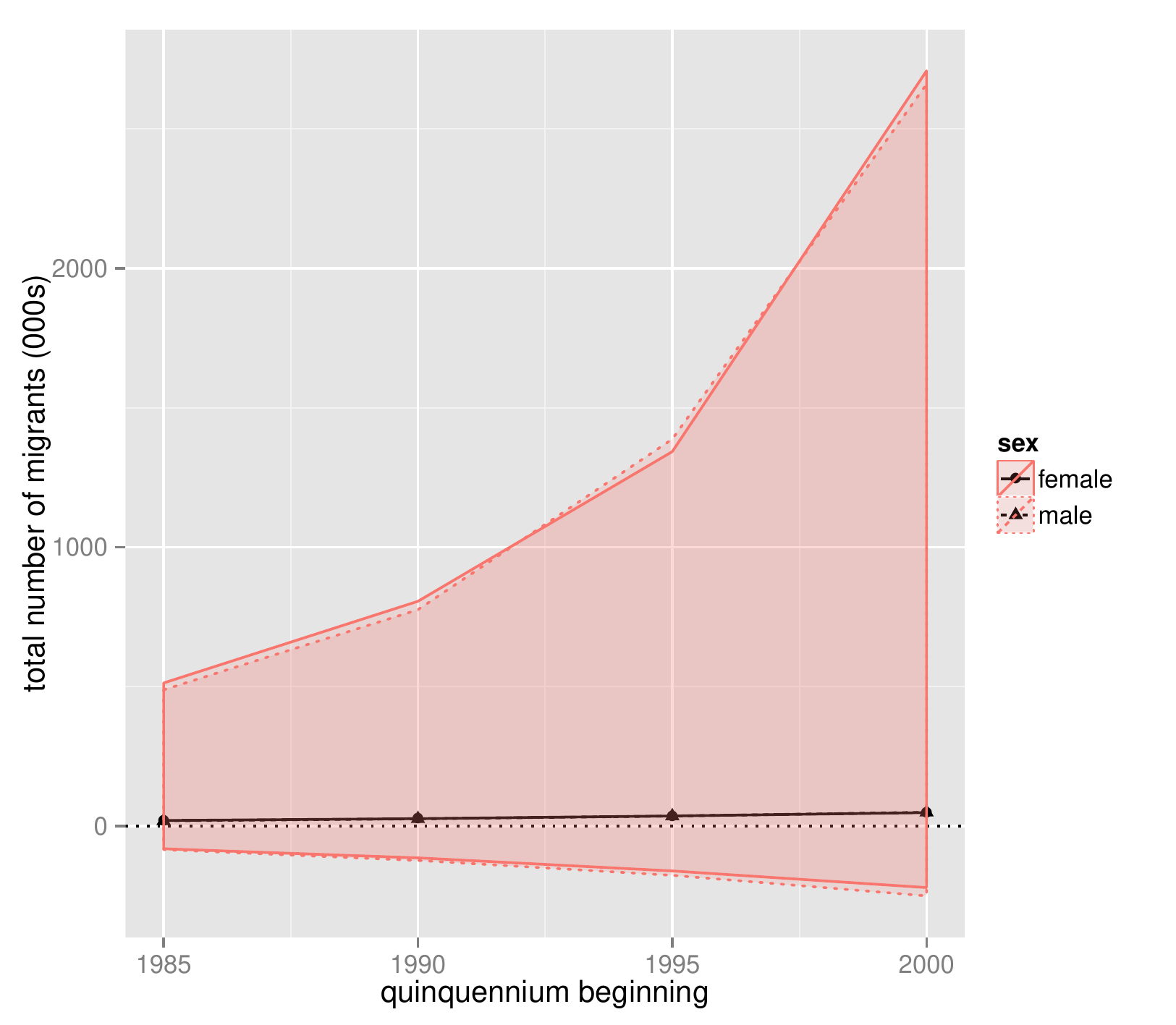}
\label{fig:laos-res-totmig-femaleVSmale-prior} 
}
  \caption{Prior and posterior medians and 95 percent credible intervals for the average annual net number of migrants for the reconstructed population of Laos, 1985--2004. \protect\subref{fig:laos-res-totmig-female}: Females; \protect\subref{fig:laos-res-totmig-male}: Males; \protect\subref{fig:laos-res-totmig-femaleVSmale}: Female and male posterior quantiles only; \protect\subref{fig:laos-res-totmig-femaleVSmale-prior}: Female and male prior quantiles only.}
  \label{fig:laos-res-totmig}
\end{figure}

Posterior distributions for the average annual net number of migrants are shown in Figure~\ref{fig:laos-res-totmig}. Posterior uncertainty is high, with a mean half-width of %
8,890. Posterior medians indicate out-migration for most of the period of reconstruction. For 1995--2005, posterior intervals for males fall completely below zero.

\GinTextWidth

\begin{figure}[tbph]
  \centering
\subfloat[]{
  \GinKARWidth{0.5\textwidth}
\includegraphics{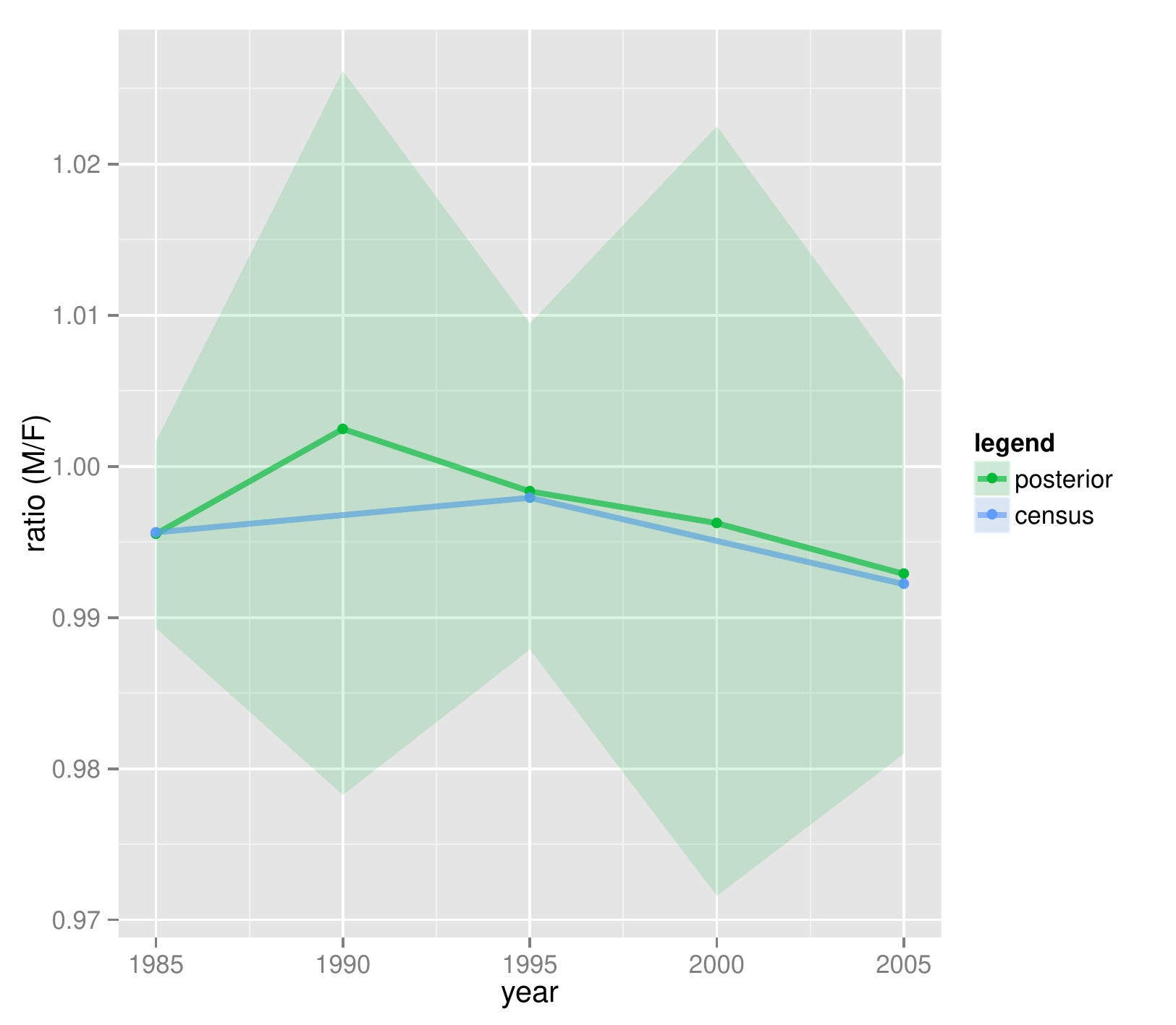}
\label{fig:laos-res-prtp}       
}
\subfloat[]{
  \GinKARWidth{0.5\textwidth}
\includegraphics{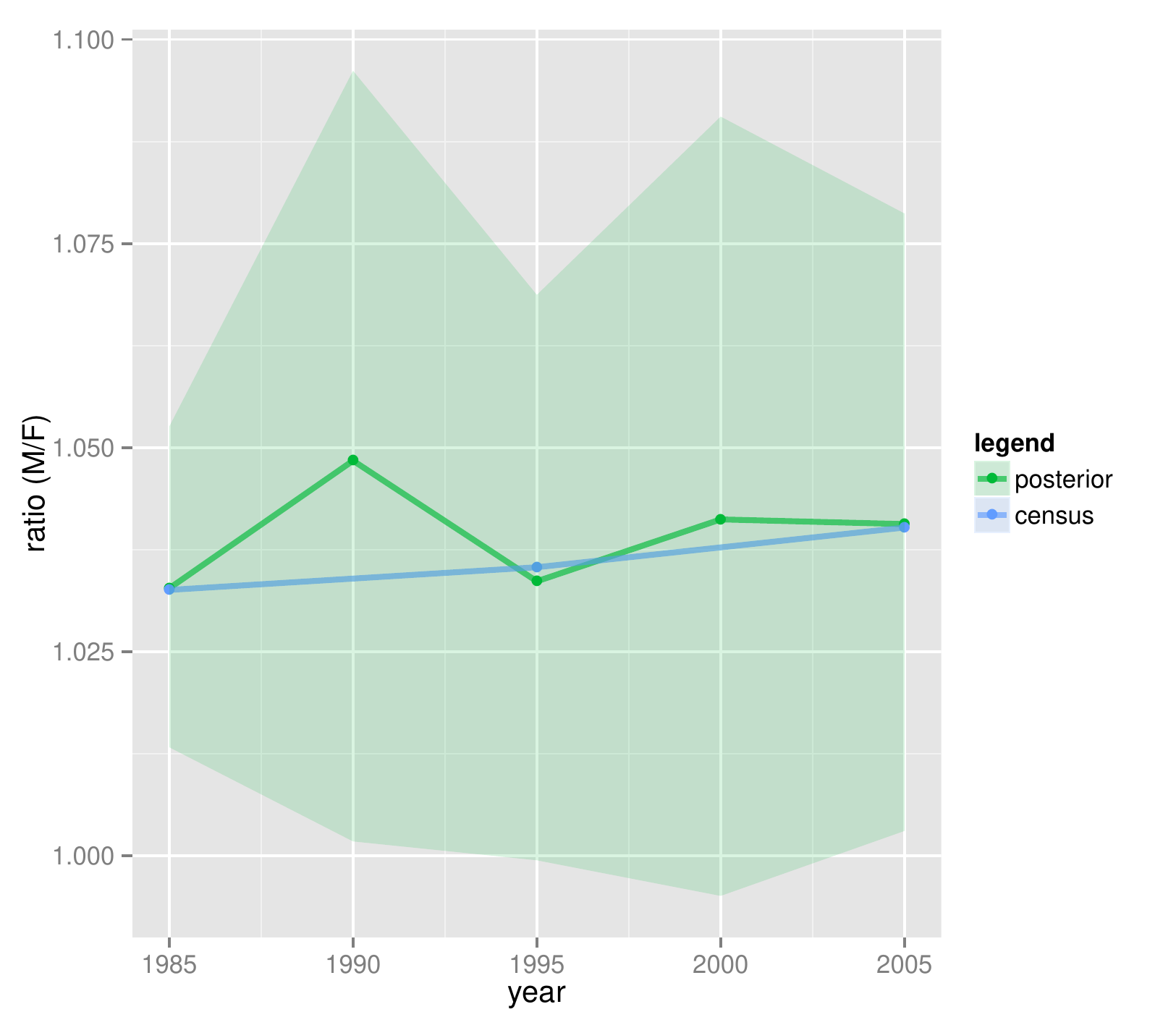}
\label{fig:laos-res-pr05}  
}

  \caption{Prior and posterior medians and 95 percent credible intervals for the reconstructed population of Laos, 1985--2005. Sex ratios (male/female) for \protect\subref{fig:laos-res-prtp}: total population; \protect\subref{fig:laos-res-pr05}: population aged 0--5; 
.}
  \label{fig:laos-res-prtp-pr05}
\end{figure}

Posterior intervals for population sex ratios are in Figure~\ref{fig:laos-res-prtp-pr05}. Posterior medians are reasonably similar to the ratios in the \ac{WPP} census counts and uncertainty is high.

\GinTextWidth


\clearpage

\setglossarysection{section}
\printglossary[title=Glossary,type=\acronymtype]

\clearpage
\printbibliography[heading=bibintoc] 

\end{document}